\documentstyle[aps]{revtex}
\begin{document}
\newcommand{\beq}{\begin{equation}}
\newcommand{\eeq}{\end{equation}}
\newcommand{\beqn}{\begin{eqnarray}}
\newcommand{\eeqn}{\end{eqnarray}}
\newcommand{\bmath}{\begin{mathletters}}
\newcommand{\emath}{\end{mathletters}}
\title{Optical sum rule violation, superfluid weight and condensation 
energy in the cuprates
}
\author{J. E. Hirsch$^{a}$ and F. Marsiglio$^{b}$ }
\address{$^{a}$Department of Physics, University of California, San Diego,
La Jolla, CA 92093-0319\\
$^{b}$Department of Physics, University of Alberta, Edmonton,
Alberta, Canada T6G 2J1}
 
\date{\today} 
\maketitle 
\begin{abstract}
The model of hole superconductivity predicts that the superfluid
weight in the zero-frequency $\delta$-function in the optical 
conductivity has an anomalous contribution from high frequencies,
due to lowering of the system's kinetic energy upon entering the
superconducting state. The lowering of
kinetic energy, mainly in-plane in origin, accounts for both the
condensation energy of the superconductor as well as an increased potential
energy due to larger Coulomb repulsion in the paired state. It leads
to an apparent violation of the conductivity sum rule, which in 
the clean limit we 
predict to be substantially larger for in-plane than for c-axis conductivity. 
However, because cuprates are 
in the dirty limit for c-axis transport, the sum rule violation is
found to be greatly enhanced in the c-direction. The model
predicts the sum rule violation to be largest in the underdoped regime and to
decrease with doping, more rapidly in the c-direction that in the 
plane. So far, experiments have detected sum rule violation in
c-axis transport in several cuprates, as well as a 
decrease and disappearance of this violation for increasing doping, but no
violation in-plane. We explore the predictions of the model
for a wide range of parameters, both in the absence and in the
presence of disorder, and the relation with current experimental
knowledge. 
\end{abstract}

\section{Introduction}
In the model of hole superconductivity the hopping amplitude for a hole of spin 
$\sigma$ hopping between sites $i$ and $j$ is given by\cite{hole1,hole2}
\beq
t_{ij}^{\sigma}=t_{ij}^h+(\Delta t)_{ij}(n_{i,-\sigma}+n_{j,-\sigma})
\eeq
with $n_{i,-\sigma}$ the occupation number for spin $(-\sigma)$ at 
site $i$. Eq. (1) leads to superconductivity at low hole concentration
due to lowering of the carrier's effective mass, or equivalently
of kinetic energy, upon pairing. In tight binding models for 
low carrier concentration the 
hopping amplitude is related to the effective mass $m^*$ through
\beq
t=\frac{\hbar^2}{2m^*a^2}
\eeq
with $a$ the lattice spacing in the given direction.

This physics leads to nontrivial consequences for the electrodynamics
of hole superconductors\cite{lond,area}. The London penetration depth in the
limit of low carrier concentration is given by
\beq
\lambda=(\frac{m^*c^2}{4\pi n_se^{*2}})^{1/2}
\eeq
with $m^*$, $e^*$ the mass and charge of the superfluid carriers, and 
$n_s$ the superfluid density. If the effective mass decreases upon
 pairing, $\lambda$  will be smaller than expected from the normal
state effective mass\cite{lond}. This leads to a violation \cite{area} of the low energy
optical sum rule (Ferrell-Glover-Tinkham sum rule\cite{fgt}) which relates
the London penetration depth (which depends on the effective mass
in the paired state) to the low frequency 'missing area'
in the optical conductivity (which is a function of the effective mass
in the normal state). Observation of such a violation in several
cuprates\cite{basov}, long after it was predicted theoretically\cite{area}, lends support to 
the model upon which the prediction was based. The London penetration depth is
determined by the weight of the zero-frequency $\delta$-function in the
optical conductivity ('superfluid weight'), which is predicted to be larger than the 
area missing
from the low frequency optical absorption, as shown schematically in Fig. 1.

The sum rule violation is a manifestation of the lowering of kinetic
energy that occurs upon pairing, and this lowering of kinetic
energy is what gives the condensation energy of the superconductor
within our model. In fact, the kinetic energy lowering
(mainly from in-plane motion) that we obtain
is much larger than the condensation energy \cite{area2}, but is compensated to
some extent by increase of Coulomb repulsion between carriers, that
are on the average closer to each other in the paired state
compared to the situation in the unpaired state. 

Anderson\cite{and1} has proposed a mechanism for high $T_c$ superconductivity based on
lowering of kinetic energy of pairs tunneling between planes (interlayer
tunneling theory, or ILT). That theory has in common with the one discussed
here that superconductivity is driven by kinetic energy lowering, but it
differs in that it only deals with c-axis transport. 
There are also other fundamental differences with
the theory discussed here. According to ILT, the weight in the
$\delta-$function for c-axis transport should account for the
condensation energy of the superconductor, and it is claimed\cite{and1} that this
is in fact the case in $La_{2-x}Sr_xCuO_4$ throughout the entire doping regime,
from underdoped to overdoped.
This implies in particular that within ILT theory the entire weight in the 
$\delta$-function reflects lowering of c-axis kinetic energy,
irrespective of whether that weight comes from low or from high 
frequency optical response. Thus, while the theory may be $consistent$ with
the observed c-axis sum rule violation\cite{chak}, it makes no definite $prediction$ on 
whether sum rule violation in the c direction 
should occur in a given doping regime. Or, perhaps
one should interpret the ILT prediction to mean that the sum rule violation
should be $100\%$ for any doping regime, which however is inconsistent
with reported observations\cite{basov}. Instead,
our theory associates with kinetic energy
lowering only the part of the $\delta-$function that comes from
high frequencies, and yields quantitative predictions for it as well as for the
total superfluid weight as a function
of doping, which we will compare here with experiment.

Furthermore, 
for $Tl_2Ba_2CuO_y$ ILT theory cannot account for the
condensation energy even using the entire weight in the zero-frequency
$\delta$-function, because it is too small by at least two orders
of magnitude \cite{thallium}. This is because ILT theory only considers kinetic energy
lowering due to c-axis motion. Instead, the theory considered here, despite using only
the fraction of the $\delta$-function weight coming from high 
frequencies, has no trouble accounting for the
condensation energy seen experimentally because it considers both c-axis 
as well as in-plane kinetic energy lowering.

In the anisotropic structure of the oxides the sum rule violation 
will naturally not be the same in all directions. We assume as
the simplest possible model, that the ratio of $t$ and 
$\Delta t$ in Eq. (1) is the same in all directions. We will show 
that under this plausible assumption our model predicts the
in-plane violation to be several times larger than the 
c-axis violation. This is in apparent contradiction with the
reported experimental observation\cite{basov} of large c-axis sum rule violation 
($\sim 50\%$) and no detectable in-plane violation($< 10\%$).However,
a natural explanation for this discrepancy arises from the fact that the
cuprates quite generally appear to be in the clean limit for in-plane
conduction and in the dirty limit for c-axis conduction. We will not
go here into a discussion of why this is the case, but there seems
to be ample experimental evidence for it\cite{marel,uchida,shiba,tajima}. 
We show that under 
those conditions the c-axis violation is very greatly enhanced and will
generally be much larger than the in-plane violation, in agreement with
observations. The effect of disorder on the sum rule violation is 
schematically shown in Fig. 1.

Concerning the doping dependence, it is important to differentiate between
relative and absolute values of kinetic energy lowering. For absolute
values, our model predicts that the kinetic energy lowering is maximum 
close to the optimally doped case, and decreases smoothly both for 
underdoped and overdoped regimes. Instead, for the degree of sum rule
violation, i.e. kinetic energy lowering relative to total superfluid
weight, the model predicts a monotonic decrease as the doping increases
(except in extremely underdoped regimes). Furthermore in the presence of disorder
the rate of decrease with doping can be greatly enhanced. Experimentally,
a rapid decrease of sum rule violation with doping has been observed\cite{basov},
and we will show that the theory is compatible with the reported
observations within reasonable assumptions.

Because of the difficulty in precisely estimating the appropiate parameters
in our model for given materials, we explore here predictions of the
model for a range of interaction parameters. This is of intrinsic
interest, and furthermore it may be relevant to materials as yet
undiscovered. Some general trends found are that for given
maximum $T_c$ the model predicts increasing sum rule violation as the
nearest neighbor repulsion increases, and as the bandwidth
decreases. The latter in turn also leads to an increasing
condensation energy. The results that we obtain are compatible with
existing observations in the cuprates for a range of parameters in
the model, and future more accurate observations should be able to
determine more precisely the parameters in the model to represent
the physics of a given cuprate.

In Sect. II we review the Hamiltonian and general formalism, and
discuss the calculation of the condensation energy. Sect. III discusses
the optical sum rule predictions, in the clean limit, and in
the presence of disorder. Sect. IV shows results for
the clean limit for the full 3-dimensional anisotropic model for a 
variety of parameters, and in sect. V we compare the predictions of the 
model with experimental results taking into account the effect of
disorder. We conclude in Sect. VI with a summary and discussion.

\section{Formalism}
The model is defined by the single band Hamiltonian\cite{hole1}
\beq
H=-\sum_{i,j,\sigma}t_{ij}^\sigma (c_{i\sigma}^\dagger c_{j\sigma}+h.c.)
+U\sum_i n_{i\uparrow}n_{i\downarrow}+\sum_{<ij>}V_{ij}n_in_j
\eeq
with $i$, $j$ sites on a three-dimensional cubic lattice, and
$t_{ij}^\sigma$ defined by Eq. (1). $c_{i\sigma}^\dagger$ creates a hole
of spin $\sigma$ in the oxygen $p\pi$ planar orbital at site $i$, and
other orbitals ($O p_\sigma$, $Cud_{x^2-y^2}$) are ignored\cite{twoband}. 
We define $t_{ij}^h$ in Eq. (1) to be $t_a^h$ or $t_c^h$ for nearest neighbor
sites in the plane or in the $c$-direction, and similarly $\Delta t_a$,
$\Delta t_c$, and assume the same anisotropy for $t$ and $\Delta t$:
\beq
\eta=\frac{t_a}{t_c}=\frac{\Delta t_a}{\Delta t_c}  .
\eeq
The effective hopping as a function of hole density $n_h$ is given by,
with $\alpha=a$ or $c$:
\beq
t_\alpha=t_\alpha ^h+\Delta t_\alpha n_h
\eeq
and the effective mass anisotropy
\beq
\frac{m_c^*}{m_a^*}=\frac{t_a}{t_c}=\eta
\eeq
is independent of doping level with this assumption. We assume isotropic nearest neighbor
repulsion for simplicity; any anisotropy in it should be much smaller than that
for the hopping amplitudes .

The formalism we use is described in Ref. \cite{hole1}.  
The results are not very dependent
on details of the band structure. To understand the behavior emerging from
the planar motion, which dominates the energetics, one can use a simple constant
density of states model,
\beq
g(\epsilon)=\frac{1}{D}
\eeq
with $D$ the bandwidth. This model illustrates well the behavior 
emerging from the planar motion\cite{area2}. 
Since we are interested in the anisotropy of various measured
properties, we will mostly use here a three-dimensional tight-binding
band structure, with a strong hopping anisotropy ($\eta = 25$, for
definiteness). The consequences of this anisotropy for various
properties have already been discussed in Refs. \cite{hole1,lond}.
In particular, we discussed an approximation in which only energy
integrations are required; anisotropy only enters through various
weighted densities of states. The results thus obtained were 
very accurate.In the rest of this paper, unless stated otherwise, we will use
this approximation \cite{approx}.

The condensation energy per site is given by
\beq
\epsilon_{cond}=\epsilon_n-\epsilon_s
\eeq
with $\epsilon_s$, $\epsilon_n$ the average energy per site at the same
temperature and for the same number of holes in the superconducting and
normal states respectively. We define the different contributions
\beq
\epsilon_{cond}=\epsilon_t+\epsilon_{\Delta t}+\epsilon_U+\epsilon_V
\eeq
arising from single particle hopping, correlated hopping,
on-site and nearest neighbor repulsion respectively. All contributions
to $\epsilon_{cond}$ in Eq. (10) except $\epsilon_{\Delta t}$ are
negative: the single particle hopping energy is lowest in the normal
state ($\epsilon_t<0$), and the Coulomb repulsions due to $U$ and $V$ increase in the
superconducting state ($\epsilon_U, \epsilon_V<0$)
because carriers in a pair are closer together
on average. The condensation energy is entirely given by the large
kinetic energy lowering due to correlated hopping ($\epsilon_{\Delta t}$)
which is compensated to some extent by the other terms in Eq. (10).
In weak coupling the condensation energy is given by the usual expression
\beq
\epsilon_{cond}=\frac{\Delta_0^2}{2}g(\epsilon_F)
\eeq
with $\Delta_0$ the energy gap and $g(\epsilon_F)$ the density of
states at the Fermi energy.

To determine the various energies that contribute to the condensation energy,
we define, as in previous work, the integrals
\beq
I_{\ell}=\int d\epsilon (-\frac{\epsilon}{D/2})^{\ell} g(\epsilon)
\frac{1-2f(E(\epsilon))}{2E(\epsilon)}
\eeq
with $f$ the Fermi function and the quasiparticle energy
$E(\epsilon)$ given by
\bmath
\beq
E(\epsilon)=\sqrt{(\epsilon-\mu)^2+\Delta(\epsilon)^2}
\eeq
\beq
\Delta(\epsilon)=\Delta_m(-\frac{\epsilon}{D/2}+c)
\eeq
\emath
and the parameters $\Delta_m$ and $c$ obtained from solution of the
BCS equations. The various contributions to the condensation
energy are given by
\bmath
\beq
\epsilon_t=2\int d\epsilon \, g(\epsilon) \, \epsilon \, 
f(\epsilon - \mu_N) -D[\frac{D}{2}I_2+\mu I_1]
\label{cond1}
\eeq
\beq
\epsilon_{\Delta t}=2K\Delta_m^2(I_1+cI_0)(I_2+cI_1)
\label{cond2}
\eeq
\beq
\epsilon_U=-U\Delta_m^2(I_1+cI_0)^2
\label{cond3}
\eeq
\beq
\epsilon_V=-W\Delta_m^2(I_2+cI_1)^2
\label{cond4}
\eeq
\emath
with
\bmath
\beq
K=2z\Delta t
\eeq
\beq
W=zV
\eeq
\emath
and $z$ the number of nearest neighbors to a site. Note that $\mu_N$ is
the chemical potential required for $n$ electrons in {\it the normal state}.

The contributions (\ref{cond2},\ref{cond3},\ref{cond4}) are useful 
in that they specify
the various  energy contributions arising from interaction terms
in the Hamiltonian. They can, however, be summed,
with the help of the gap equation\cite{hole1}, to give a single result
\beq
\epsilon_{\rm int} \equiv \epsilon_{\Delta t} + \epsilon_U + \epsilon_V
= \int d\epsilon \, g(\epsilon) \, \Delta^2(\epsilon) \,
\frac{1-2f(E(\epsilon))}{2E(\epsilon)},
\label{free}
\eeq
in agreement with the usual expression for the internal energy contribution.

\section{Optical sum rule}

The real part of the optical conductivity for light polarized in
direction $\delta$ is given by\cite{mahan}
\beq
\sigma_{1\delta}(\omega,T)=
\frac{\pi}{\hbar \Omega Z}\sum_{n,m}\frac{e^{-\beta E_n}-e^{-\beta E_m}}{E_m-E_n}
<n|J_\delta|m><m|J_\delta|n>\delta(\omega-\frac{E_m-E_n}{\hbar})
\eeq
where $n,m$ run over all eigenstates of the system with energies
$E_n,E_m$, $Z$ is the partition function, $\Omega $ the volume,
 and $J_\delta$ the component of 
the paramagnetic current operator in the $\delta$ direction. To derive the
current operator for the tight binding model Eq. (4) we define the
polarization operator 
\beq
\vec{P}=e\sum_i\vec{R_i}n_i
\eeq
and obtain the current operator from its time derivative
\beq
\vec{J}=\frac{d\vec{P}}{dt}=\frac{i}{\hbar}[H,\vec{P}]
\eeq
yielding for the component in the $\delta$ direction
\beq
J_\delta=\frac{ie a_\delta}{\hbar}\sum_{i,\sigma}t_{ij}^\sigma [c_{i+\delta,\sigma}^\dagger
c_{i\sigma}-c_{i\sigma}^\dagger c_{i+\delta,\sigma}].
\eeq
Note that the hopping amplitudes in Eq. (20) have the operator dependence
given by Eq. (1), but because density operators commute with each other the form 
Eq. (20) is the same as in the ordinary tight binding model with constant
hopping amplitude. Similarly the commutator of the current and
polarization operators yields
\beq
[J_\delta,P_\delta]=-\frac{ie^2}{\hbar}a_\delta^2<-T_\delta>
\eeq
where $a_\delta$ is the lattice spacing in the $\delta$ direction and
$T_\delta$ is the part of the kinetic energy arising from hopping
processes in the $\delta$ direction:
\beq
T_\delta=-\sum_it_{i,i+\delta}^\sigma [c_{i\sigma}^\dagger c_{i+\delta,\sigma}+h.c.].
\eeq

Using Eq. (19) we can write
\beq
\frac{<n|J_\delta|m><m|J_\delta|n>}{E_m-E_n}=
\frac{i}{2\hbar}[<n|J_\delta|m><m|P_\delta|n>-
<n|P_\delta|m><m|J_\delta|n>].
\eeq
Substituting in Eq. (17), integrating over frequency and summing over
intermediate states yields the 
'partial' conductivity sum rule for our model:
\beq
\int_0^{\omega_m} d\omega\sigma_{1\delta}(\omega,T)=\frac{\pi^2a_\delta^2 e^2}
{2\hbar^2 \Omega}<-T_\delta>
\eeq
which formally looks the same as in the usual tight binding model\cite{maldague}.  The
high frequency cutoff $\omega_m$ in Eq. (24) indicates that transitions
to higher energy states not described by our Hamiltonian Eq. (4) are excluded.
If we were to extend the integral to infinity instead, the usual conductivity
sum rule follows:
\beq
\int_0^\infty d\omega\sigma_{1\delta}(\omega,T)=\frac{\pi e^2n}{2m},
\eeq
where $n$ is the electron density and $m$ the bare electron mass. For the
tight binding model with a nearly empty or a nearly full band it is
easily seen that Eq. (24) takes the form Eq. (25) with $m$ replaced by
the effective mass $m^*$ given by Eq. (2).

More generally, the complex frequency-dependent conductivity for
light polarized in direction $\delta$ is given by\cite{mahan}
\beq
\sigma_\delta(\omega)=\frac{i}{\omega}[\Pi_\delta (\omega)-
\frac{e^2 a_\delta ^2}{\hbar^2 \Omega}<T_\delta>]
\eeq
with $\Pi_\delta$ the complex current-current correlation function, with spectral
representation
\beq
\Pi_{\delta}(\omega)=
\frac{1}{ \Omega Z}\sum_{n,m}\frac{e^{-\beta E_n}-e^{-\beta E_m}}{E_m-E_n+\hbar \omega +i\delta}
<n|J_\delta|m><m|J_\delta|n>
\eeq
and the London Kernel, that gives the penetration depth $\lambda_\delta$, is 
given by
\beq
K_\delta=\frac{1}{\lambda_\delta^2}=\frac{4\pi\omega}{c^2}\sigma_{2\delta}
\eeq
with $\sigma_2$ the imaginary part of the conductivity. Hence,
\beq
K_\delta=\frac{4\pi}{c^2}\Pi_{1\delta }-
\frac{4\pi e^2 a_\delta ^2}{c^2\hbar^2 \Omega }<T_{1\delta}> \equiv K_{1\delta}+K_{2\delta}
\eeq
with $\Pi_{1\delta}$ the real part of the current-current
correlation function. $K_{1\delta}$ and $K_{2\delta}$ are the paramagnetic
and diamagnetic London Kernels.

In the superconducting state at temperature $T$ the real part of the
conductivity is
\beq
\sigma_{1\delta}^s=D_\delta(T)\delta(\omega)+\sigma_{1\delta}^{s,reg}(\omega,T).
\eeq
The superfluid weight, $D_\delta(T)$, gives rise,
through a Kramers-Kronig relation, to a  $1/\omega$ contribution
to the imaginary part of the conductivity and hence to the London Kernel
\beq
K_\delta=\frac{8D_\delta}{c^2}.
\eeq
Integrating Eq. (30) and using Eq. (24) yields
\beq
D_\delta (T)+\int_{0^+}^{\omega _m}\sigma_{1\delta}^s(\omega,T)=
\frac{\pi e^2a_\delta^2 e^2}{2\hbar^2 \Omega }<-T_\delta>_{s,T}    .
\eeq
On the other hand in the normal state at temperature $T_1$
\beq
\int_{0}^{\omega_m}\sigma_{1\delta}^n(\omega,T_1)=
\frac{\pi e^2a_\delta^2e^2}{2\hbar^2 \Omega } <-T_\delta>_{n,T_1}    
\eeq
and Eqs. (32) and (33) yield
\beq
D_\delta(T)=\int_{0^+}^{\omega _m}d\omega 
[\sigma_{1\delta}^n(\omega,T_1)- \sigma_{1\delta}^s(\omega,T)]+
\frac{\pi e^2a_\delta^2}{2\hbar^2 \Omega }[<-T_\delta>_{s,T}-<-T_\delta>_{n,T_1}]
\equiv \delta A_l+\delta A_h .
\eeq
The first term in Eq. (34) is the 'low frequency missing area' $\delta A_l$ that
arises from the opening of the superconducting energy gap, as discussed by 
Ferrell, Glover and Tinkham. It is $not$ related to $changes$ in kinetic
energy in going into the superconducting state,
but reflects simply the kinetic energy of the carriers in the
normal state. In conventional superconductors, only this term is expected
to contribute to the superfluid weight. The second term in Eq. (34) was
predicted to exist in high $T_c$ materials in 1992 \cite{area}, and experimental
evidence for its existence was found in 1999\cite{basov}. The qualitative
behavior of $\sigma _1(\omega)$ and the different contributions to 
the superfluid weight are shown schematically in Figure 1.

Integrating Eqs. (30) and
(33) to infinity instead should yield the same answer according to the
'global' sum rule Eq. (25), so that we have also
\beq
D_\delta(T)=\int_{0^+}^{\infty}d\omega 
[\sigma_{1\delta}^n(\omega,T_1)- \sigma_{1\delta}^s(\omega,T)]
=\delta A_l+\delta A_h
\eeq
and
\beq
\delta A_h=
\int_{\omega_m}^{\infty}d\omega 
[\sigma_{1\delta}^n(\omega,T_1)- \sigma_{1\delta}^s(\omega,T)]=
\frac{\pi e^2a_\delta^2}{2\hbar^2 \Omega}[<-T_\delta>_{s,T}-<-T_\delta>_{n,T_1}]
\eeq
so that the change in optical absorption at high frequencies is
given by the change in kinetic energy. The states involved in the optical
transitions that contribute to Eq. (36) are $not$ in the Hilbert space
where the Hamiltonian Eq. (4) is defined. However, a more general
Hamiltonian can be found\cite{polar} that both contains these states and
yields Eq. (4) as a low energy effective Hamiltonian when these states are
projected out.

Expressions for the kinetic energies are given in Refs. \cite{lond,area};
we reproduce them here for completeness:
\bmath
\beq
<T_\delta>=<T_\delta^t>+<T_\delta^{\Delta t}>
\eeq
\beq
<T_\delta^t>=-2(t_\delta^h+n \Delta t_\delta)\sum_kcosk_\delta
(1-\frac{\epsilon_k-\mu}{E_k}(1-2f(E_k)))
\eeq
\beq
<T_\delta^{\Delta t}>=-\frac{4\Delta t_\delta}{N}
\sum_{k,k'}(cosk_\delta+cosk'_\delta)\frac{\Delta_k}{2E_k}
\frac{\Delta_{k'}}{2E_{k'}}(1-2f(E_k))(1-2f(E_{k'})).
\eeq
\emath
Eq. (36) shows that if there is a change in the carrier's kinetic
energy in going from the normal state at temperature $T_1$ to the
superconducting state at temperature $T$, there will be an apparent
violation of the conductivity sum rule. Such a change arises in our
model from the pair contribution to the kinetic energy, Eq. (37c), which is
zero above $T_c$ and becomes non-zero (negative) below $T_c$ as
the superconducting state develops.

The superfluid weight and missing areas are related to the London
Kernel $K_\delta$ by
\beq
D_\delta=\delta A_l^\delta+\delta A_h^\delta=\frac{c^2}{8}K_\delta
\eeq
\beq
K_\delta=K_{1\delta}+K_{2\delta} .
\eeq
The diamagnetic part of the London Kernel $K_2$ is given by
\beq
K_{2\delta}=\frac{4\pi e^2a_\delta^2}{\hbar^2 c^2 \Omega }
[<-T_\delta^t>+<-T_\delta^{\Delta t}>] .
\eeq
The low frequency missing area is given by the London paramagnetic
Kernel and the single particle part of the kinetic energy
\beq
\delta A_l=\frac{c^2}{8}[K_{1\delta}+
\frac{4\pi e^2a_\delta^2}{\hbar^2 c^2 \Omega}<-T_\delta^t>],
\eeq
and the high frequency missing area by the remaining part of the London
diamagnetic Kernel
\beq
\delta A_h=\frac{\pi e^2a_\delta^2}{2\hbar^2 \Omega}<-T_\delta^{\Delta t}>.
\eeq
>From Eqs. (27) and (29), the spectral representation of 
the paramagnetic London Kernel is
\beq
K_{1\delta}=\frac{8\pi}{c^2 \Omega Z} \sum_{n,m}
\frac{e^{-\beta E_n}}{E_n-E_m}|<n|J_\delta|m>|^2.
\eeq
Finally, we define the sum rule violation parameter in
direction $\delta$
\beq
V_\delta=\frac{\delta A_h^\delta}{\delta A_l^\delta+\delta A_h^\delta}
\eeq
which quantifies the relative amount of sum rule violation.

\subsection{Clean limit}

In the absence of disorder the paramagnetic London Kernel is easily
evaluated and yields
\beq
K_{1\delta}=\frac{32\pi e^2t_\delta a_\delta^2}{\hbar^2c^2 \Omega }
\frac{1}{N}\sum_k sin^2 k_\delta(\frac{\partial f}{\partial E_k}).
\eeq
It is easily seen\cite{lond} that at $T=T_c$, $K_{1\delta}$ exactly cancels
the single-particle kinetic energy in Eq. (41), hence $\delta A_l=0$.
Similarly $\delta A_h=0$ since $\Delta_k=0$. Hence the superfluid
weight goes to zero at $T_c$ as expected. In the limit of zero
temperature, $K_{1\delta}$ goes to zero and
the low frequency missing area Eq. (41) is given by the expectation
value of the single particle kinetic energy
\beq
\delta A_l=\frac{\pi e^2 a_\delta ^2}{2\hbar^2 \Omega }<-T_\delta^t>.
\eeq
Hence the sum rule violation parameter is simply
\beq
V_\delta=\frac{<T_\delta^{\Delta t}>}
{<T_\delta^{t}>+<T_\delta^{\Delta t}>}\eeq
and can be calculated from Eq. (37).

As we will see in the next section, the in-plane sum rule violation is generally 
larger than the c-axis one in the clean limit. From Eq. (47), this will be
the case if
\beq
\frac{<T_a^{\Delta t}>}{<T_c^{\Delta t}>} > \frac{<T_a^{t}>}{<T_c^{t}>}
\eeq
holds, that is, if the anisotropy in the anomalous part of the kinetic energy
is larger than that in the single particle part. This is indeed the case and
can be understood both from a strong and a weak coupling argument, as discussed
in the following.

From Eq. (37) it would appear that the anisotropy in both $<T_\delta^t>$ and 
$<T_\delta ^{\Delta t}>$ is given  by $\eta$, Eq. (5), since they have as
prefactor $t_\delta$ and $\Delta t_\delta$ respectively. However this argument
is misleading. In strong coupling, the anomalous kinetic energy is proportional to
\cite{lond}
\beq
<T_\delta^{\Delta t}> \sim \frac{(\Delta t_\delta)^2}{U}
\eeq
corresponding to second order processes where a hole hops onto a site
already occupied by another hole, with an energy cost $U$. In contrast,
$<T_\delta ^t>$ is dominated by first order processes, hence is proportional
to $t_\delta$. As a consequence, the anisotropy in $<T_\delta ^t>$ will be closer
to $\eta$ and that in $<T_\delta ^{\Delta t}>$ closer to $\eta^2$.

From a weak coupling point of view we can also understand the different
anisotropy Eq. (48) from the expressions for the kinetic energies Eq. (37).
The contributions to the sum over $k$, $k'$ in Eq. (37c) are dominated by
values of $k$, $k'$ in the vicinity of the Fermi surface. Except for 
extremely small doping, the Fermi surface for the anisotropic band
structure has a "cigar" form extending over all values of $k_z$ but only a small
range of $k_x$, $k_y$ close to the origin. For $<T_c^{\Delta t}>$, the factors of
$cos k_z$, $cos k'_z$ lead to cancellations because they extend over both positive and
negative value, e.g. for $k_z$ and $(\pi -k_z)$, while for $<T_a^{\Delta t}>$ the 
factors $cos k_x$, $cos k'_x$ have always the same sign. Hence the anisotropy
in $<T_\delta ^{\Delta t}>$ will in general be substantially larger than $\eta$,
except for very low doping where the Fermi surface is just small pockets 
around the points $(0,0,+/-\pi)$. For $<T_\delta ^t>$ instead, Eq. (37b),
the contributions to the sum over $k$ do not come only from points around the
Fermi surface but from all points inside the Fermi surface. There is also a 
cancellation here between positive and negative values of $cos k_z$ and hence
the anisotropy in $<T_\delta ^t>$ is also larger than $\eta$. However the
cancellation is less complete here because there is more phase space inside the Fermi
surface for $k_z$ than there is for $(\pi -k_z)$, for $k_z <\pi /2$.
Hence the anisotropy in $<T_\delta ^t>$ is smaller than that in 
$<T_\delta ^{\Delta t}>$,
as given by Eq. (48), leading to larger in-plane than c-axis sum rule violation.

\subsection{Effect of disorder}

In the presence of disorder the weight in the $\delta$-function is decreased
and hence the penetration depth increases. Qualitatively this can be seen from 
the Drude form of the optical conductivity
\beq
\sigma_1(\omega)=\frac{ne^2\tau}{m^*}\frac{1}{1+\omega^2\tau^2}
\eeq
with $\tau$ the scattering time. Upon entering the superconducting state
the optical absorption at frequencies of order $\hbar \omega<2\Delta$ is suppressed
due to the opening of the gap; as the disorder increases, the weight in that
frequency range decreases by a factor of order $\Delta/(\hbar/\tau)$, and
the penetration depth at low temperatures
is given approximately by\cite{tink}
\beq
\frac{1}{\lambda^2_\delta}=(\frac{1}{\lambda_\delta^{clean}})^2
{1 \over 1 + \frac{\hbar/\tau_\delta} {\pi\Delta} }
\equiv\frac{1}{(\lambda_\delta^{clean}})^2 p_\delta
\eeq
The detailed calculation within BCS theory is given in Ref. 21 for the
jellium model, and in Ref. 22 for arbitrary impurity scattering rate.
The diamagnetic London
Kernel, given by the expectation value of the single particle kinetic energy,
is assumed to be unaffected by disorder. Similarly we expect the expectation
value of the pair contribution to the kinetic energy, $<T_\delta^{\Delta t}>$, to 
be unaffected by disorder as long as it is weak enough not to cause
pairbreaking. Under those conditions
 disorder will only affect the paramagnetic London Kernel, and
hence only the low frequency missing area. Since $\delta A_l$ in Eq. (44)
is reduced by disorder and $\delta A_h$ is unaffected, the sum rule
violation will increase, as was shown schematically in Figure 1.
In the presence of sum rule violation
Eq. (51) becomes
\beq
\frac{1}{\lambda^2_\delta}=(\frac{1}{\lambda_\delta^{clean}})^2 
[p_\delta+V_\delta^{clean}(1-p_\delta)]
\eeq
and the sum rule violation in the presence of disorder is given by
\beq
V_\delta=\frac{\delta A_h^\delta}{\delta A_l^\delta \times p_\delta+\delta A_h^\delta}
=V_\delta^{clean}(\frac{\lambda_\delta}{\lambda_\delta^{clean}})^2
\eeq
where the missing areas are understood to be those in the clean limit. The disorder
parameter $p_\delta$ can be written as
\beq
p_\delta=[(\frac{\lambda_\delta^{clean}}{\lambda_\delta})^2-
V_\delta^{clean}]\frac{1}{1-V_\delta^{clean}}
\eeq
so that it can be obtained from our calculated penetration depth and sum
rule violation in the clean limit together with the observed value of the
penetration depth.

There is substantial experimental evidence that the effect of disorder is
substantially stronger for c-axis transport than for in-plane transport in the
cuprates in the underdoped regime, i.e. $\tau_c/\tau_a<<1$, and that transport 
in the planes can be understood within the clean limit\cite{marel,uchida,shiba}.
As the doping is increased, transport in the c-direction is found to become
more coherent, i.e. $\tau_c$ increases. The penetration depth in the c-direction
will hence be increased compared to its clean limit value, especially in the
underdoped regime, and consequently from Eq. (53) the sum rule violation in the
c-direction will also be increased. This effect is much larger than the
anisotropy in the violation in the clean limit discussed in the previous
section. Hence in the presence of disorder the sum rule violation in the 
c-direction will dominate. As the doping increases we will find
that the sum rule violation in the c-direction decreases rapidly, both because
the system is approaching the clean limit and because the intrinsic violation in the
clean limit also decreases with doping.

\section{Results in the clean limit}

Much of the behavior of our model is determined by the in-plane physics,
that dominates the energetics. Furthermore the properties of the model are
not very sensitive to the detailed density of states, and hence it is possible
to learn about many of the properties by  considering simply a two-dimensional
model with constant density of states, as done for example in
Refs. \cite{area} and \cite{area2}. Here, to compare the 
behavior of in-plane and
c-axis properties we will consider the full anisotropic 
three-dimensional model at the outset. The in-plane properties of that
model are very similar to those of the simpler two-dimensional model.

For definiteness we will assume an on-site Hubbard repulsion
$U=5$ eV, and an anisotropy in hopping and in correlated hopping
$\eta=25$ (Eq. 5). We will show results for a set of bandwidths
spanning the weak to strong coupling regime, 
$D=1.5$ eV, $1$ eV, $0.5$ eV, $0.1$ eV, and nearest neighbor repulsion
$V=0$ and $V=0.65$ eV. We expect the actual value of $V$ in the
cuprates to be somewhere in-between those two values. The magnitude
of the hopping interaction $\Delta t$ is chosen to yield a
maximum $T_c$ of $37.5$ K or $90$ K. The first is appropiate
for $LaSrCuO_4$,  the
second for  $Tl2201$, $Hg1201$ and $YBCO123$ structures.

Figures 2 (a) and (b) show results for the critical temperature versus
doping for nearest neighbor repulsion $V=0$ and maximum $T_c$ of
$37.5$ K and of $90$ K respectively, for the set of bandwidths considered.
It can be seen that the results are very similar for all cases considered,
except that as the bandwidth decreases the width of the peak increases
slightly. On the other hand the condensation energy, shown in Fig. 3,
is strongly dependent on the bandwidth and increases as the bandwidth
decreases, as would be expected from Eq. (11). 

The in-plane sum rule violation Eq. (44) is shown in Fig. 4. In the
purely two-dimensional model it is a monotonically decreasing
function of doping\cite{area2}, whereas in the three-dimensional structure it
decreases at very low densities in weak coupling, as the density of states goes to 
zero. The sum rule violation decreases rapidly as the bandwidth
increases, and is larger for the cases corresponding to higher $T_c$.

The behavior of the sum rule violation parameter in the c-direction is
similar to the in-plane one but  smaller in magnitude, as 
shown in Fig. 5, in accordance with the discussion in the
previous section. The decrease at low densities is somewhat less
pronounced than for the in-plane case. In Fig. 6 we plot the sum rule
violation anisotropy, $V_a/V_c$. As the bandwidth increases the 
sum-rule violation anisotropy increases, and the 
c-axis sum rule violation can be up to a factor of $5$ smaller 
than the in-plane violation for the parameters considered here.

Next we consider the effect of nearest neighbor repulsion $V$. The
critical temperature versus doping, shown in Fig. 7, shows somewhat
larger dependence on bandwidth than the case $V=0$ but is 
otherwise similar. The condensation energy, Fig. 8, is somewhat 
decreased compared to the case $V=0$ (Fig. 3), (for parameters
chosen to yield the same $T_c^{max}$), particularly as the
bandwidth becomes small. On the other hand the in-plane sum rule violation,
Fig. 9, increases compared to the case $V=0$ (Fig. 4) for all the
different bandwidths. The anisotropy in the sum rule violation,
Fig. 10, shows similar magnitude and doping dependence as the case $V=0$ (Fig.
6).

These results indicate that the in-plane sum rule violation will be
easiest to detect in the underdoped regime, and in cuprates where the
condensation energy 
is large, due to a large density of states (small bandwidth), and where
the nearest neighbor Coulomb repulsion is appreciable.

In addition to the sum rule violation, which involves the ratio of
kinetic energies, it is useful to consider the behavior of the
kinetic energies themselves. Figure 11 shows the behavior of the
in-plane kinetic energies versus doping for one case: single-particle
contribution (a), pair contribution (b) and total (c). The pair contribution
is proportional to the anomalous part of the $\delta-$function response
coming from high frequencies, Eq. (36), while the single particle contribution
gives the regular part of the $\delta-$function coming from low frequencies,
Eq. (46), {\it in the clean limit}. In the presence of disorder the latter
contribution will be reduced, while the former one is expected to
remain the same. Finally, the total kinetic energy will be proportional
to the total weight in the $\delta-$function and hence the
inverse squared penetration depth in the clean limit only.

It can be seen from Fig. 11 that the pair contribution to the 
kinetic energy is maximum approximately at the same doping where
$T_c$ is maximum (optimally doped) for the larger bandwidths,
and at even higher doping for small bandwidth. This is in contrast to
the sum rule violation parameter that decreases monotonically from
the underdoped through the overdoped regime. That is, we predict that
the anomalous kinetic energy contribution should persist well into
the overdoped regime. Experimentally this effect may be difficult to
detect because the normal contribution increases rapidly with doping
and will strongly dominate the superfluid weight in the overdoped
regime. This effect will be even more pronounced if, as we
expect, the effect of disorder becomes less pronounced as the doping
increases. This will be further discussed in the following section.

The behavior of the kinetic energies in the z-direction is similar,
as seen in Fig. 12. In Fig. 13 we show the anisotropy in the
normal and
anomalous parts of the kinetic energy. In accordance with the discussion
in the previous section, both are larger than the anisotropy in 
the hopping, $\eta$, and they increase with dooping. Fig. 14 shows the
resulting anisotropy
in the total kinetic energy, which in the clean limit is inversely
proportional to the square of the anisotropy in the penetration depths.
For all except the smallest bandwidth, 
the doping dependence obtained is opposite to what is observed 
experimentally e.g. in $LaSrCuO$, where the anisotropy in penetration
depths is found to decrease as the doping increases. We attribute 
this discrepancy to a variation of the effect of disorder with
doping\cite{uchida}, as will be discussed in the next section.

Within a simple two-fluid picture with parabolic bands
one would expect the anisotropy
in the kinetic energies or the clean limit squared penetration depths
to be given by the effective mass anisotropy,
\beq
\frac{<T_a>}{<T_c>}=\frac{\lambda_c^2}{\lambda_a^2})_{clean}=
\frac{m^*_c}{m^*_a}=\frac{t_a}{t_c}
\eeq
which for the case considered here is $25$. From Fig. 14 it can be seen that
this value is in fact only approached in the weak coupling regime (large
bandwidth) and for small doping. As the doping increases the anisotropy
increases rapidly, and for the case of strong coupling (small bandwidth) the 
anisotropy also increases in the underdoped regime. In fact in the strong
coupling limit the anisotropy in the kinetic energies will approach the
square of the band structure anisotropy. The anisotropy in penetration depths
will furthermore be increased even further due to the effect of disorder,
as will be discussed in the next section.

Finally, it is interesting to consider the various different 
contributions to the condensation energy. Figure 15 shows results
for a two-dimensional case, for parameters appropiate to mimic
the in-plane behavior of the three-dimensional model for parameters
corresponding to the case of Figures (12-14) for bandwidths $1$ eV and
$0.1$ eV. All contributions to
the condensation energy are negative except the one corresponding
to the correlated hopping term, which is about $50$ times larger than
the condensation energy. Quite generally the kinetic energy
lowering in our model due to the pair hopping contribution is
much larger than the condensation energy and is 
partially compensated by an increase in Coulomb repulsion in the
paired state, because carriers in the pair are on average closer
to each other than in the normal state. Note from Fig. 13 that
the kinetic energy lowering from c-axis motion is a small fraction
of the kinetic energy lowering from in-plane motion,
in contrast to the prediction of the interlayer tunneling theory\cite{and1}.

In summary, we have seen in this section that for a given value of 
maximum $T_c$ the model can yield a fairly wide range of values of
condensation energy and sum rule violation depending on parameters in
the Hamiltonian. Still, the systematics with doping is always the same, 
as is the fact that the in-plane sum rule violation is always 
larger than the c-axis one. The magnitude of the 
condensation energy for given maximum $T_c$ decreases monotonically
as the bandwidth increases (i.e. density of states decreases) for any doping. 
In Fig. 16 we show the bandwidth dependence for the optimally doped case
and $T_c^{max}=90K$. As seen in Fig. 16, the effect of nearest neighbor
repulsion is to decrease somewhat the condensation energy. Both
the in-plane and interplane sum-rule
violation decrease as the bandwidth increases,
and the effect of nearest neighbor repulsion is to increase the
degree of sum-rule violation, as shown in Fig. 17. The violation is
always larger in-plane than out of plane, and the anisotropy in the
violation increases as the bandwidth increases except in the very
overdoped regime.

\section{Comparison with experiment}
It is generally accepted that in-plane transport in the cuprates is
described by the clean limit, and here we adopt this point of view. To
determine the parameters appropiate to the different materials we use
results for the condensation energy obtained by Loram\cite{loram}
from analysis of specific heat data. Loram reports a maximum
condensation energy $U_0=3.6 J/g-at$ for $YBCO$, $U_0=2.8 J/g-at$ for $YBCO$
with $20\%$ $Ca$ substituting for $Y$, $U_0=2 J/g-at$ for 
$Bi2212$, and $U_0=1.3 J/g-at$ for $LaSrCuO_4$. If we assume that the
condensation energy is dominated by the physics of the planar
oxygens, we have per planar oxygen a condensation energy $\epsilon_c$
given by
\bmath
\beq
\epsilon_c(\mu eV)=U_0 (\frac{Joule}{g-at})\times 5.18\times \frac{N_1}{N_2}
\eeq
\beq
N_1 = number\ of\ atoms\ in\ formula\ unit
\eeq
\beq
N_2=number\ of\ planar\ CuO_2's\ in\ formula\ unit
\eeq
\emath
yielding $\epsilon_c=121\mu eV$ for $YBCO$ ($N_1=13$, $N_2=2$),
$\epsilon_c=78\mu eV$ for $Bi2212$ ($N_1=15$, $N_2=2$),
$\epsilon_c=94\mu eV$ for $Y_{0.8}Ca_{0.2}Ba_2Cu_3O_{6+y}$,
and $\epsilon_c=47\mu eV$ for $LaSrCuO$ ($N_1=7$, $N_2=1$). Assuming
a value for the nearest neighbor repulsion we can then extract the
required value of the bandwidth by inspection of Fig. 16.

It can be seen that the maximum condensation energy for the various
materials with $T_c^{max} \sim 90$ K coincide within a factor of 2. 
Differences may be due to contributions to the condensation energy from
other atoms in the structure besides the $CuO_2$ units. We will
assume that a proper value for the bandwidth in our model to describe
a $T_c^{max} \sim 90$ K material is $D=0.5$ eV, which yields a maximum
condensation energy in the range of values given above, for reasonable
values of the nearest neighbor repulsion. Figure 18 shows the condensation
energy versus doping for values of the nearest neighbor repulsion
$V=0$ and $V=0.65$ eV. The maximum condensation energy is 
$\epsilon_c=104\mu eV$ and $\epsilon_c=90\mu eV$ respectively.
For $LaSrCuO$, the condensation energy reported by Loram is somewhat larger 
than expected if
the same bandwidth is assumed as for the $90K$ materials, since
Eq. (11) (assuming equal gap ratios) would predict a condensation
energy $\sim 6$ times smaller for $LaSrCuO$. If Loram's values are accurate
it would imply within our model a bandwidth of $D\sim 0.25$ eV for 
$LaSrCuO$, which would give rise to a substantial sum rule violation 
even in the clean limit, as seen in the previous section.

\subsection{Effect of disorder}

We choose the case of $YBCO$ for a detailed comparison with
experiment, since experimental 
results for the sum rule violation for several values of doping\cite{basov}
are available only for this material.
>From band structure calculations\cite{allen} we extract for the band 
structure anisotropy $t_a/t_c=10$. Figure 19 shows the calculated values
for the sum rule violation in the $a$ and $c$ directions in the clean limit 
for this case. For definiteness we will consider in what follows the case
of zero nearest neighbor repulsion.
For the $c$ direction we also show the experimental values measured
by Basov et al\cite{basov}. We chose to assign to the samples in
Basov's experiments the value of doping that would give rise to
the same critical temperature in our model as seen experimentally.
The results in Fig. 19b show that experiments exhibit a much faster
decrease in the sum rule violation with doping than our clean-limit
calculation predicts.

However, as mentioned earlier there exists substantial experimental evidence
pointing to the fact that transport in the c-direction is described by
the dirty rather than the clean limit\cite{marel,uchida,tajima,shiba}. 
Furthermore, transport\cite{takenaka}
as well as optical\cite{uchida} experiments indicate that the c-axis
scattering rate decreases rapidly as the doping increases. This has been
modeled theoretically\cite{levin} taking into account diagonal and 
off-diagonal disorder, and assuming that Coulomb effects cause both 
the in-plane and interplane hopping amplitudes to increase with doping.
This latter assumption is in fact the basis of our model, i.e.
Eq. (1). These experiments and calculations suggest that the c-axis
transport evolves from the dirty towards the clean limit as the doping 
increases. As we show in what follows, this is in fact consistent with
the predictions of our model, and leads to a faster rate of decay of
the c-axis sum rule violation with doping than in the clean limit,
consistent with observations. 

The penetration depths at zero temperature are related to the kinetic energies
by
\bmath
\beq
\lambda_a(\AA)=4638 (d(\AA))^{1/2}\frac{1}{(T_a(meV))^{1/2}}
\eeq
\beq
\lambda_c(\AA)=4638 (\frac{a(\AA)^2}{d(\AA)})^{1/2}\frac{1}{(T_c(meV))^{1/2}}.
\eeq
\emath
The in-plane kinetic energy obtained from our model for the parameters
under consideration and optimal doping is $T_a=6.6$ meV (per oxygen site). 
It is not completely
clear what the parameter $d$ in Eq. (57a) should be. If we assume that
only the oxygens in the CuO planes contribute to the penetration depth,
we should take for $d$ one-half of the c-axis lattice constant $d=11.68 \AA$, which 
yields from Eq. (57a) an in-plane penetration depth $\lambda_a\sim 4300 \AA$. This is 
substantially larger than the
observed penetration depth, $\lambda_a\sim 1400 \AA$ \cite{panag}, and suggests that 
carriers
from other atoms in the structure also contribute to the superfluid weight.
We will simply treat $d$ in Eq. (57a) as a parameter to be determined to fit the
observed penetration depth in the optimally doped case. For Eq. (57b) we will
use the YBCO lattice constants $a=3.84 \AA$ and $d=11.68 \AA$.

Figure 20 shows calculated values of the penetration depth and experimental 
observations\cite{panag,basov}. Again we infer the appropiate values of n for 
the experimental
data by comparison of experimental and calculated $T_c/T_c^{max}$. The
in-plane calculated penetration depth for small doping becomes somewhat
larger than the experimental one. The discrepancy may arise from contribution
to the superfluid weight from carriers in other bands with weaker carrier
concentration dependence with doping than the oxygen band described by our
model\cite{twoband}.

The calculated penetration depth in the c direction increases much more slowly than the 
experimental one as the doping decreases. We attribute this to an 
enhanced effect of disorder in the c-axis transport in the underdoped regime.
This is in fact consistent with interpretation of optical experiments
in LaSrCuO\cite{uchida} indicating a stronger c-axis scattering rate in the underdoped
regime, as well as with transport experiments in YBCO\cite{takenaka}. 
By comparison of calculated and observed penetration depth
anisotropies, shown in Fig. 21, we extract the doping dependence of the
c-axis scattering rate in Eq. (51):
\beq
(1 +  \frac{\hbar/\tau_c}{\pi \Delta})^{-1} \equiv 
p_c=\frac{(\lambda_c/\lambda_a)^2_{theory, clean limit}}
{(\lambda_c/\lambda_a)^2_{experiment}} .
\eeq
This assumes that the penetration depth in the plane is described
approximately by the clean limit result given by our calculation. The c-axis
scattering rate thus obtained has a strong doping dependence.

We can hence calculate the doping dependence of the c-axis sum rule
violation in the presence of disorder, as determined by Eq. (53):
\beq
V_c=V_c^{clean}(\frac{\lambda_c}{\lambda_c^{clean}})^2=
V_c^{clean}p_c^2
\eeq
and plot the results in Fig. 22, together with experimental 
observations\cite{basov}. The reported experimental error in the sum rule
violation parameter is approximately $10\%$\cite{basov2}. It can be seen
that the theoretical results now show a much stronger variation
with doping than in the absence of disorder, and are closer to
the experimental observations. A similar analysis was recently
given for $LaSrCuO_4$\cite{houston}.

\subsection{Kinetic energy lowering}

If instead of considering the relative amount of sum rule violation we consider
the absolute amount of kinetic energy lowering, we need not worry about
the effect of disorder. The amount of kinetic energy in the zero-frequency
$\delta$-function is given by
\beq
Q_\delta=\frac{2\hbar^2 v}{\pi e^2 a_\delta ^2}D_\delta=
\frac{\hbar^2 c^2 v}{4 \pi e^2 a_\delta ^2}\frac{1}{\lambda_\delta^2} .
\eeq
with v the volume per $Cu O_2$ $planar$ $unit$.
The kinetic energy lowering is obtained from the experimentally measured
penetration depth and sum rule violation as
\beq
\Delta Q_\delta=Q_\delta V_\delta.
\eeq
In particular, for the c-direction
\bmath
\beq
\Delta Q_c=Q_cV_c
\eeq
\beq
Q_c=\frac{\hbar^2 c^2 a^2}{4 \pi e^2 d}\frac{1}{\lambda_c^2}
\eeq
\emath
For YBCO, with two $CuO_2$ planes per unit cell, we 
take $d$ as half the unit cell dimension in the c-direction, and Eq. (62b) gives
the kinetic energy per $CuO_2$ planar unit.

Basov and coworkers reported values for $V_c$ for $YBCO$ for several
dopings, for $Tl2201$ for an optimally doped and one overdoped sample,
and for slightly underdoped $LaSrCuO$, together with the corresponding
c-axis penetration depths\cite{basov}. Furthermore, Basov reports an
error of approximately $10\%$ for the reported value of $(1-V_c)$\cite{basov2}.
We can then obtain the experimentally observed absolute value of kinetic
energy lowering from Eq. (62), and the associated error from
\beq
\delta (\Delta Q_c)\sim 0.1 Q_c
\eeq
assuming the relative error in the measured penetration depth is much smaller than in 
the sum rule violation. Table I summarizes Basov's results and the
resulting values for kinetic energy lowering from Eqs. (62) and (63).

In our model, the kinetic energy lowering is given by the expectation value
of the correlated hopping term $<T_\delta ^{\Delta t}>$. Since there are
two $O$ atoms per $CuO_2$ unit, we have
\beq
\Delta Q_c)_{theory}=2 <T_\delta ^{\Delta t}>
\eeq
since our calculated kinetic energy lowering is per $O$ atom.

We estimate hopping anisotropies from band structure calculations to be
approximately  $t_a/t_c=10$ for $YBCO$ \cite{allen}, $25$ for 
$LaSrCuO$ \cite{allen}, and $50$ for $Tl2201$ \cite{pickett}. In the latter it should be 
noted this is estimated to be the average anisotropy for all bands,
while for the $Cu-O$ band alone it is estimated to be about $600$.
According to
our estimates from the previous subsection we assume bandwidths
$D=0.5$ eV for $YBCO$ and $Tl2201$, and $D=0.25$ eV for $LaSrCuO$. 
In Fig. 23 we plot the observed and calculated values of kinetic energy
lowering. For $LaSrCuO$, we use a non-zero nearest neighbor repulsion
in order to fit the rather large value of kinetic energy lowering
observed. It can be seen that our calculation gives reasonable agreement
with experimental measurements in the underdoped regime.
In the optimally doped and overdoped regimes 
our calculation predicts a significant kinetic energy lowering, but unfortunately
experimental errors are at present too large to confirm or rule out our predictions.

\section{Summary and conclusions}

The sum rule violation considered here has also been discussed by
other workers. Kim\cite{kim} considered the role of impurity scattering
in c-axis transport for a $d_{x^2-y^2}$ gap and concluded that the
superfluid weight could be both larger or smaller than the missing
area in the conductivity depending on parameters. Ioffe and
Millis\cite{millis} argued that the explanation of Basov's observations
lies in the interplay of phase coherence, quantum and thermal fluctuations,
and scattering processes.  Neither of these treatments predicts an
in-plane sum rule violation, in contrast to our model. Kim and
Carbotte\cite{carbotte} found that if there is coherent interlayer
coupling there should be no c-axis sum rule violation; their model 
however did not include a correlated hopping term. Furthermore they
found that for incoherent interlayer coupling there should be 
sum rule violation, however of opposite sign to that observed. Within
their model (without a $\Delta t$ term) they found that in-plane sum
rule violation (of either sign depending on parameters) can occur only
if the electronic density of states has fine structure on the scale of
the superconducting gap.

In summary, we have studied here the predictions of the model of hole superconductivity
for the condensation energy and for quantities related to the optical
sum rule, for a wide range of parameters. The model predicts a violation
of the Ferrell-Glover-Tinkham sum rule that can range from a few percent
to close to $100\%$ depending on the parameters in the Hamiltonian. For given
maximum $T_c$, the most important parameter determining the magnitude of
sum rule violation is the bandwidth, or density of states: larger
density of states gives rise to larger sum rule violation, as well as to 
larger condensation energy. Comparison of calculated and measured
condensation energy for a given system allows for a determination of the
parameters in the Hamiltonian appropiate for that system.

The transition to the superconducting state is driven by lowering of 
kinetic energy for all dopings in this model. Quite generally, the 
lowering of kinetic energy is one to two orders of magnitude larger than
the superconducting condensation energy. Because the condensation energy
peaks approximately at the same doping as $T_c$, the model predicts
a substantial kinetic energy lowering in the overdoped regime.
For the anisotropic structures of the cuprates, the contribution to 
the condensation energy from in-plane kinetic energy lowering is
two to three orders of magnitude larger than that of interplane motion.

Furthermore the model predicts that in the clean limit the in-plane
sum rule violation should be a few times larger than the interplane one.
The fact that the opposite has been reportedly observed so far\cite{basov} 
(large c-axis violation, no in-plane violation) leads us to conclude that there is
a significant effect of disorder in c-axis transport, that suppresses the
low frequency spectral weight and hence allows for easier
detection of the anomalous high frequency spectral weight. This assumption is in
fact consistent with a variety of other experimental 
observations\cite{marel,uchida,shiba,tajima}, that have
led several workers to conclude that c-axis transport is described by the
dirty limit and planar transport by the clean limit. Using this assumption
our model can explain the large difference between observed c-axis and
in-plane sum rule violations.

Furthermore, comparison of our calculated and measured c-axis penetration
depths led us to conclude that the effect of disorder in c-axis transport 
increases substantially in the underdoped regime. This conclusion is also
in fact consistent with other independent experimental 
observations\cite{uchida,takenaka}. This fact
leads to a much more rapid doping dependence of the sum rule violation than that
obtained in the clean limit, which resembles the experimental
observations\cite{basov}.

Experimental papers have suggested that the sum rule  violation effect disappears
in optimally doped or overdoped regime, and concluded from this that there 
is no anomalous kinetic energy lowering in those regimes\cite{basov}. We argue that this
conclusion is flawed. Because the 'normal' contribution to the superfluid weight
increases rapidly as the doping increases, it can easily mask the anomalous
part of the superfluid weight, which has, according to our calculation,
a much slower doping dependence and decreases slowly in the overdoped regime. 
In experimental papers\cite{basov} it is always the fractional sum rule violation that is
plotted, rather than the absolute value of kinetic energy lowering. This is
presumably due to the fact that large error bars prevent meaningful extraction
of the latter quantity. However it should then be recognized that an apparent vanishing
of the fractional sum rule violation within error bars does not imply a 
vanishing of the anomalous kinetic energy lowering.

There is another fundamental reason to reject this experimental conclusion.
If there is indeed the unusual phenomenon of kinetic energy lowering,
contrary to ordinary BCS theory, it is logical to conclude that whatever
mechanism is causing the kinetic energy lowering drives the transition to
superconductivity, and that the lowering of kinetic energy 
is responsible for the condensation energy of the superconductor. At present 
both our calculations and
experiments\cite{loram} indicate that the condensation energy in the cuprates as a function of
doping peaks in the optimally doped or even overdoped regimes. If there
is no kinetic energy lowering in the optimally doped or overdoped regimes,
as proposed by Basov, it would imply that a different mechanism explains
superconductivity in those regimes. Moreover, the kinetic energy lowering
mechanism would apply for optimally doped $Tl_2Ba_2CuO_{6+\delta}$
 but not to optimally doped
$YBa_2Cu_3O_x$ according to Basov's point of view. Clearly, while such a scenario
would not be impossible, it does not appear to be very plausible.

In this paper we have not discussed the temperature dependence of these
effects. So far, no experimental results have been reported for the 
temperature dependence of missing areas and sum rule violation. 
We have discussed elsewhere for selected cases the temperature dependence of
the real part of the conductivity\cite{coherence}, London penetration depth\cite{lond}
and high and low frequency missing areas\cite{area} within our model. Once experimental
results for these quantities become available it will be possible to 
provide detailed comparison with the theory.

In conclusion we note that the mechanism of hole superconductivity 
discussed here is the only mechanism of superconductivity so far proposed
that both predicts an optical sum rule violation due to kinetic
energy lowering, and allows for quantitative
evaluation of the doping and temperature dependence of the effect. Existing
experiments confirm the existence of the c-axis sum rule violation, and suggest
that kinetic energy lowering also occurs in in-plane transport\cite{uchida2,basov}, 
in agreement with the predictions of the model.
Future more accurate experiments should be able to provide more stringent
tests of the theory. Furthermore, the model of hole superconductivity is the only
one proposed so far that provides an explanation for the origin of 
the high frequency spectral weight that appears in the zero-frequency
delta-function\cite{polar}. It remains a challenge 
for other theories to provide
explanations for these unusual experimental observations. 

\acknowledgements
The authors are grateful to D. Basov and J. Loram for stimulating
discussion and for sharing their experimental results prior to
publication. One of us (FM) acknowledges the Natural Sciences and
Engineering Research Council (NSERC) of Canada and the Canadian
Institute for Advanced Research (CIAR) for support.

\widetext
\begin{table}
\caption{Experimental results for c-axis transport, from Ref. 6. }
\begin{tabular}{|l|c|c|c|c|c|c|}
Material& $T_c(K)$ &$ T_c/T_c^{max}$  & $\lambda_c(A)$ & $1-V_c$ & 
 $Q_c (\mu eV)$ &$\Delta Q_c (\mu eV)$\cr
\tableline
$YBa_2Cu_3O_{6+\delta}$& & &  & & & \cr
$\delta=6.5$  & 50 & 0.53  & 77,350 & 0.2 & 9.08 & 7.26+/- 0.18 \cr
$\delta=6.6$  & 60 & 0.64  & 63,400 & 0.37 & 13.5 & 8.51+/- 0.50 \cr
$\delta=6.7$  & 65 & 0.70  & 51,500 & 0.87 & 20.5 & 2.66+/- 1.8 \cr
$\delta=6.8$  & 80 & 0.86  & 35,000 & 0.77 & 44.3 & 10.2+/- 3.4 \cr
$\delta=6.85$  & 85 & 0.91 & 30,940 & 1 & 56.7 & 0+/- 5.7 \cr
$\delta=6.9$  & 90 & 0.96  & 15,400 & 1 & 229& 0+/- 23 \cr
$\delta=6.95$  & 93.5 & 1  & 10,300 & 1 & 512 & 0+/- 51 \cr
\hline
$Tl2201$& & &  & & & \cr
opt. doped  & 81 & 1  & 119,000 & 0.5 & 1.99 & 1.0+/- 0.20 \cr
overdoped  & 32 & 0.40  & 110,000 & 0.9 & 2.33 & 0.23+/- 0.21 \cr
\hline
$LaSrCuO_4$  &32 & 0.85 & 50,000 & 0.4 & 18.6 & 11.2+/- 0.7 \cr
\end{tabular}
\end{table}
\begin{figure}
\caption{Sketch of the real part of the conductivity in the superconducting (dashed lines)
and
normal state (solid lines). The delta function at the origin is the superfluid
weight $D$ that determines the London penetration depth.
Additional weight is present in the delta function 
in the superconducting state that originates at high frequency in the normal state. 
Note that in
the dirty limit the contribution to the delta function originating
at low frequencies is 
reduced, and hence the additional weight from high frequencies
represents a substantially larger fraction of the 
delta function. This figure is discussed in detail in section III.}
\end{figure}

\begin{figure}
\caption{Critical temperature $T_c$ vs hole concentration $n$ for
various bandwidths. In all cases we used $U = 5$ eV, $V = 0$, and $\eta = 25$
for the band anisotropy. The parameter $\Delta t$ was determined to give (a)
$T_c^{\rm max} = 37.5$ K and (b) $T_c^{\rm max} = 90$ K. 
For bandwidths $D=1.5eV, 1 eV, 0.5 eV, 0.1 eV$ the values of $\Delta t$
used, in $eV$, are (a) $\Delta t= 0.295, 0.263, 0.215, 0.141 $ and 
(b) $\Delta t= 0.318, 0.286, 0.240, 0.173 $ respectively}
\end{figure}

\begin{figure}
\caption{Condensation energy, $\epsilon_{\rm cond}$ per planar oxygen, vs. hole concentration
for the same parameters as in Fig. 2. The condensation energy decreases with
increasing bandwidth, and peaks at roughly the same doping level at which $T_c$ peaks.}
\end{figure}

\begin{figure}
\caption{The in-plane sum rule violation, $V_a$, Eq. (44), vs. hole concentration, for the
same cases as in the two previous figures. Except for a small doping level near
zero, the violation parameter decreases to zero as hole doping increases.
The magnitude of the violation increases as the bandwidth decreases,and is larger when the 
maximum $T_c$ is larger.}
\end{figure}

\begin{figure}
\caption{The out-of-plane sum rule violation, $V_c$, vs. hole concentration, for the
same cases as in the three previous figures. The violation is smaller than the in-plane
violation, but otherwise very similar.}
\end{figure}

\begin{figure}
\caption{The sum rule violation anisotropy, $V_a/V_c$, which results from the previous
two figures, as a function of hole doping. The anistropy tends to increase as the
bandwidth increases, reaching a value of 5 for the cases considered here.
The anisotropy is smaller for the case of larger $T_c^{max}$.}
\end{figure}

\begin{figure}
\caption{Critical temperature $T_c$ vs hole concentration $n$ for
various bandwidths, as in Fig. 2, but now for $V = 0.65$ eV. As before, $U = 5$ eV
and $\eta = 25$ for the band anisotropy. The detailed dependence on bandwidth is
more pronounced than with $V= 0$. Again the parameter $\Delta t$ was determined to
give (a) $T_c^{\rm max} = 37.5$ K and (b) $T_c^{\rm max} = 90$ K. 
For bandwidths $D=1.5eV, 1 eV, 0.5 eV, 0.1 eV$ the values of $\Delta t$
used, in $eV$, are (a) $\Delta t= 0.517, 0.503, 0.486, 0.470 $ and 
(b) $\Delta t= 0.533, 0.518, 0.500, 0.483 $ respectively.}
\end{figure}

\begin{figure}
\caption{Condensation energy vs. hole concentration, as in Fig. 3, but with the same
parameters as in Fig. 7 (i.e. with $V=0.65$ eV). The condensation energy is decreased
compared to the $V=0$ case in Fig. 3.}
\end{figure}

\begin{figure}
\caption{The in-plane sum rule violation, $V_a$, vs. hole concentration, for
$V= 0.65$ eV. The magnitude of the violation is significantly increased in
comparison to the $V=0$ case (Fig. 4).}
\end{figure}

\begin{figure}
\caption{The sum rule violation anisotropy, $V_a/V_c$, for $V= 0.65$ eV. The
results are similar to the $V=0$ case (Fig. 6).}
\end{figure}

\begin{figure}
\caption{The in-plane kinetic energy (per planar oxygen)
vs. doping for the case with $T_c^{\rm max} = 90$
K and $V = 0.65$ eV. We show (a) the single-particle contribution, (b) the pair
contribution, and (c) the total. The single-particle contribution increases with
increasing bandwidth and dominates the pair contribution, which decreases with
increasing bandwidth. Note that the pair contribution peaks close to the maximum $T_c$, 
while the single particle contribution increases monotonically with doping. }
\end{figure}

\begin{figure}
\caption{The out-of-plane kinetic energy vs. doping for the same case as in Fig. 11. The
results are similar to the in-plane kinetic energy.}
\end{figure}

\begin{figure}
\caption{The anisotropy of the single particle (a) and 
pair contribution (b) to the kinetic energy for the same
case as in Fig. 11.}
\end{figure}

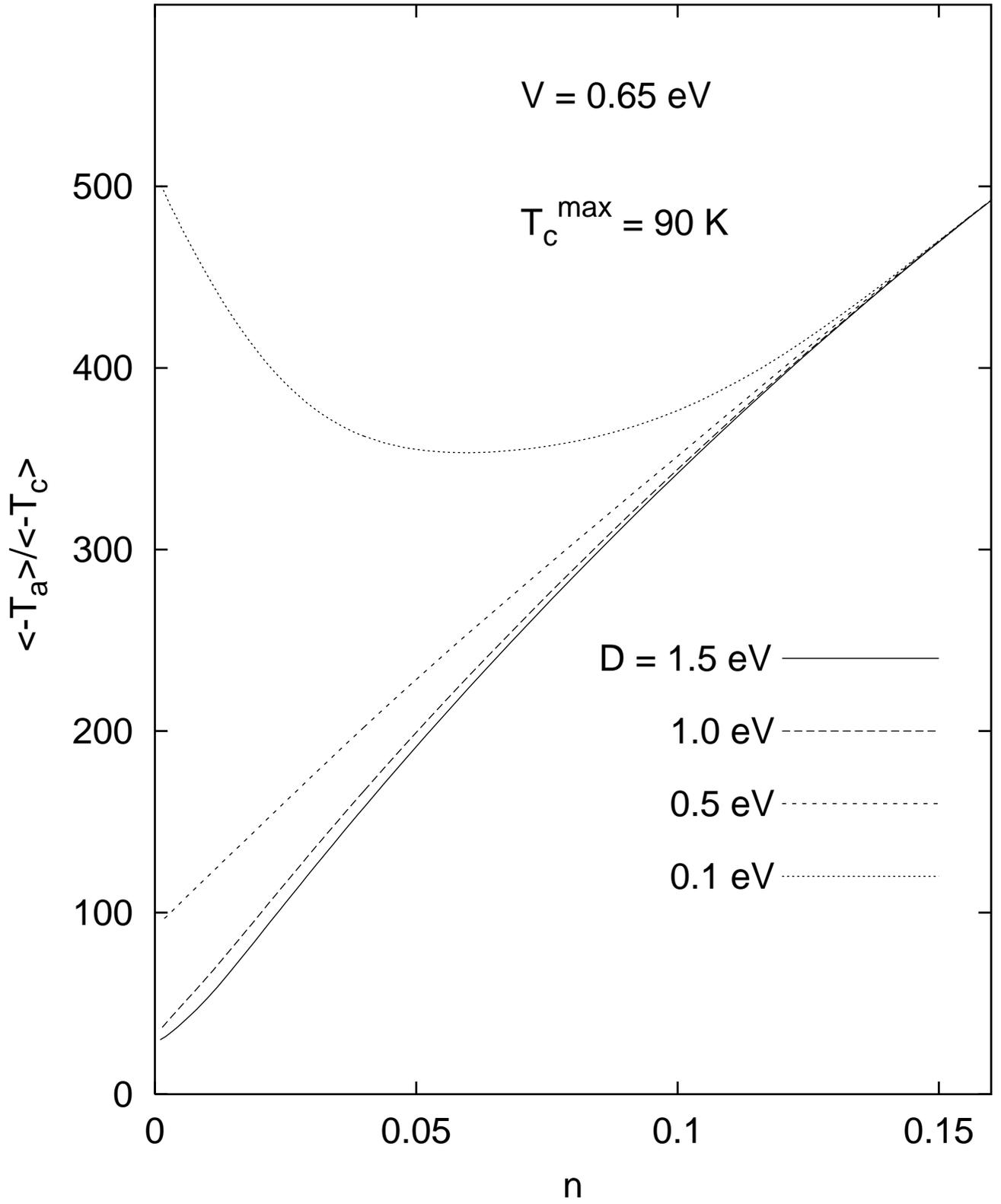
\begin{figure}
\caption{The anisotropy of the total kinetic energy for the same case as in
Fig. 11. The anisotropy is proportional to $\eta$ at low hole densities in
the weak coupling limit. At higher densities, the anisotropy is proportional
to $\eta^2$ in weak coupling and rises rougly in proportion to hole density. 
In strong coupling, the anisotropy is proportional to $\eta^2$ for all 
hole densities.}
\end{figure}

\begin{figure}
\caption{Various contributions to the condensation energy, using two cases from
Figs. (11-14), one with bandwidth (a) $D=1.0$ eV, and one with bandwidth (b) $D=0.1$ eV.
For simplicity we used a two-dimensional case, with a constant density of states, and
$\Delta t$ re-determined to give $T_c^{\rm max} = 90$ K. The parameters in the
two-dimensional model used to yield the same density of states at the Fermi level
and condensation energy
as the three-dimensional model for these cases are: (a) $D=1.6 eV$, $\Delta t=0.51 eV$
and (b) $D=0.25 eV$, $\Delta t=0.48 eV$.}
\end{figure}

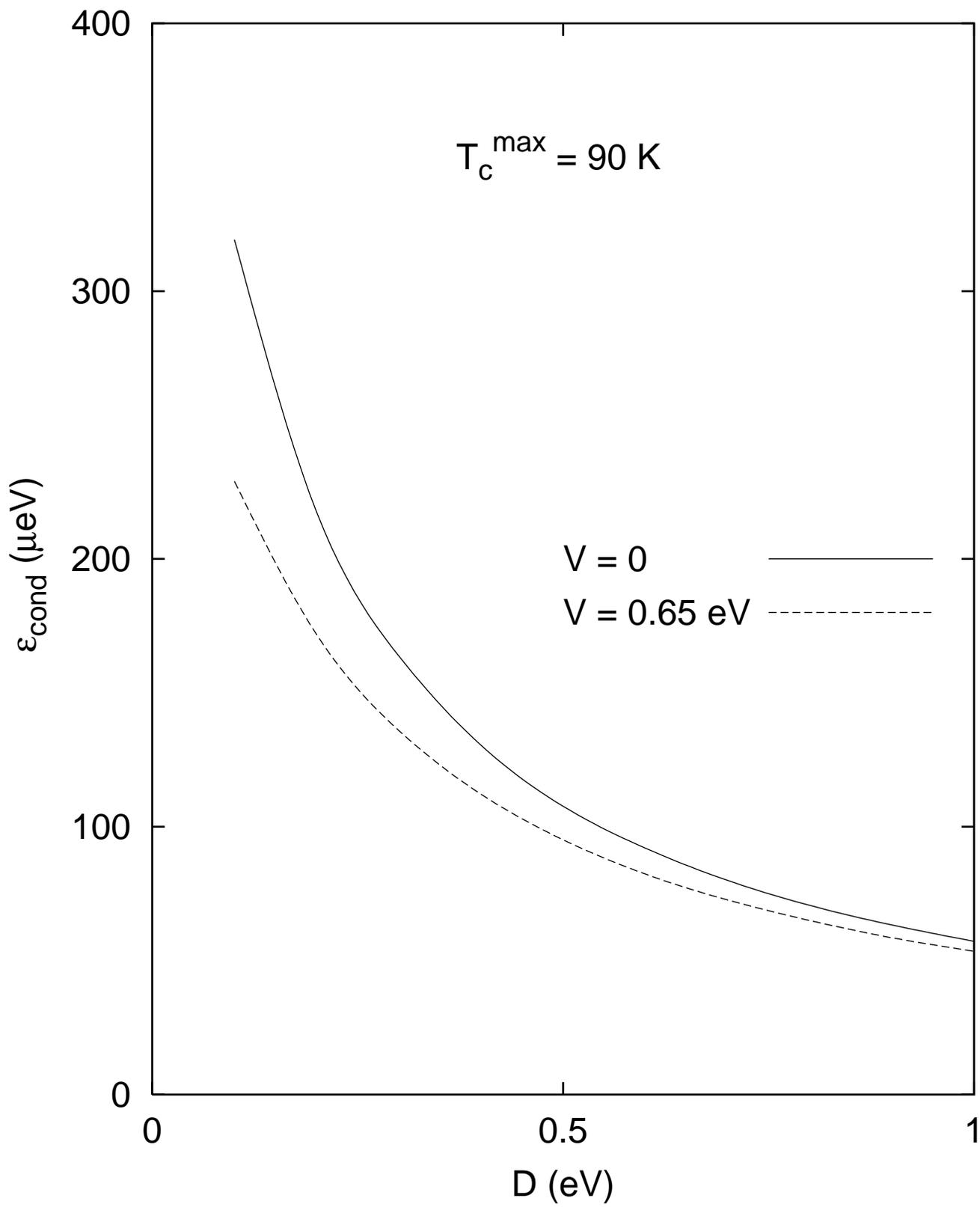
\begin{figure}
\caption{Condensation energy vs. bandwidth for two values of $V$ for the
optimally doped case (with
$T_c^{\rm max} = 90$ K). A non-zero nearest neighbour repulsion tends to
decrease the condensation energy for a fixed value of $T_c^{max}$.}
\end{figure}

\begin{figure}
\caption{The (a) in-plane, and (b) out-of-plane sum rule violation as a
function of bandwidth, for the same two cases as in Fig. 16. The sum rule
violation tends to increase with increasing $V$.}
\end{figure}

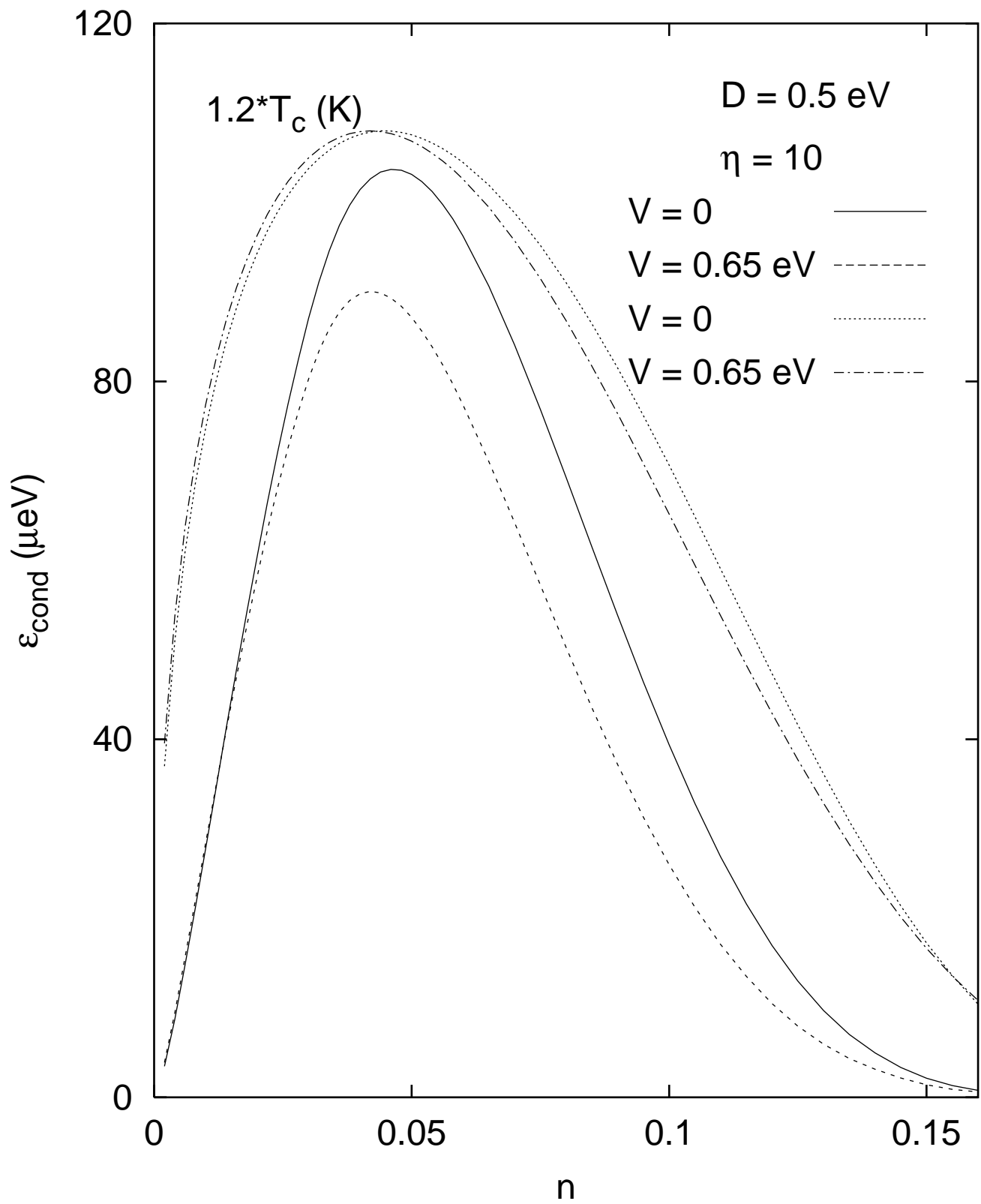
\begin{figure}
\caption{The condensation energy vs. doping for a specific case: $D = 0.5$ eV,
$U = 5$ eV, $T_c^{\rm max} = 90$ K, for two values of $V$. $\Delta t=0.241 eV$ for
$V=0$, $\Delta t=0.501 eV$ for
$V=0.65 eV.$ The $T_c$ curves
are also shown to facilitate a comparison of peak positions.}
\end{figure}

\begin{figure}
\caption{The (a) in-plane, and (b) out-of-plane sum rule violation as a function
of doping for the two cases in Fig. 18. In Fig. (19b) we include for comparison
the experimental data from $YBCO$ from Basov {\it et al.} \protect\cite{basov}.}
\end{figure}

\begin{figure}
\caption{The (a) in-plane, and (b) c-axis, penetration depth, as a function of hole
doping, using the case from the previous two figures with $V = 0$. As the doping
decreases the experimental data \protect\cite{panag} in (b) increases
much more quickly than the theoretical clean limit result. }
\end{figure}

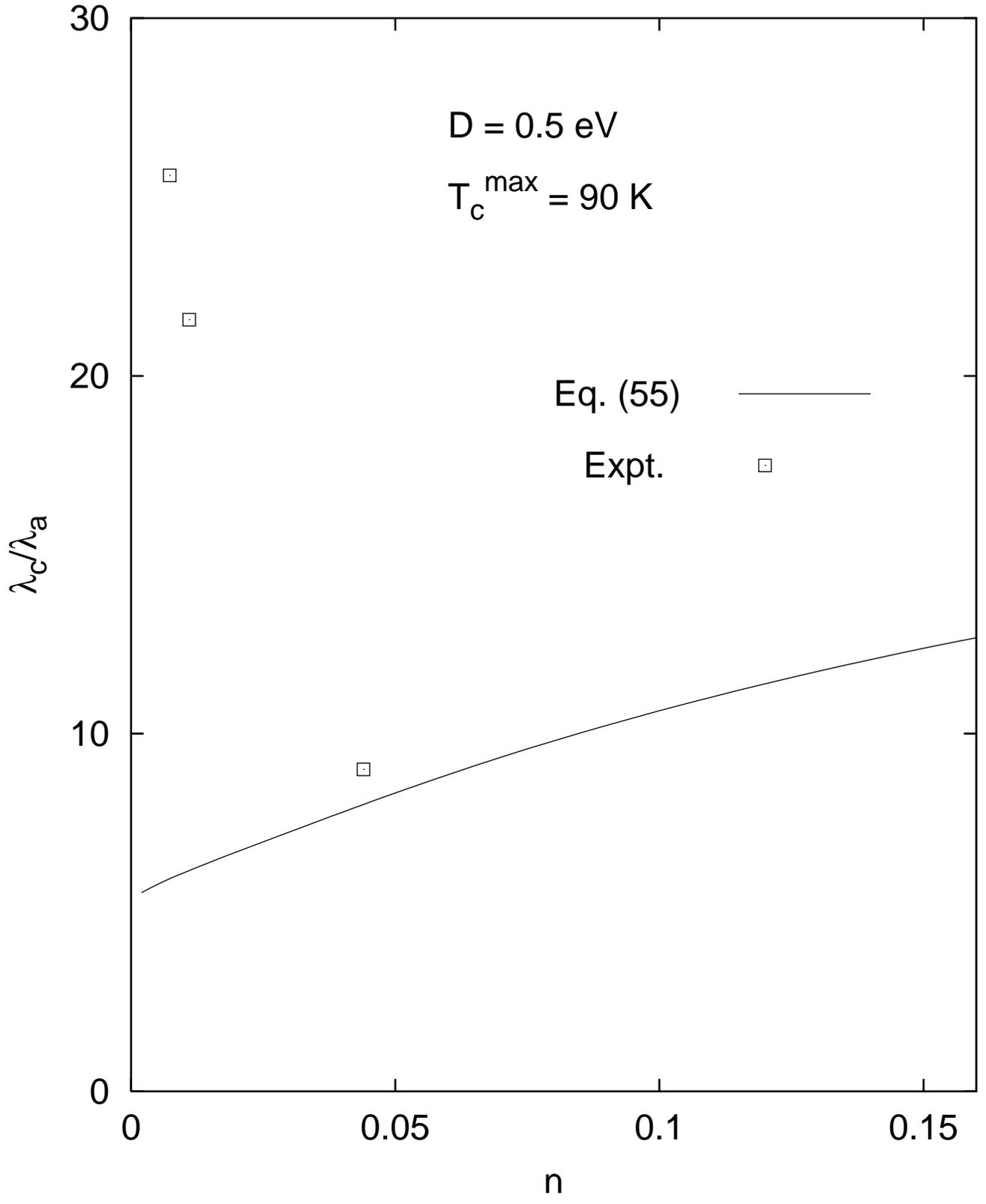
\begin{figure}
\caption{The ratio of the penetration depths vs. hole doping, along with the
experimental results \protect\cite{panag}, used to determine the impurity parameter,
$p_c$ (Eq. (56)).}
\end{figure}

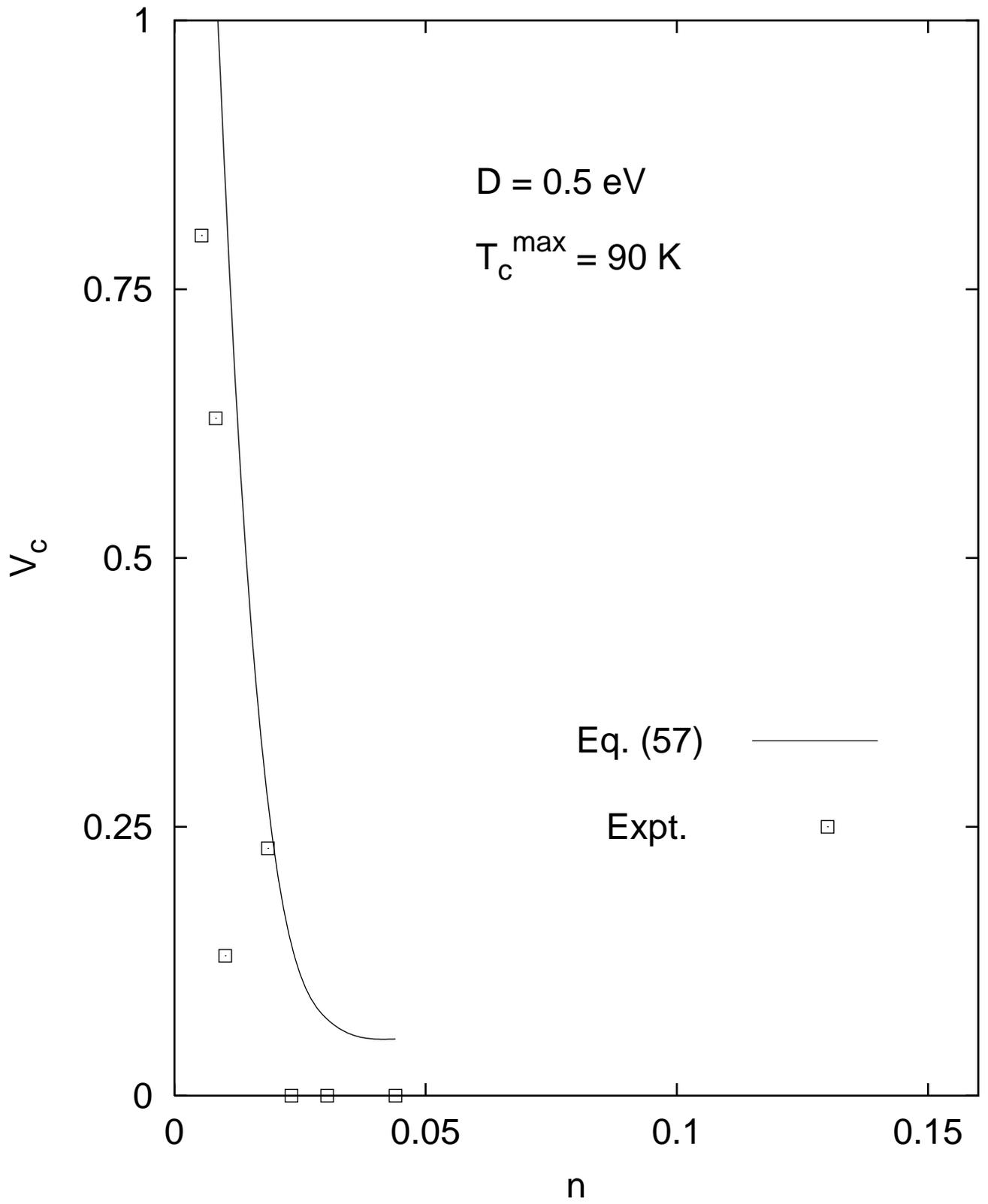
\begin{figure}
\caption{The c-axis sum rule violation vs. doping, together with the experimental
results \protect\cite{basov}. The agreement with experiment is satisfactory, once
the stronger c-axis impurity scattering and its doping dependence is accounted for.}
\end{figure}

\begin{figure}
\caption{The anomalous contribution to the out-of-plane kinetic energy, for
(a) YBCO, (b) Tl2201, and (c) LSCO, plotted along with data from Basov
\protect\cite{basov}. 
Note that the error bars become very large as the doping increases
(particularly in (a), where two of them exceed the
page size). Representative values of the band structure anisotropy are shown in
each case. The results are qualitatively consistent with the data, within the error.
$\Delta t=0.240 eV$ for (b), $0.477 eV $ for (c).}
\end{figure}


\begin{references}
\bibitem{approx} Actually, in many cases the approximation alluded to (discussed
in detail in the third of Refs. \cite{hole1}) is exact (for example, when $V = 0$).
\end{references}
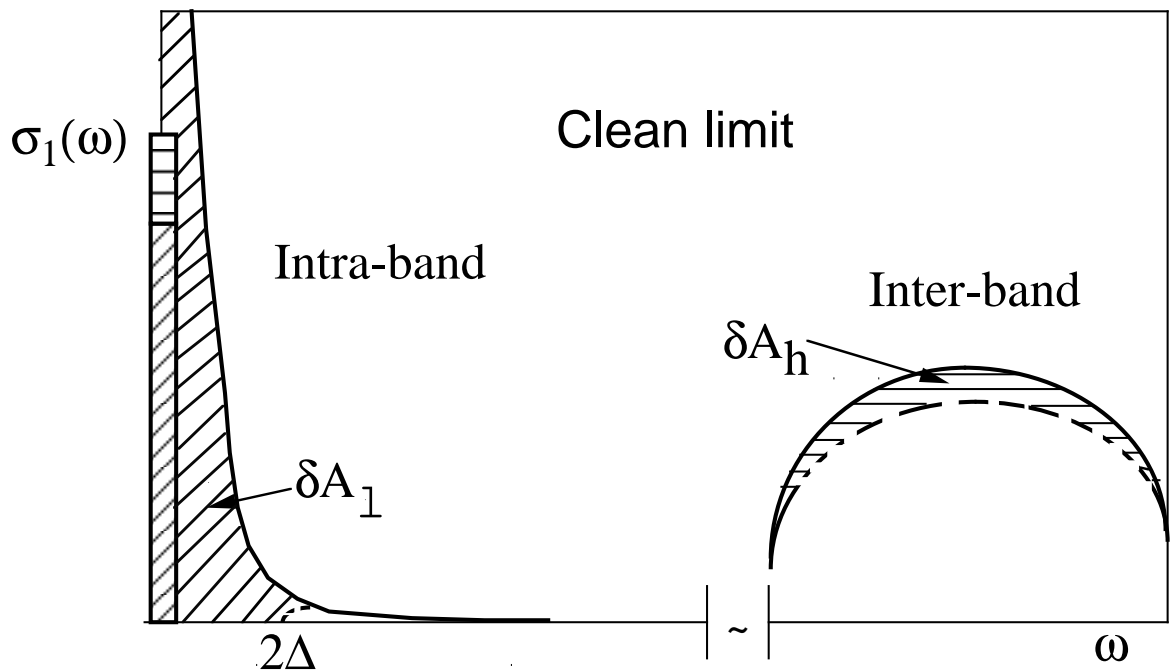
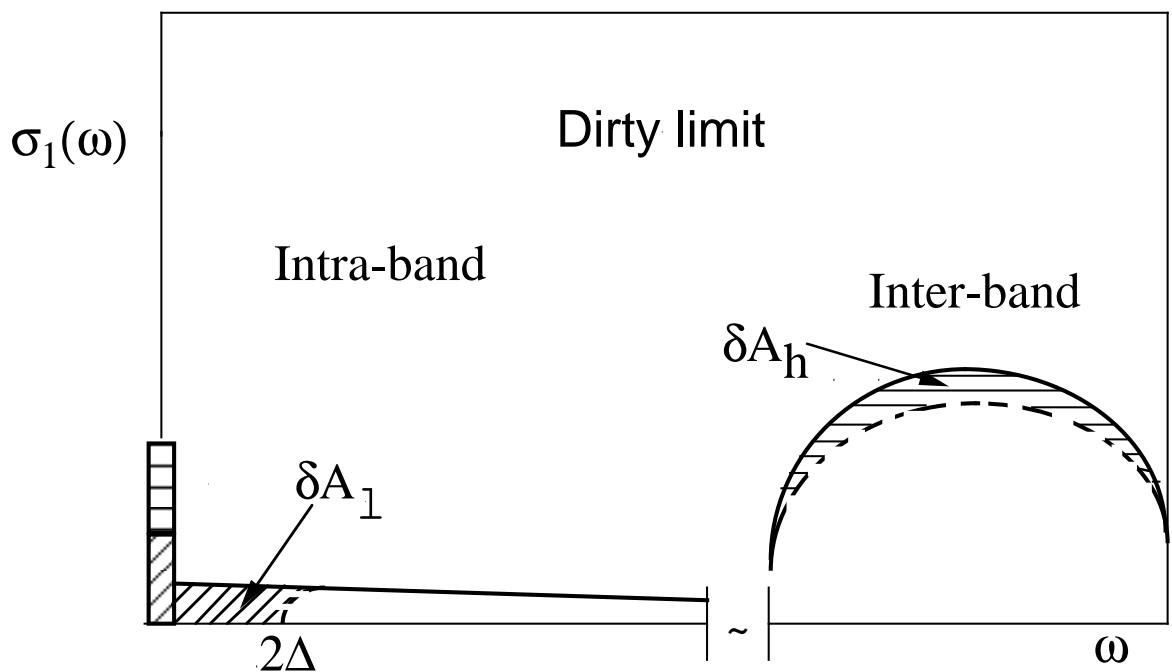
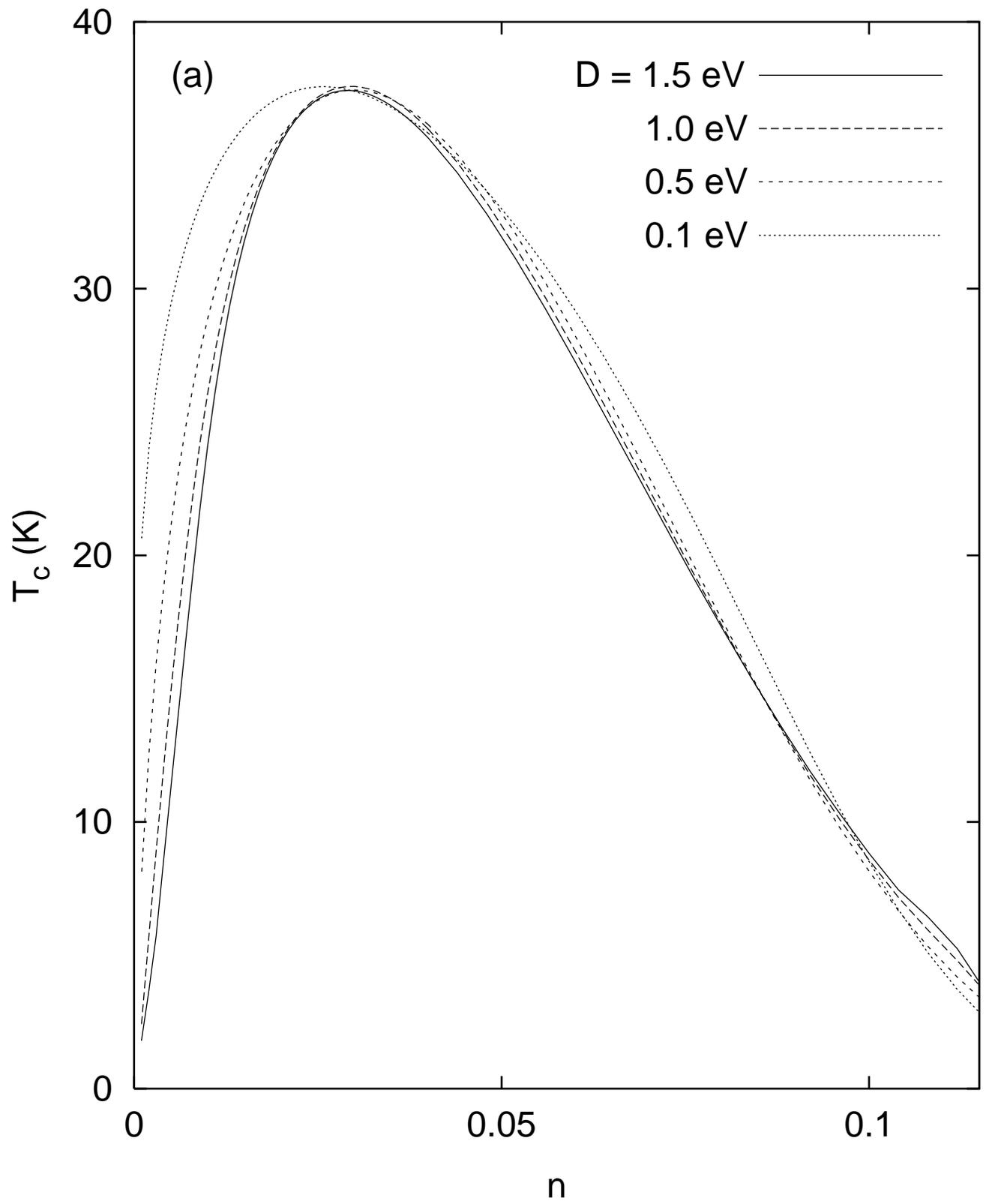
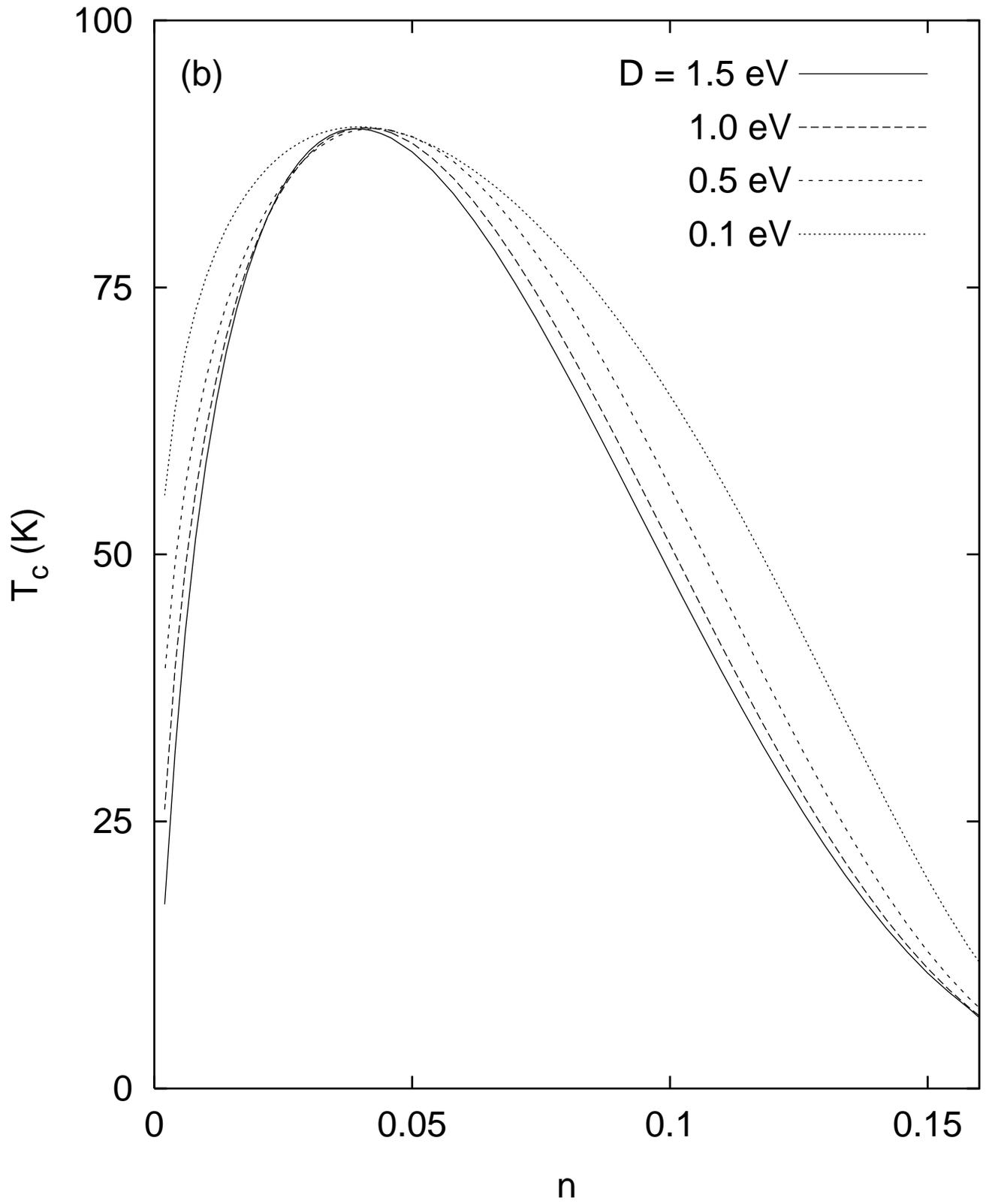
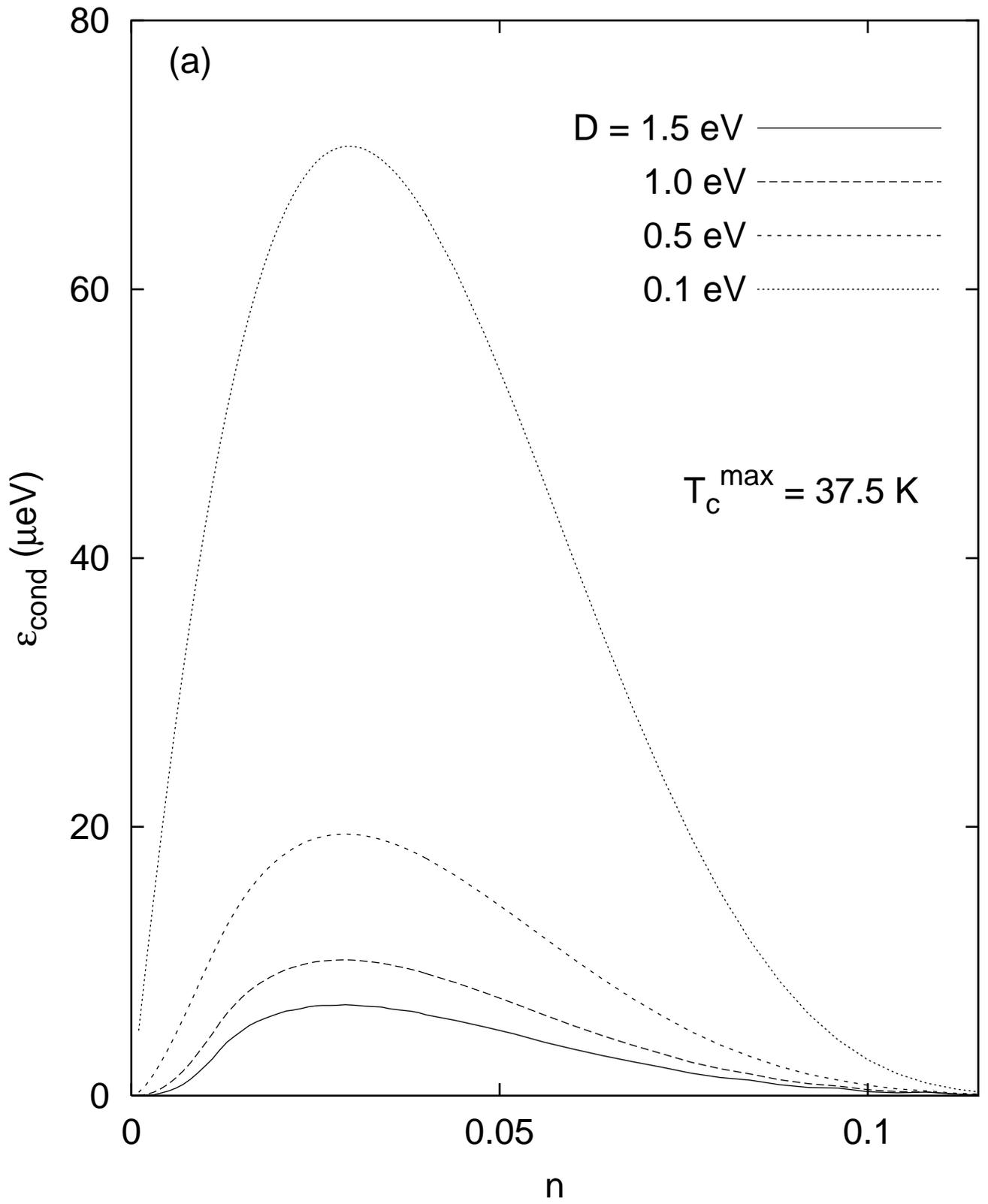
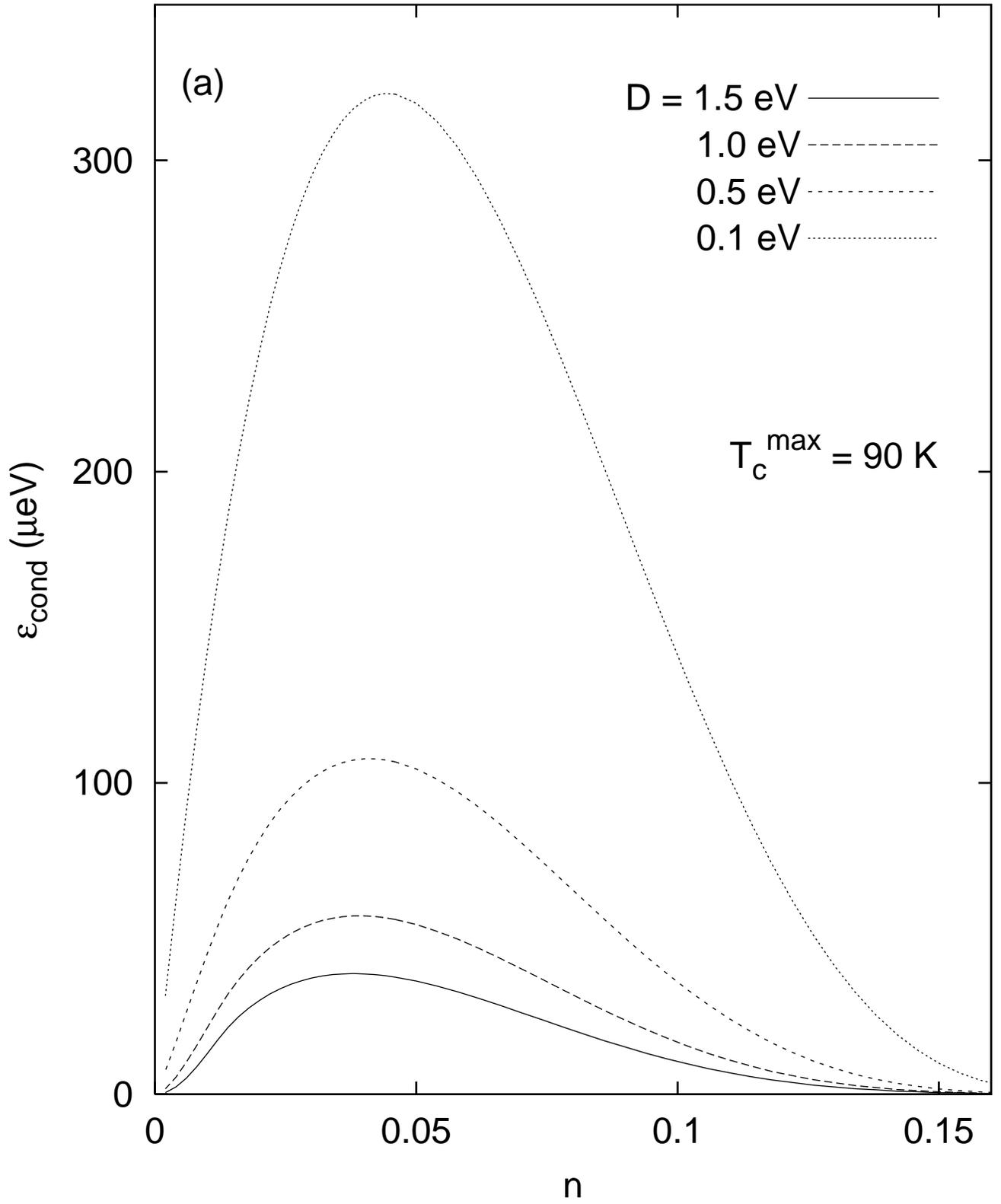
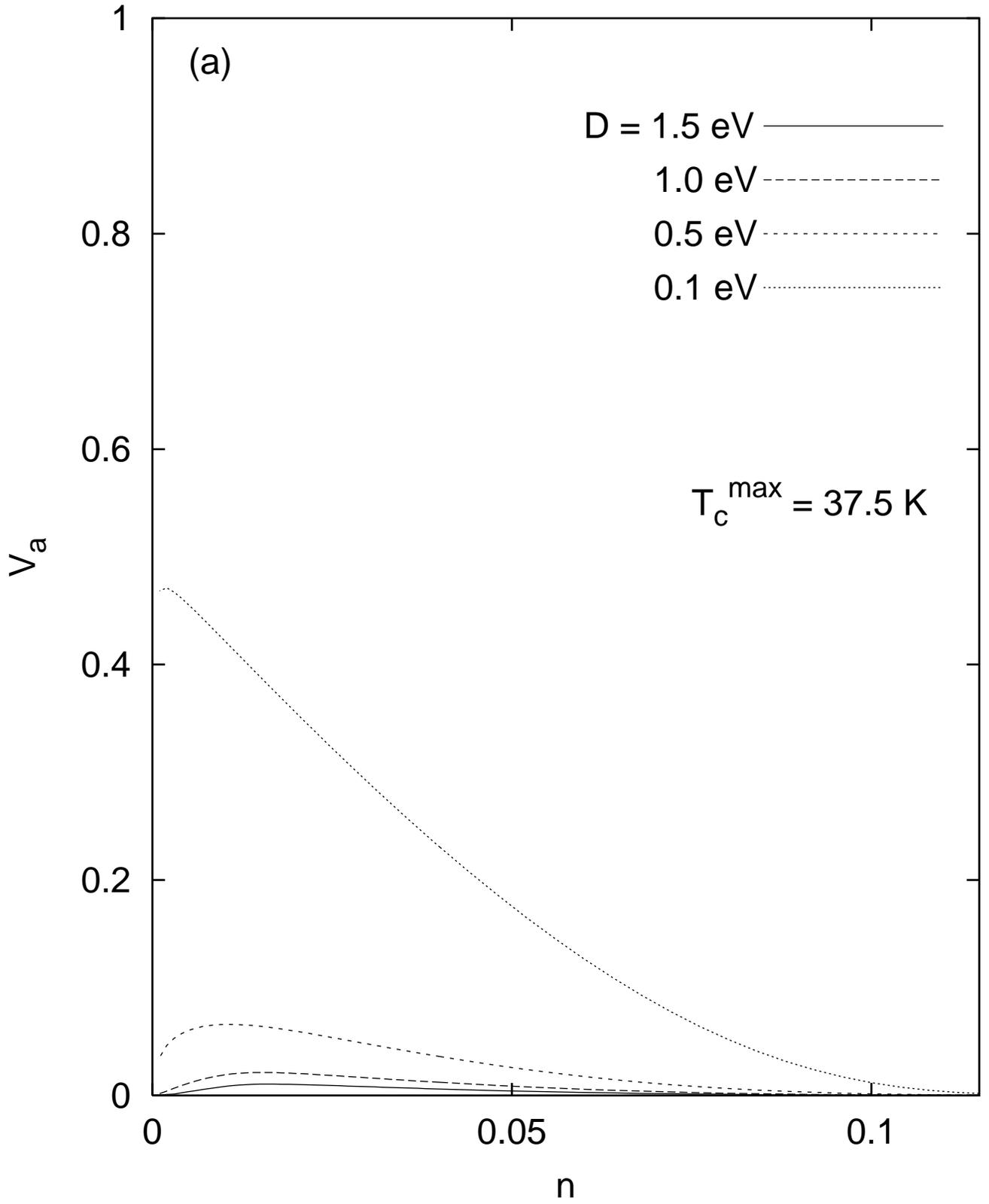
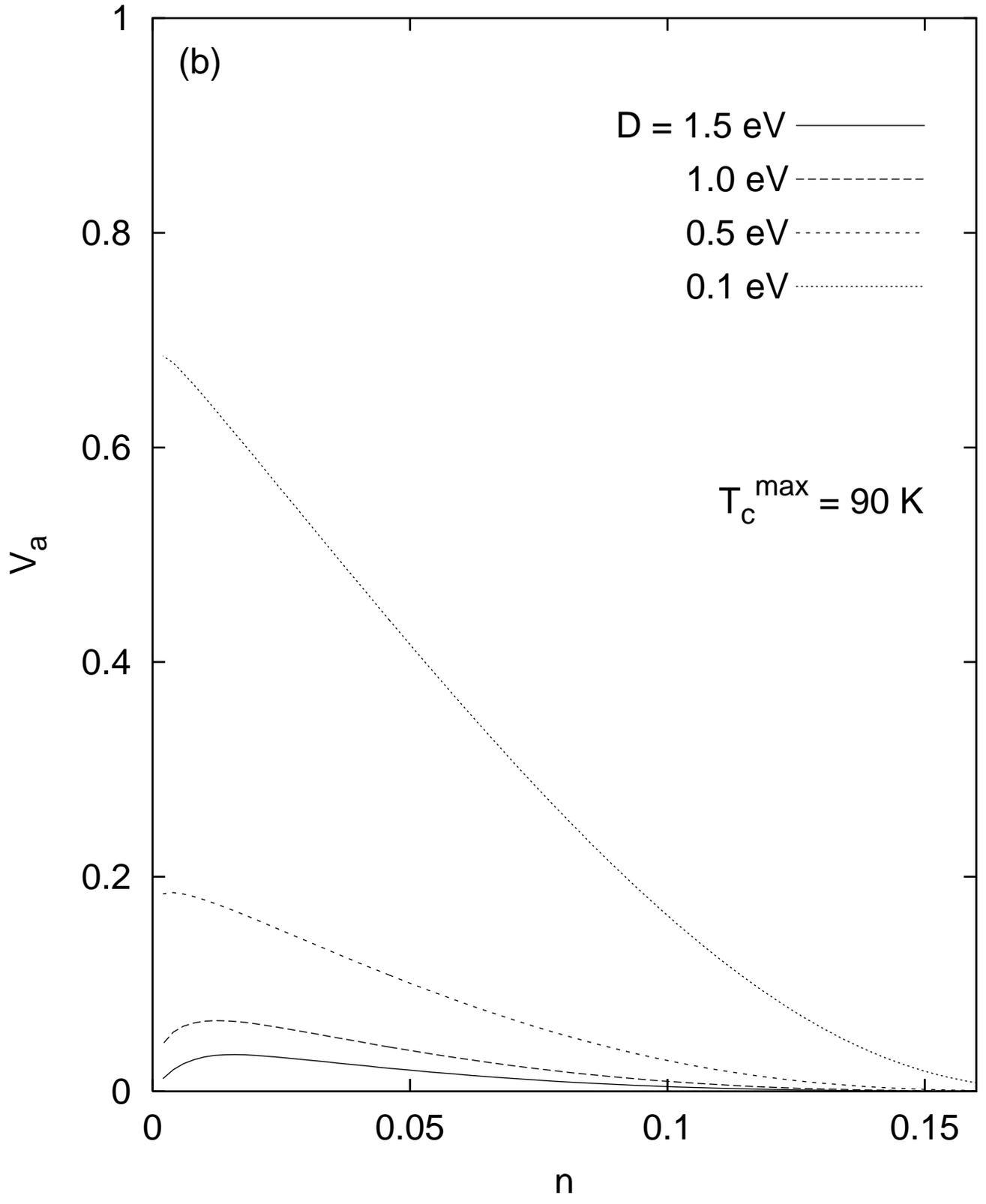
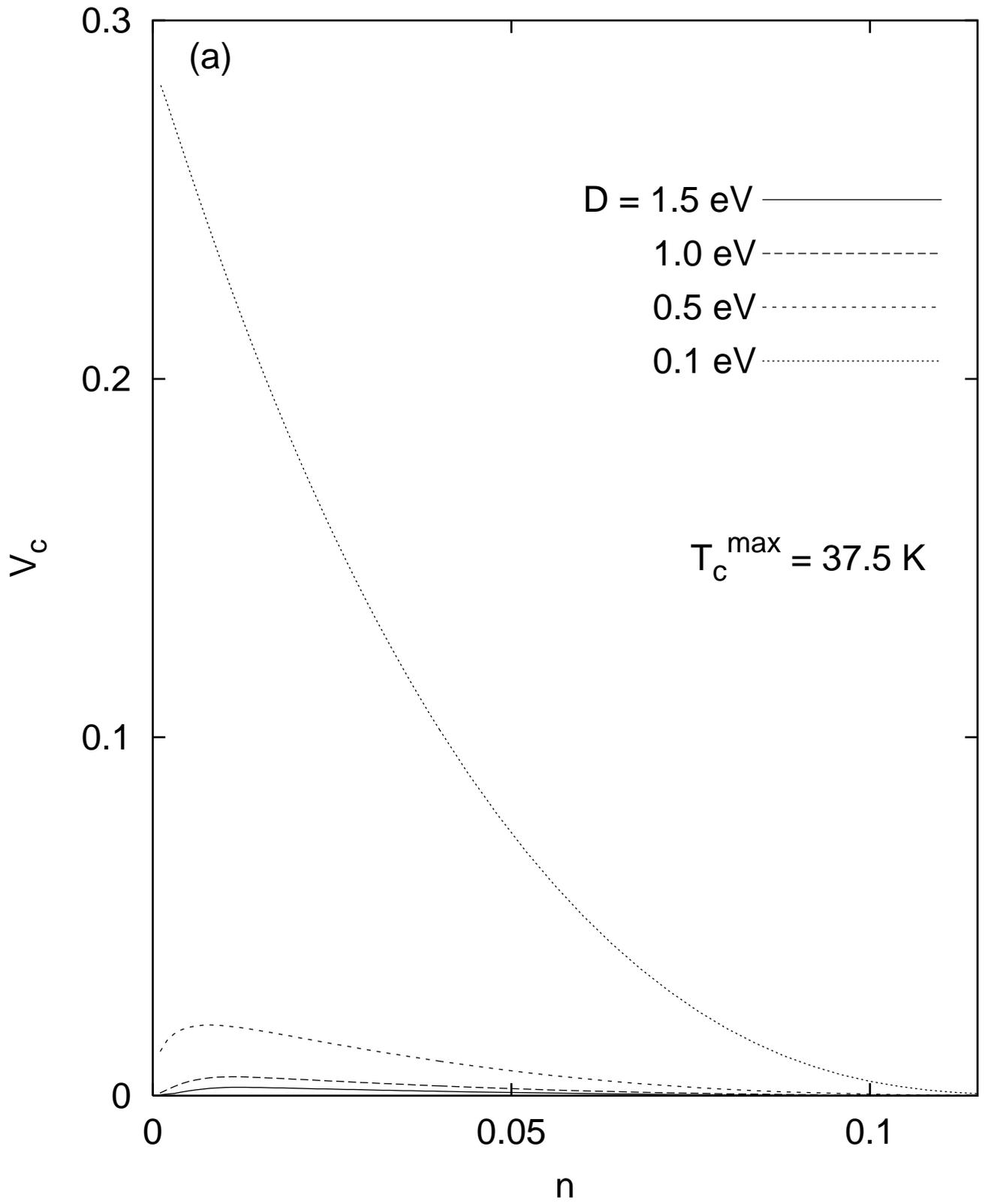
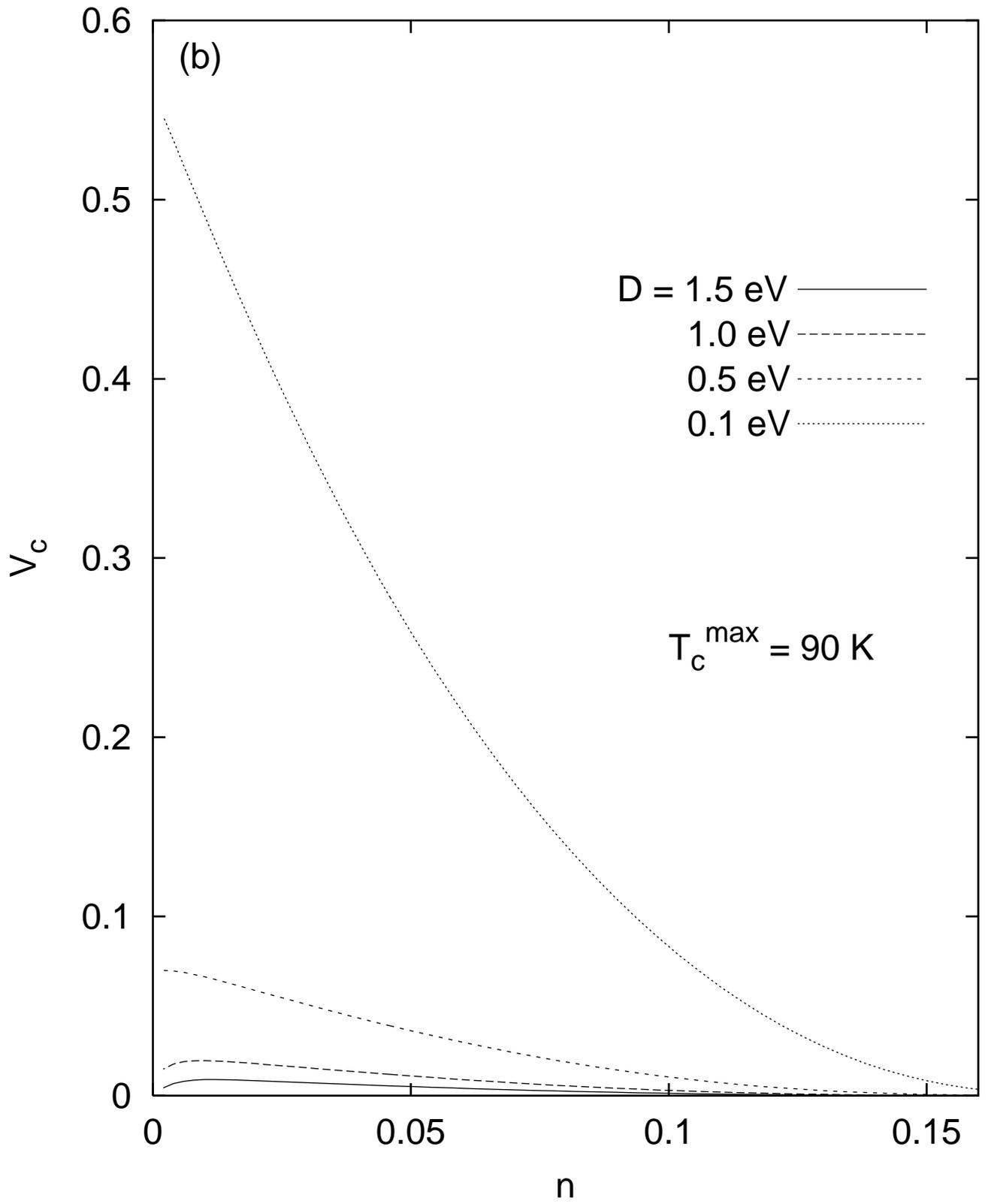
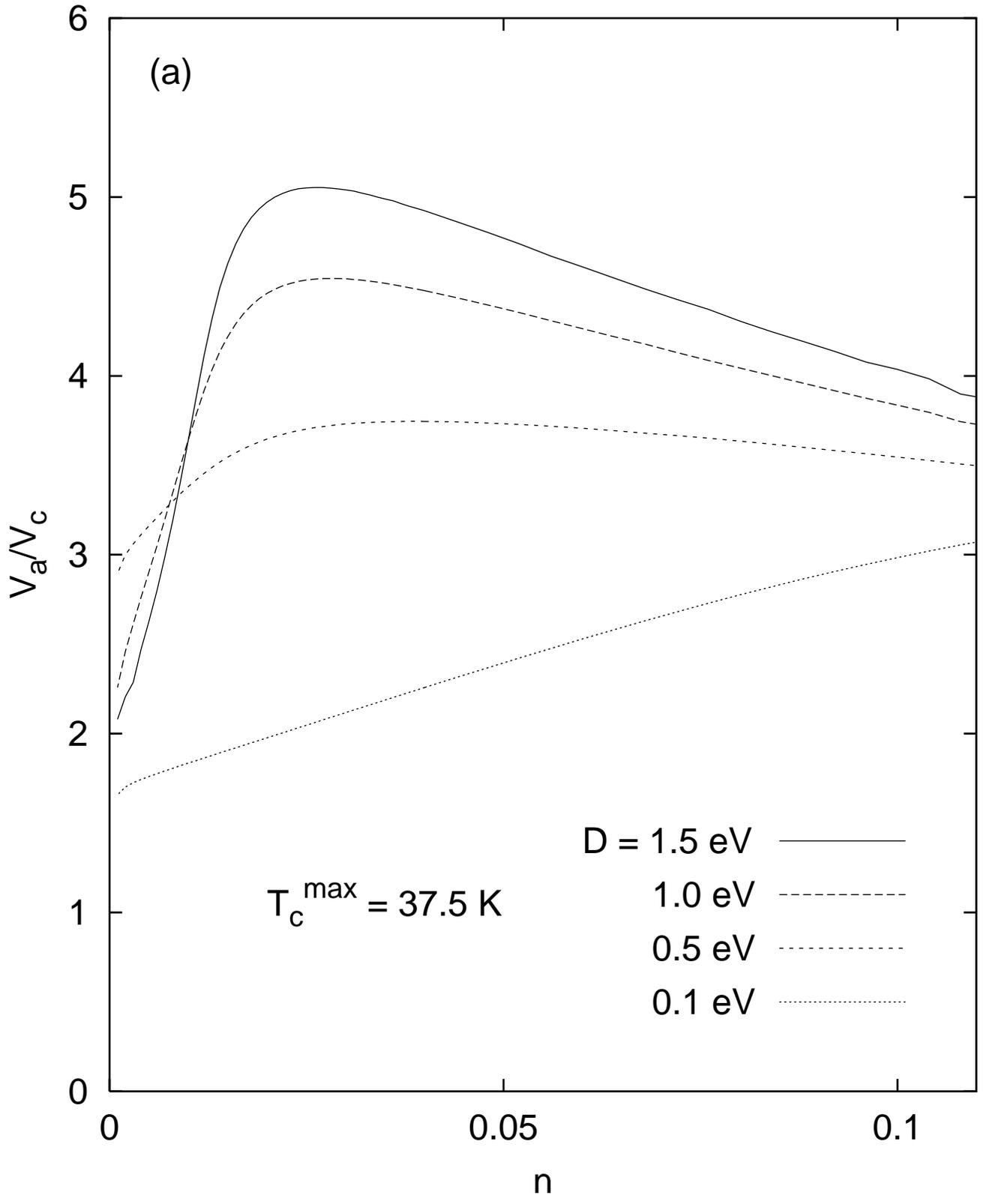
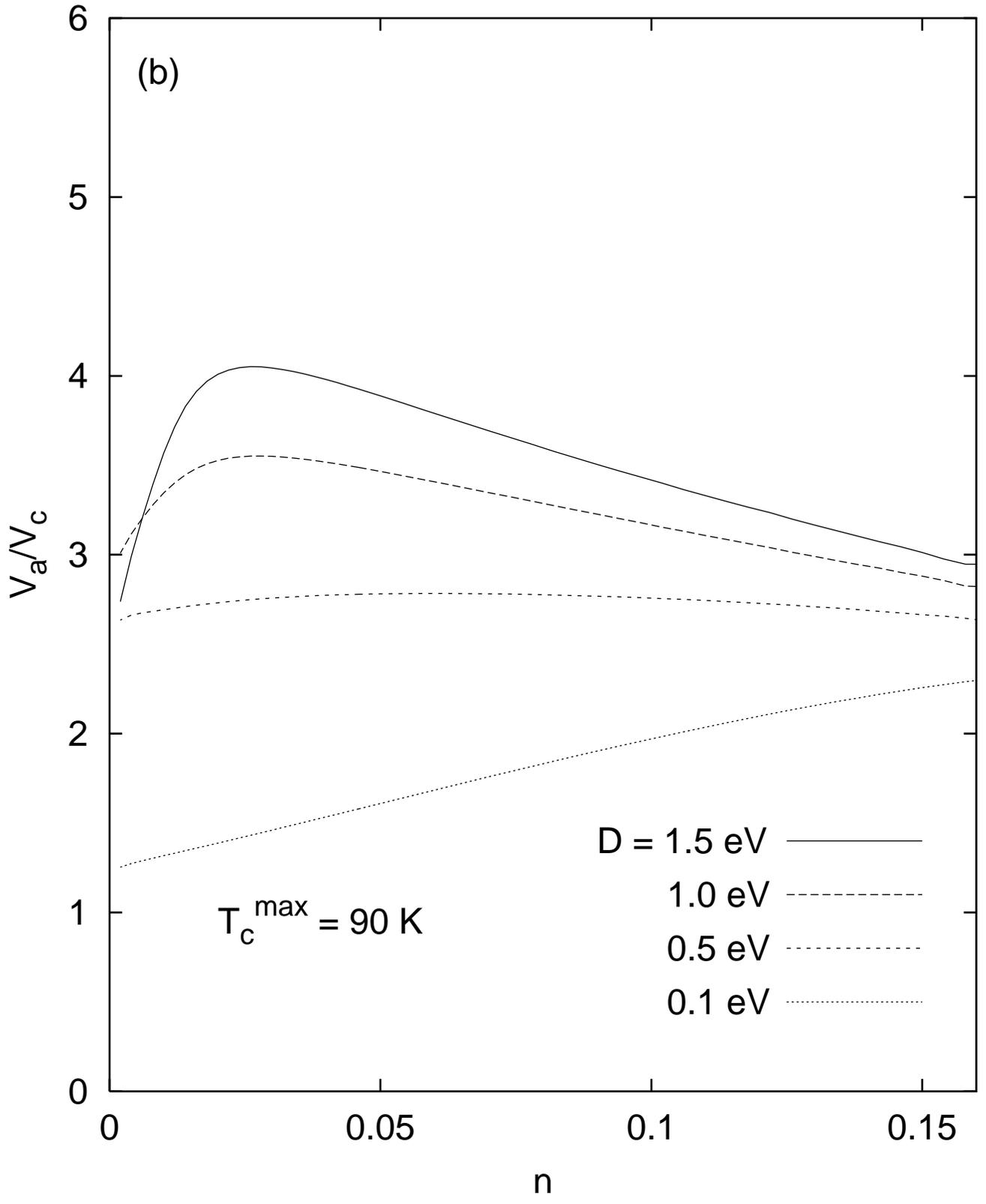
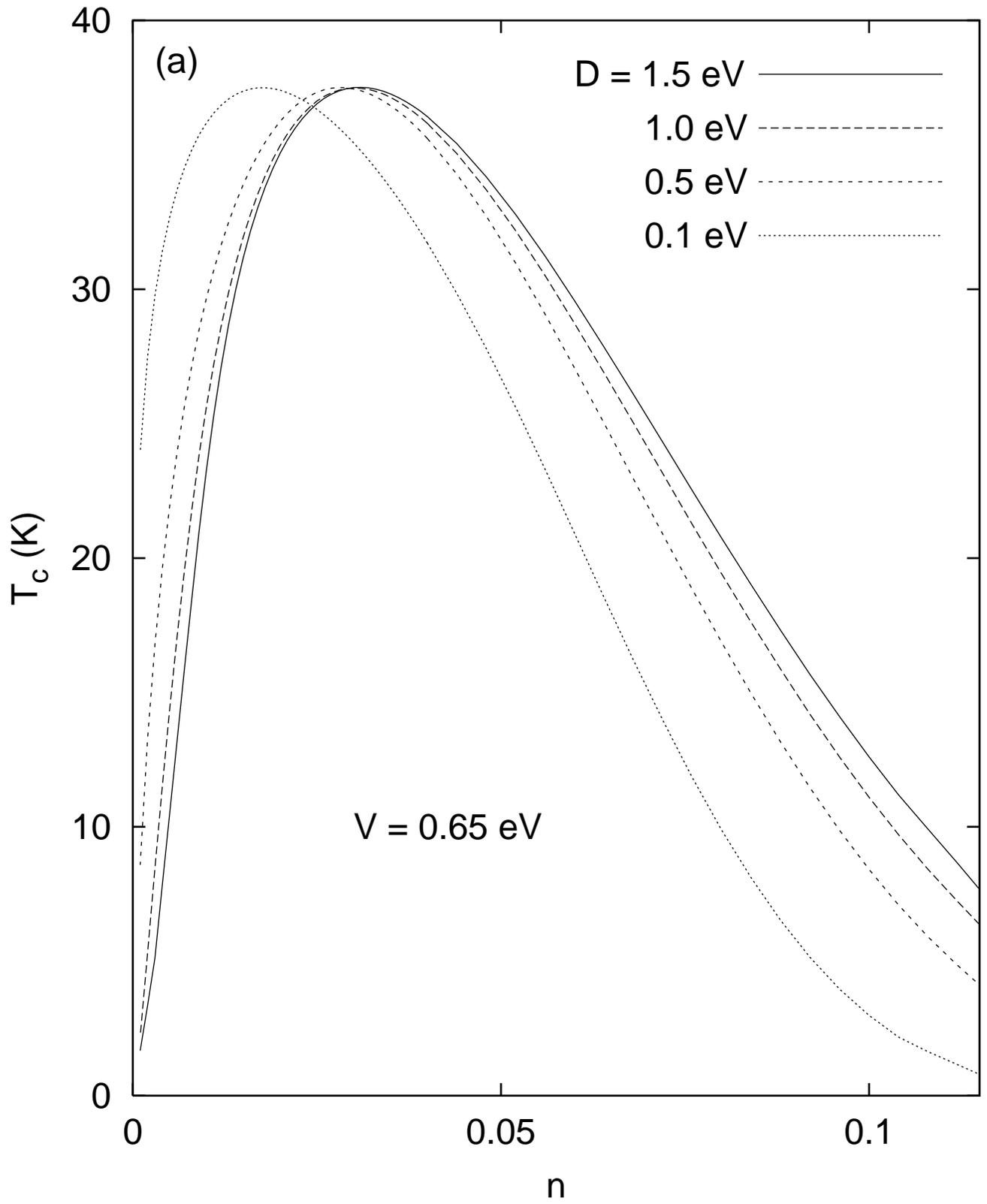
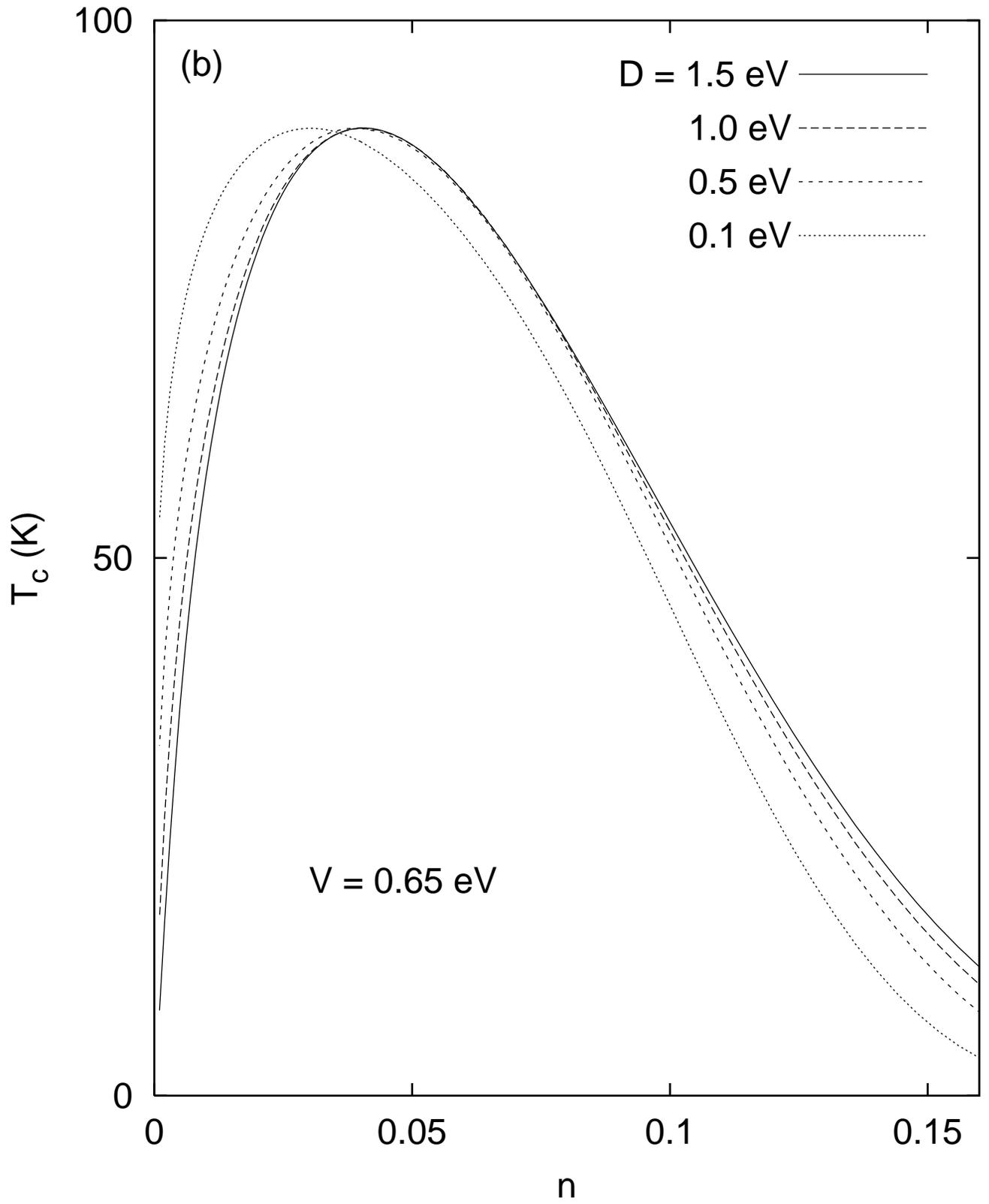
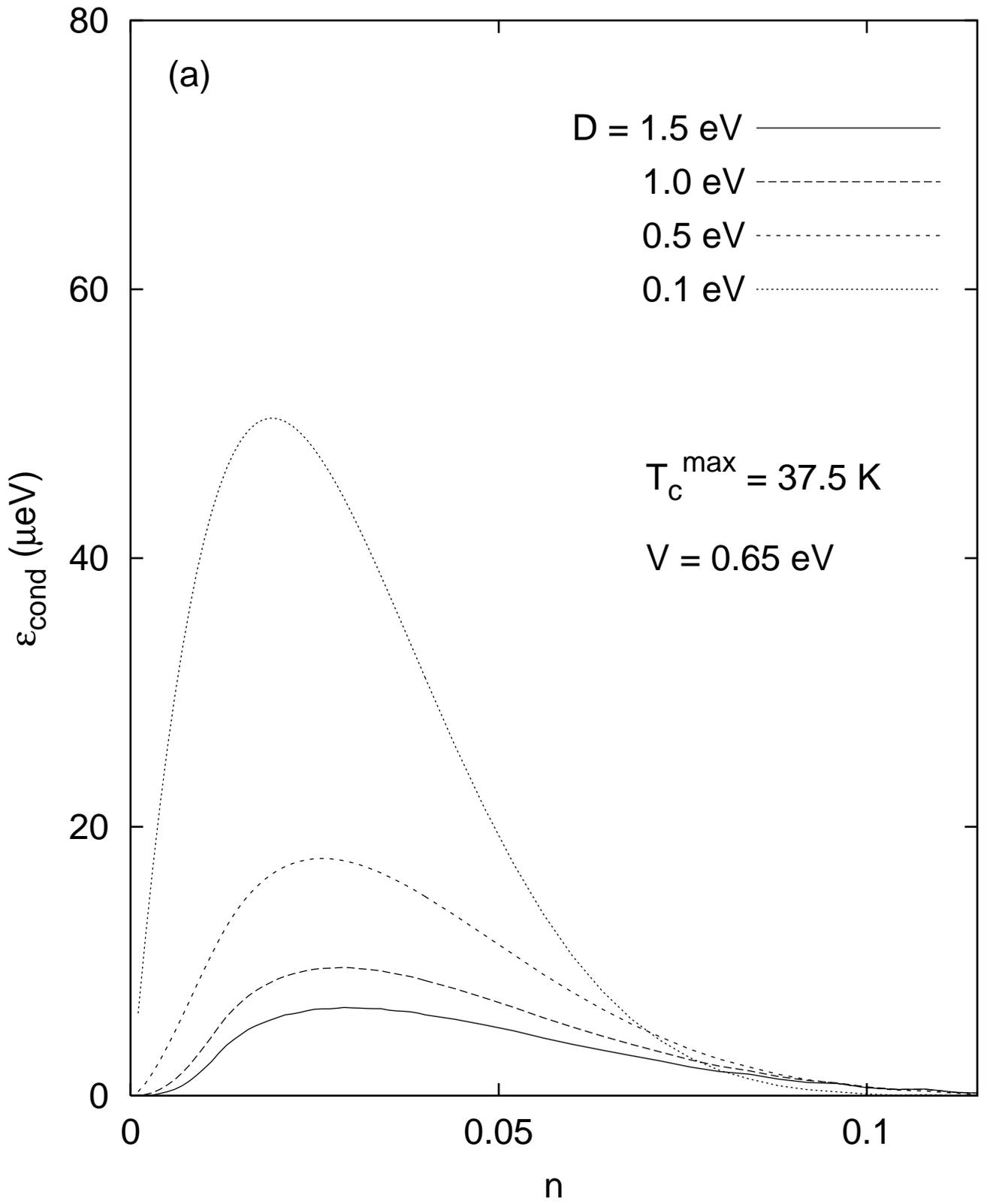
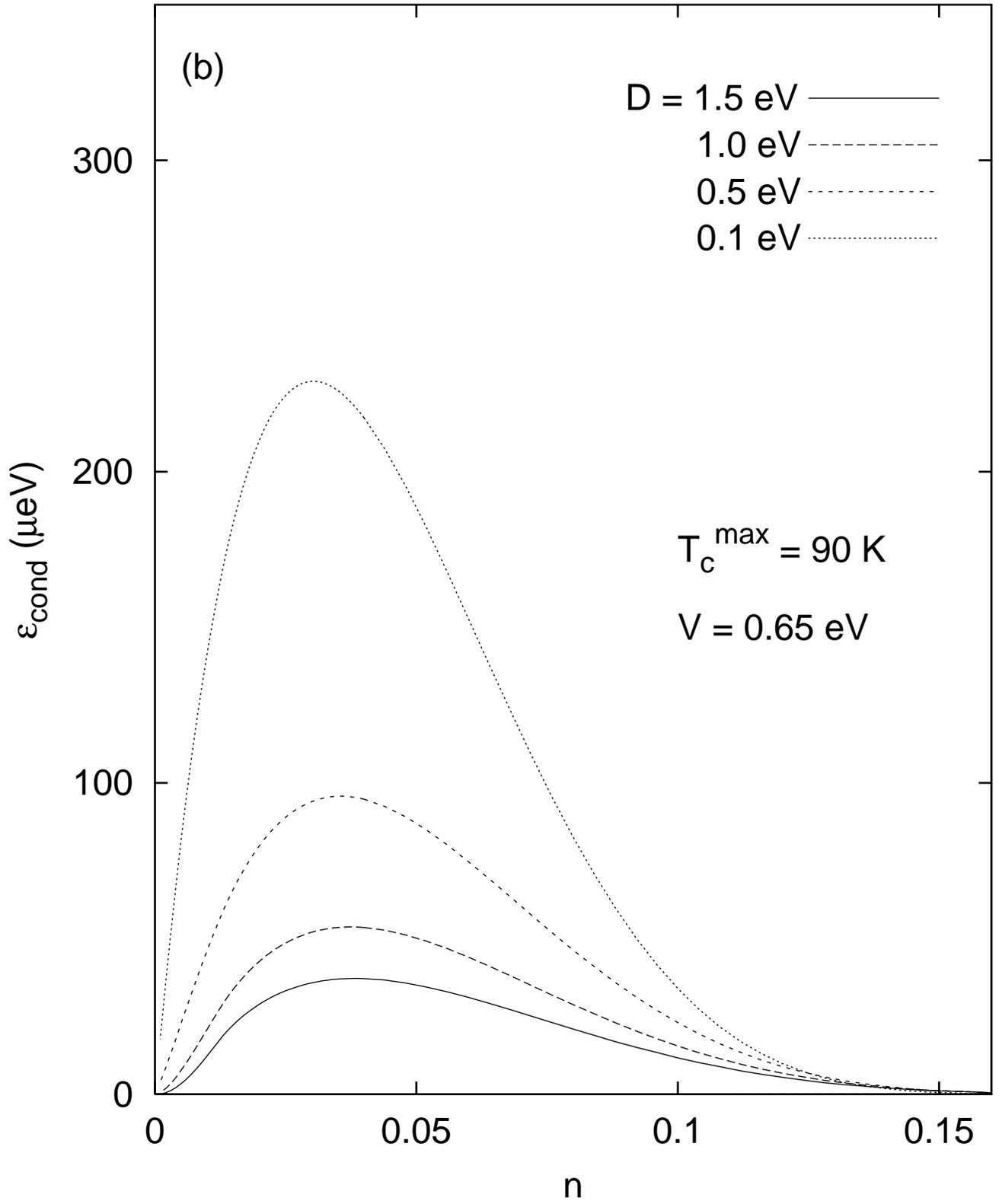
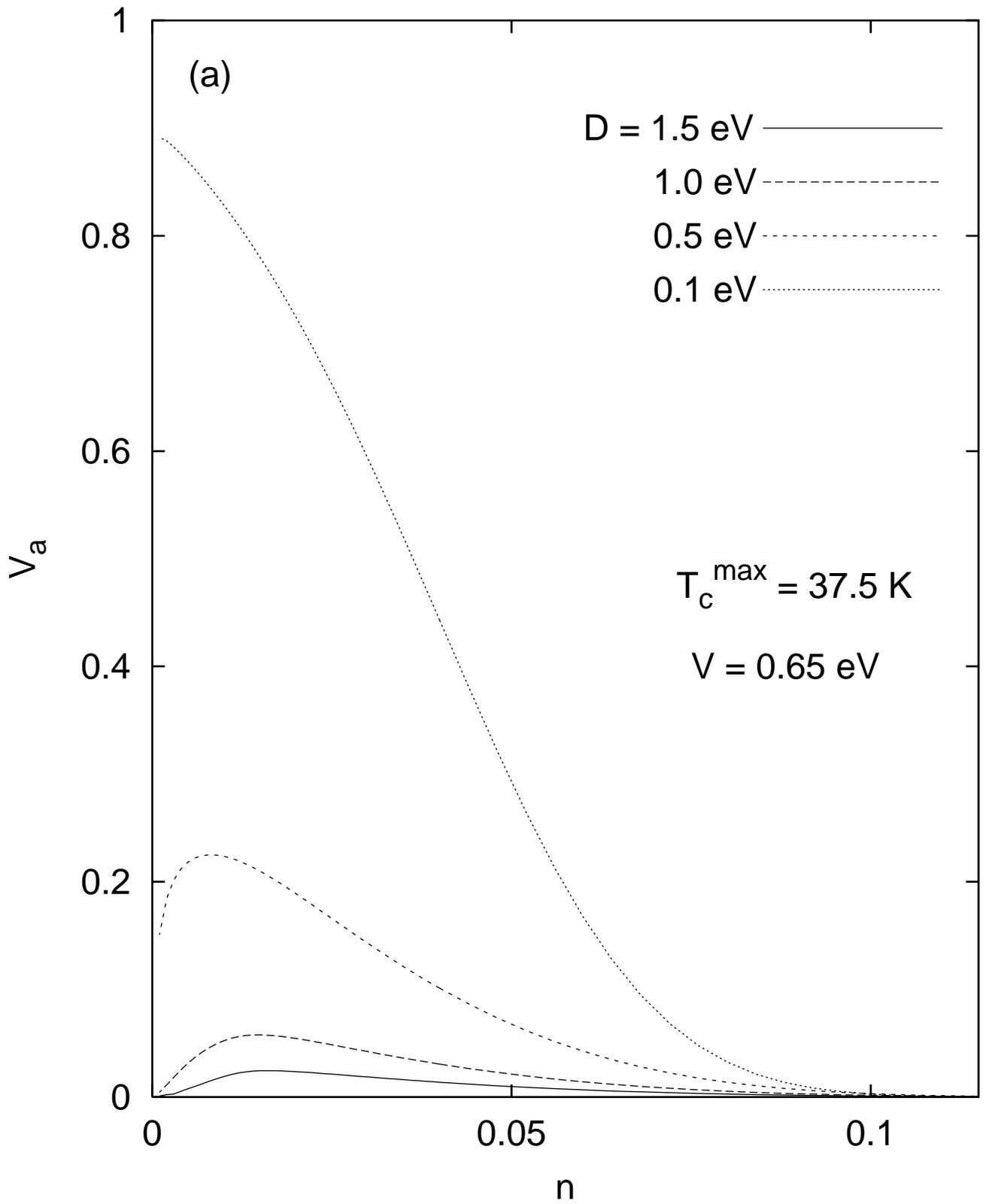
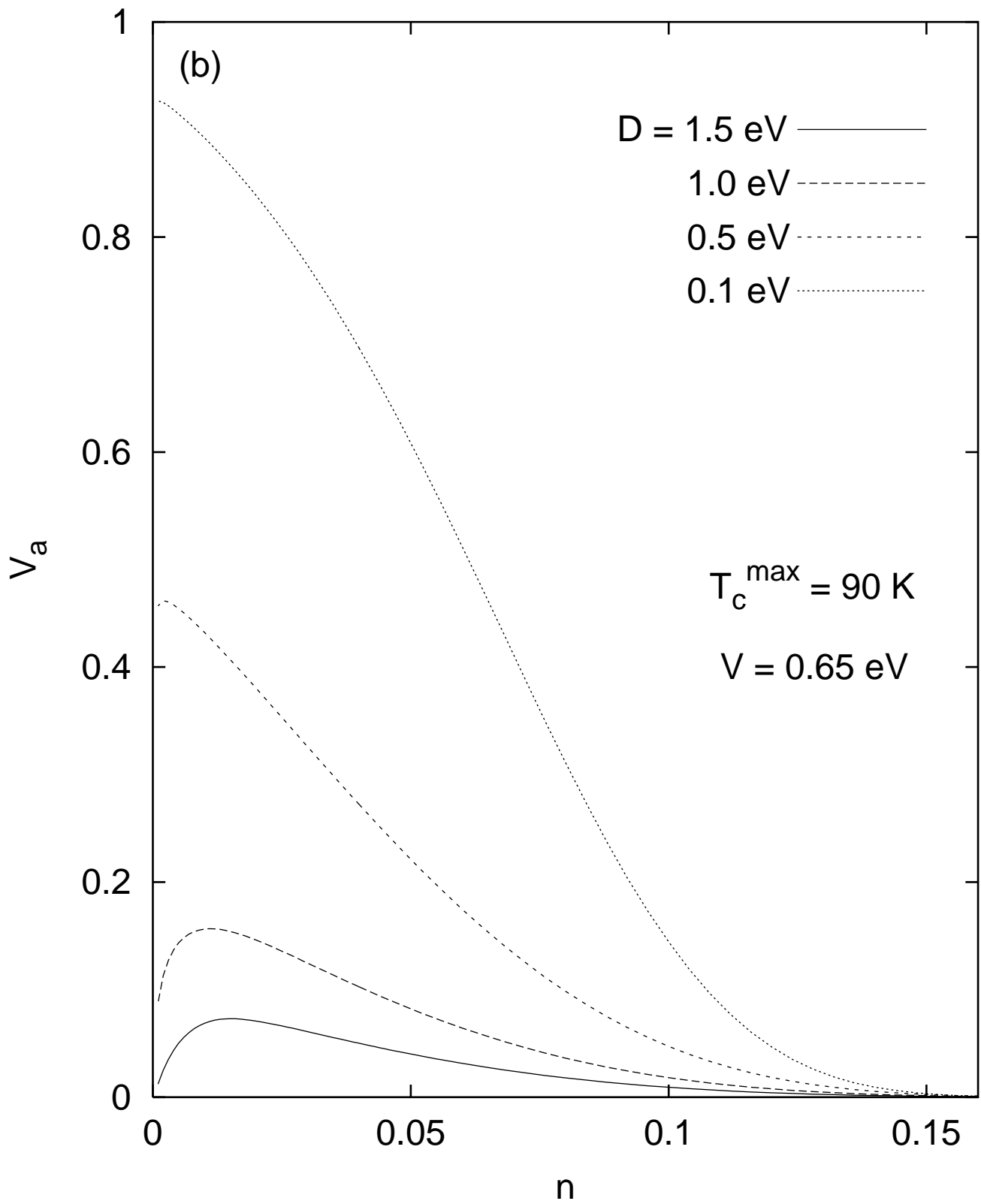
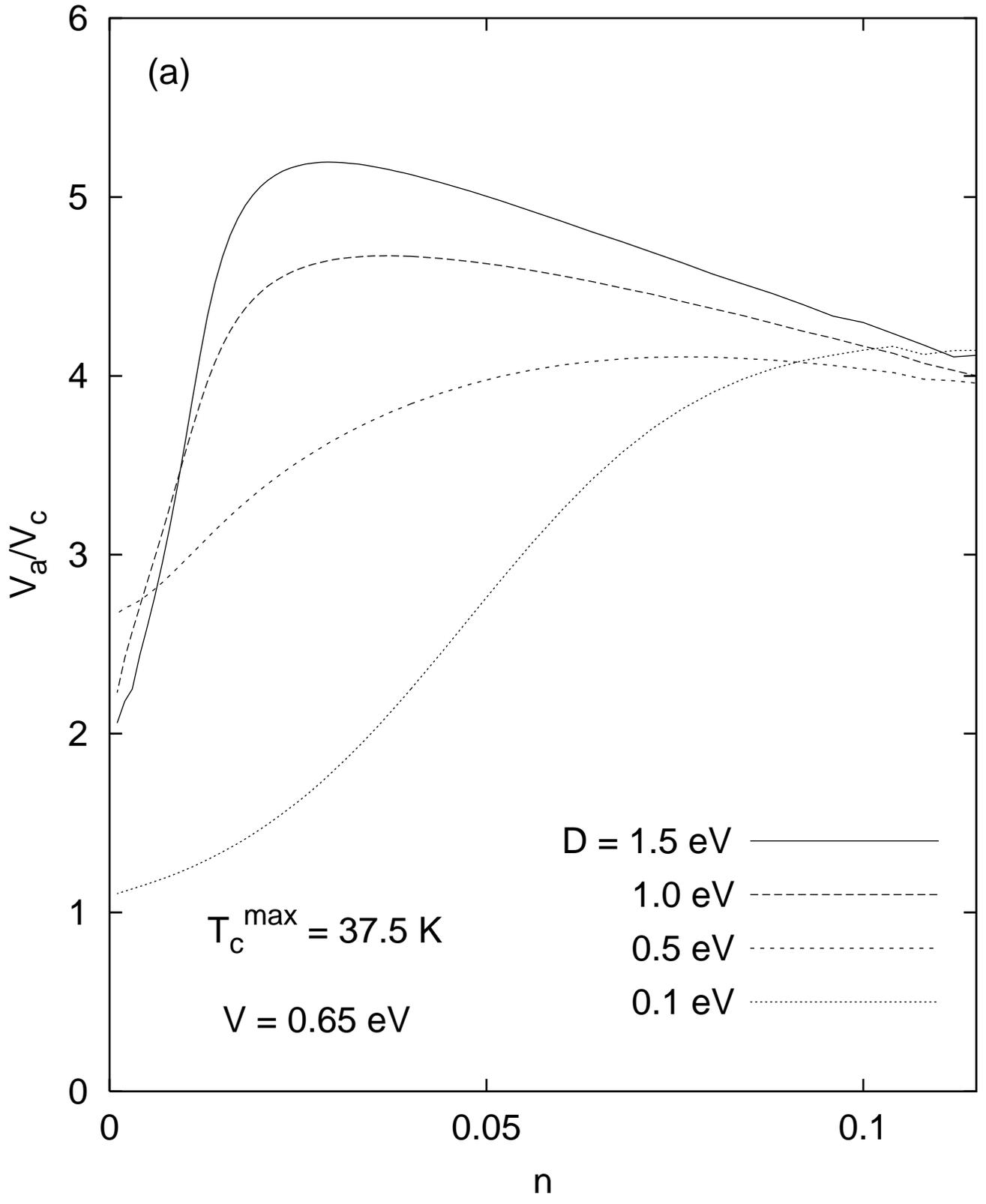
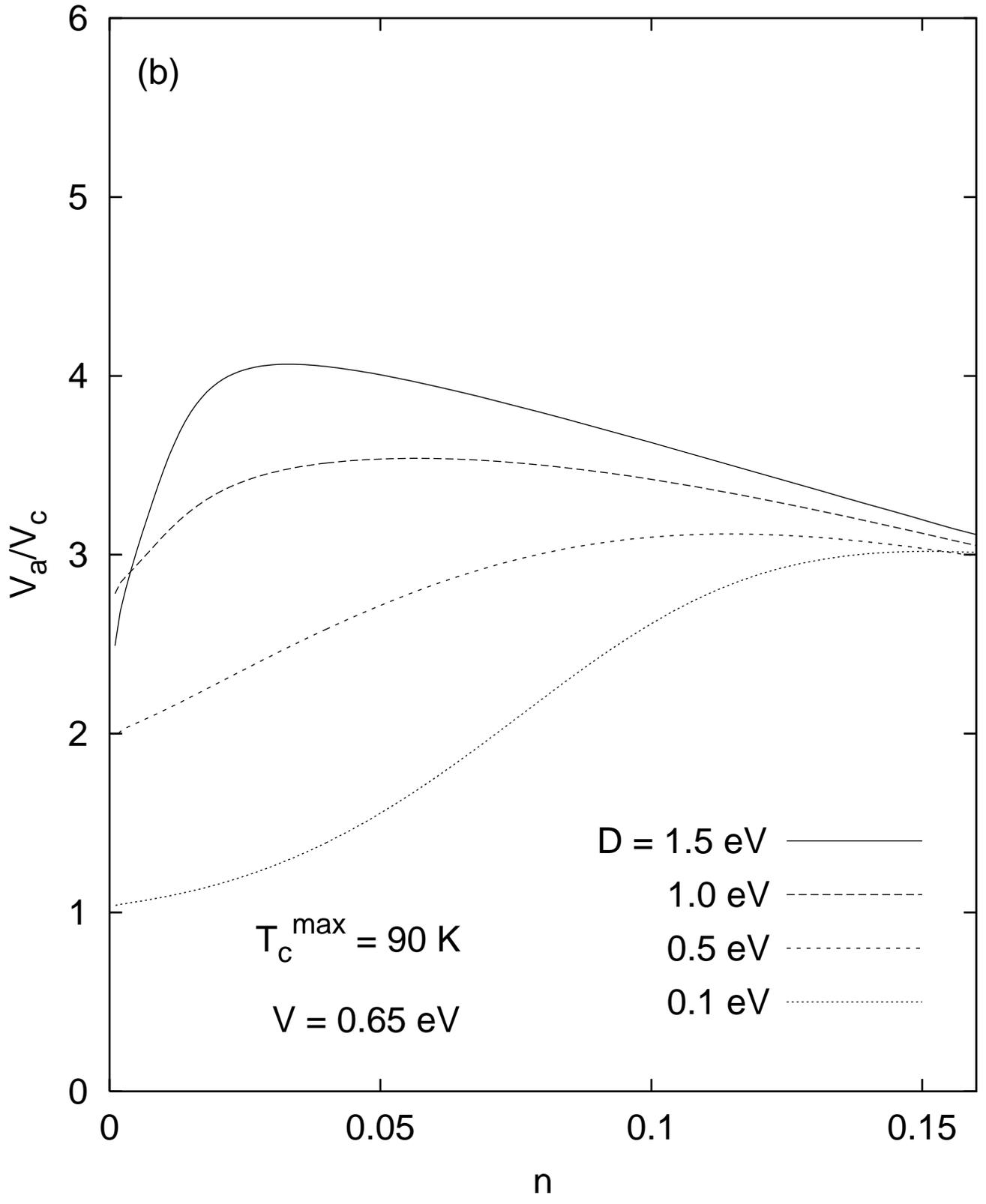
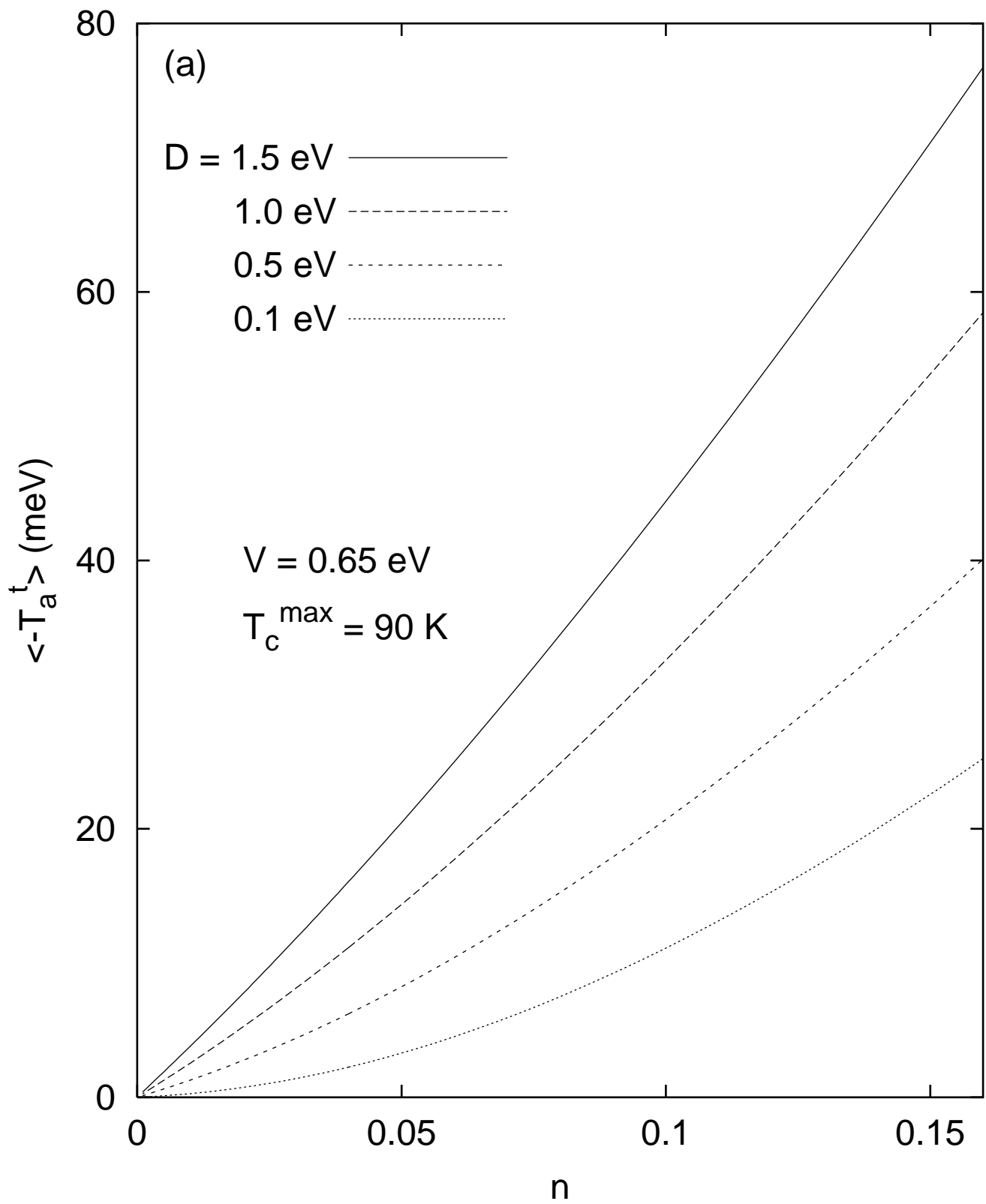
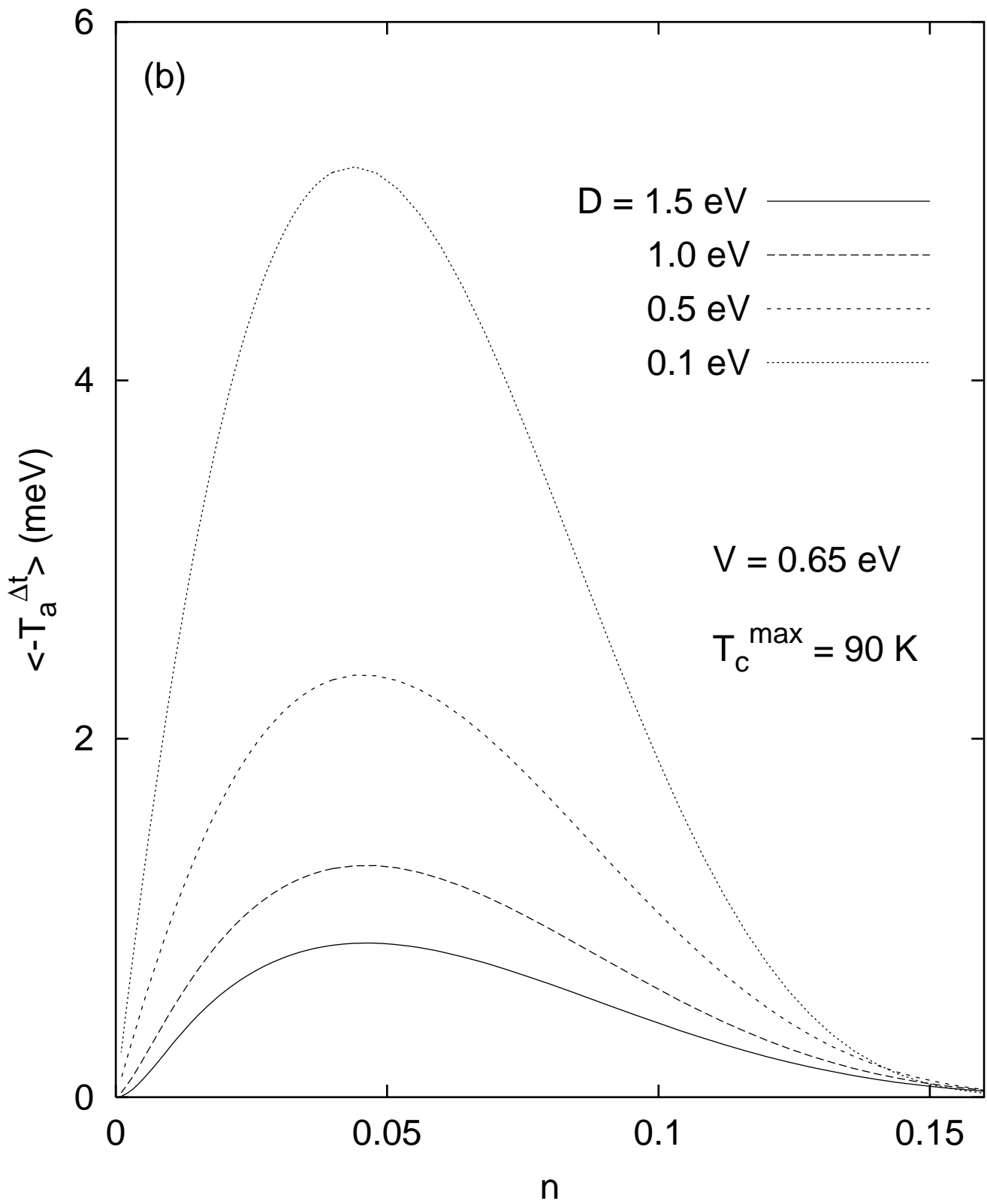
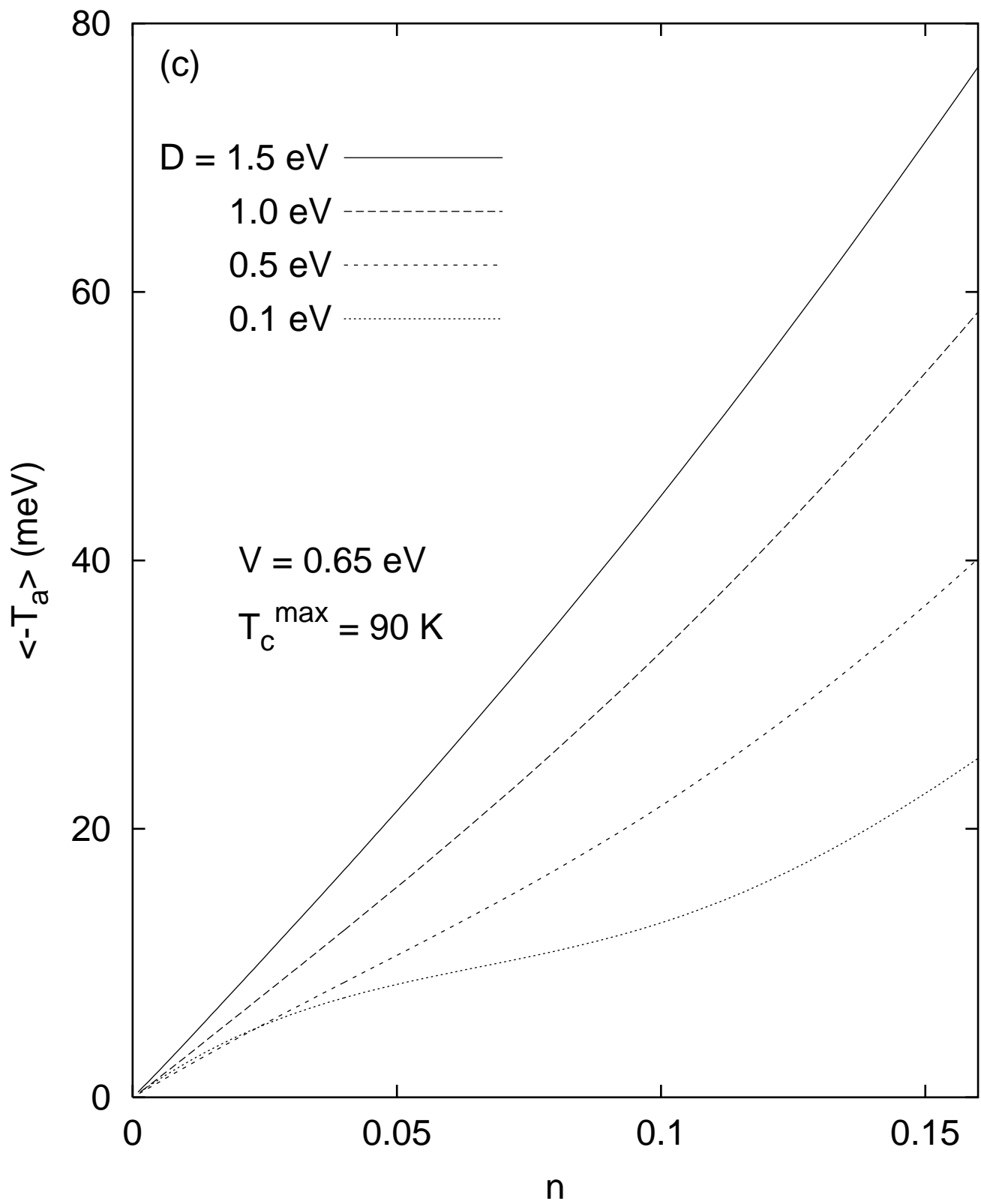
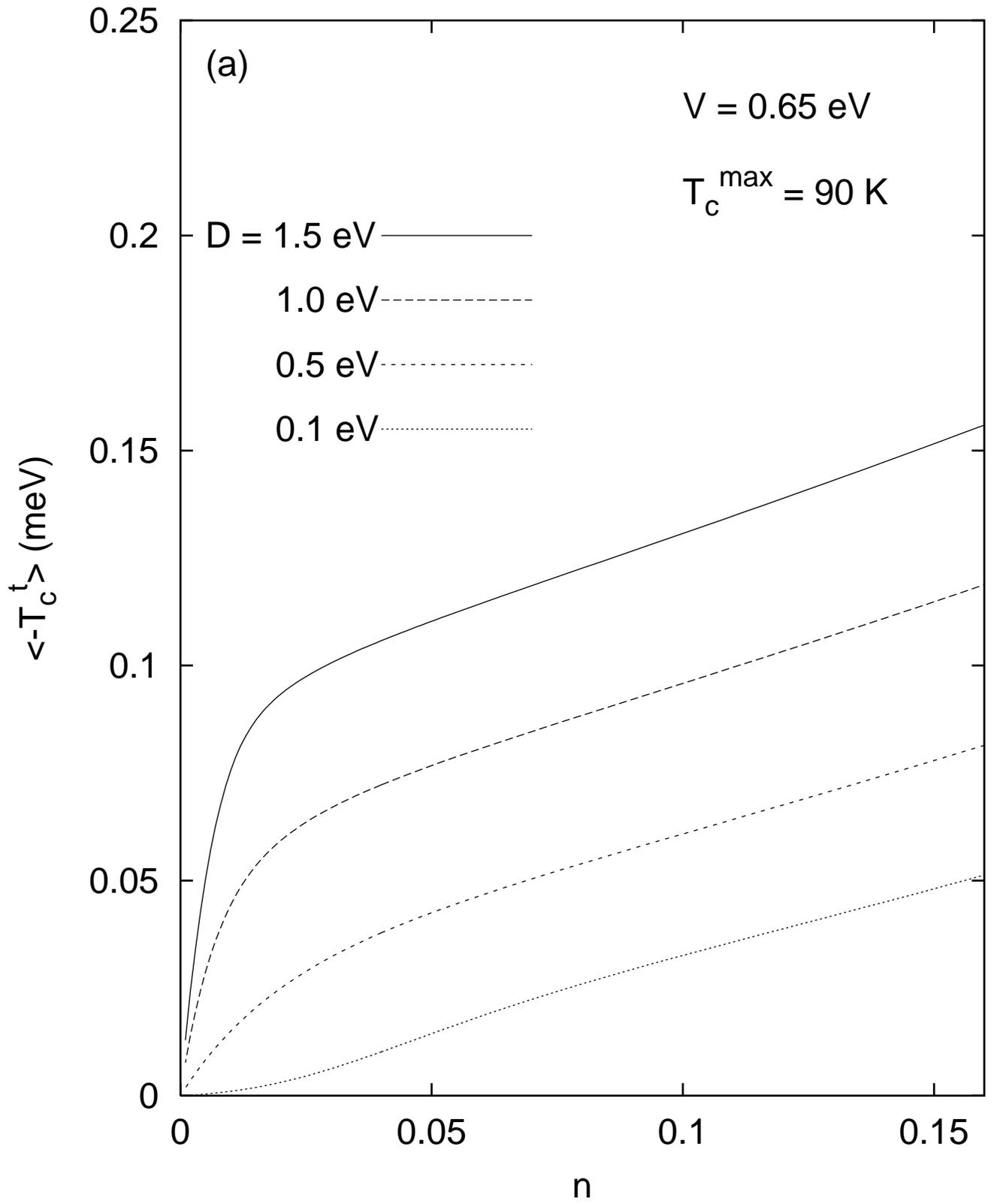
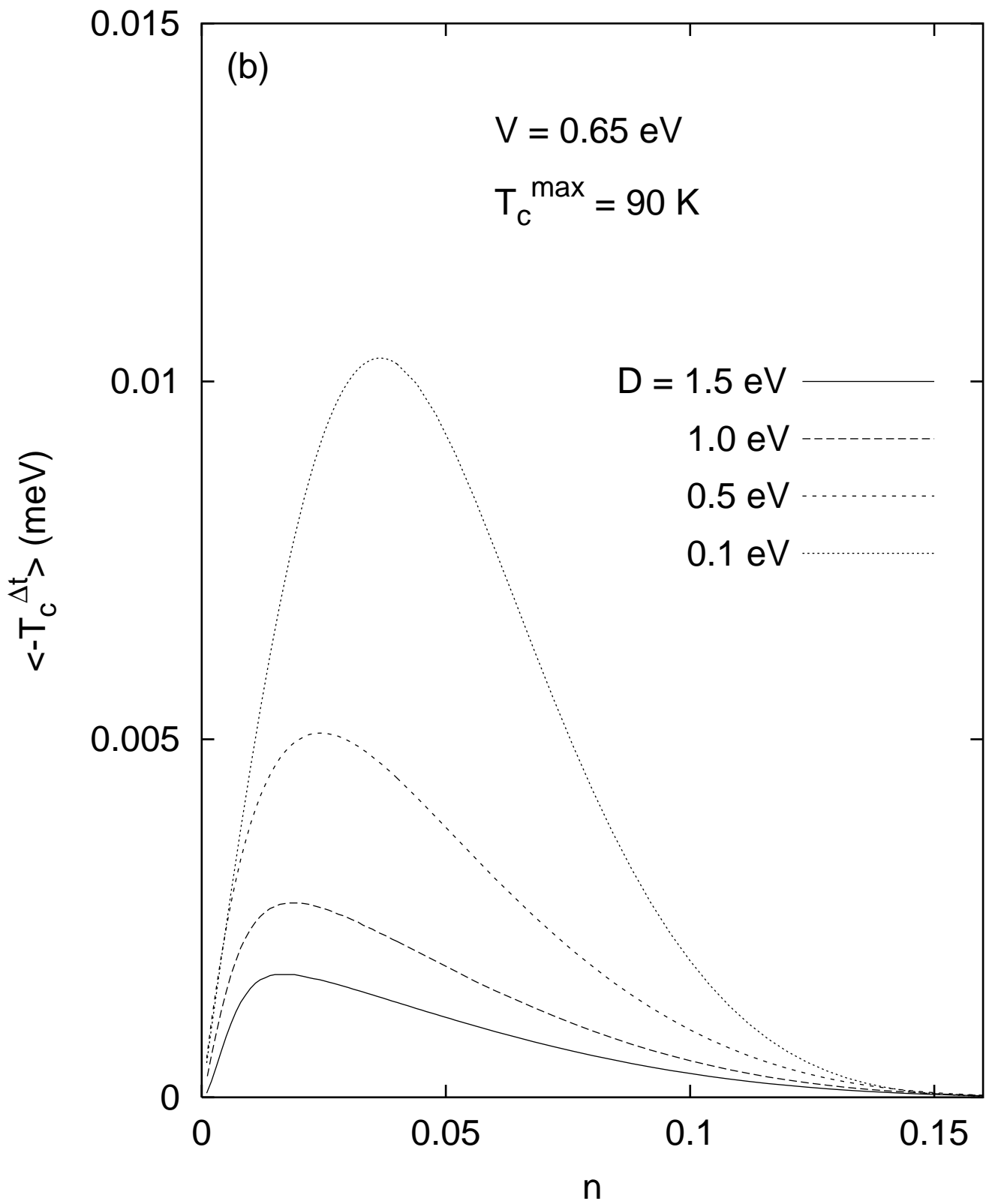
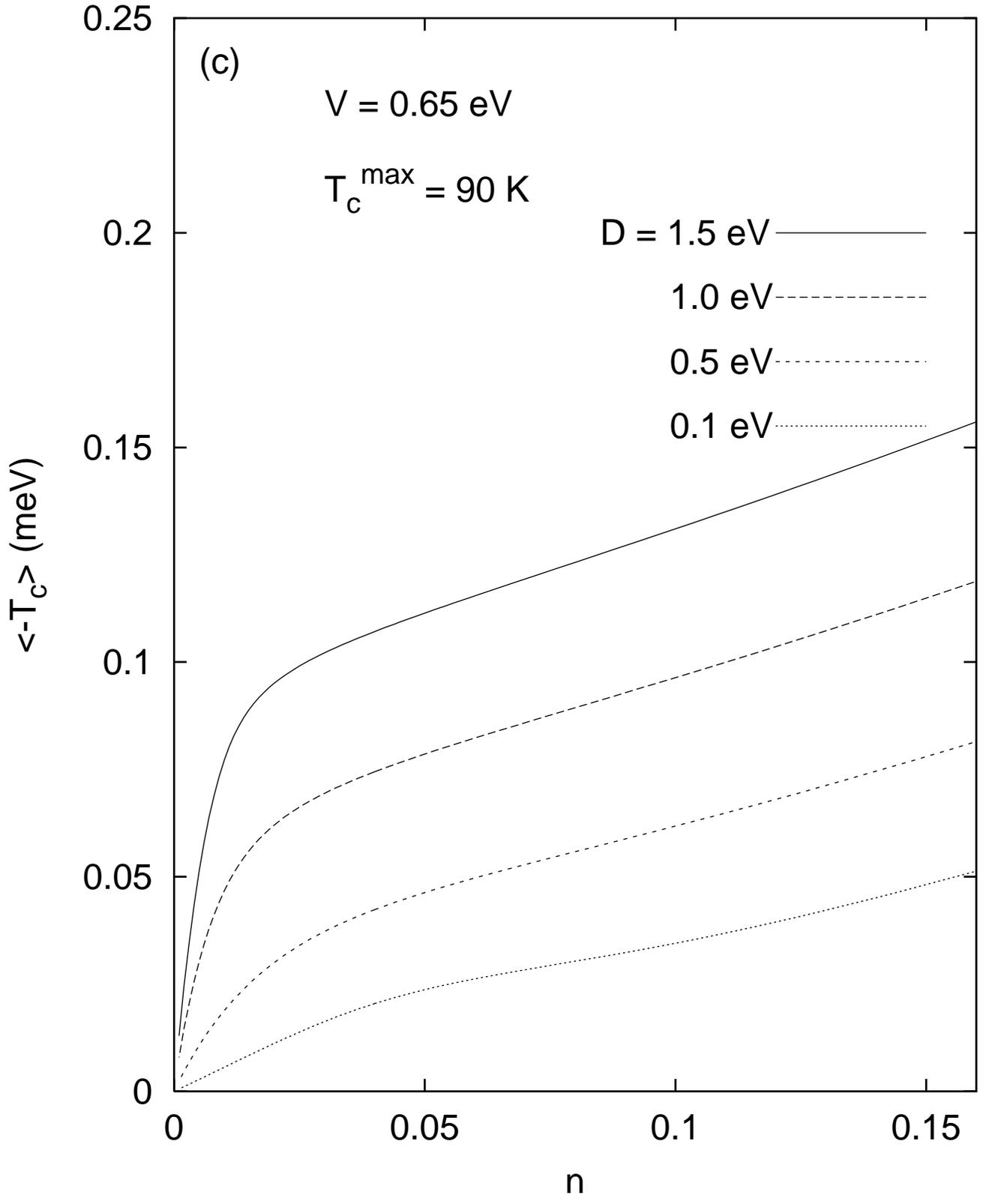
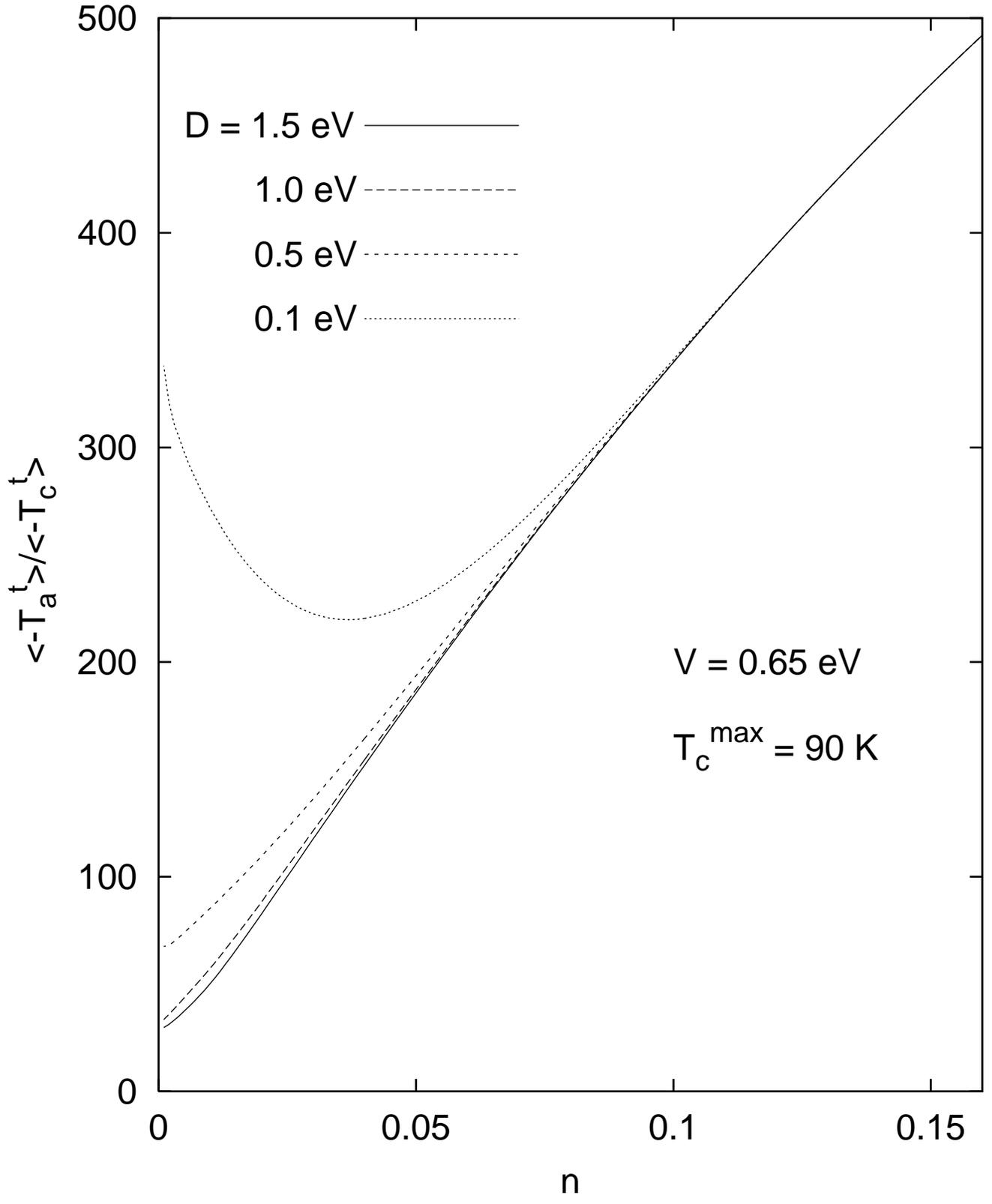
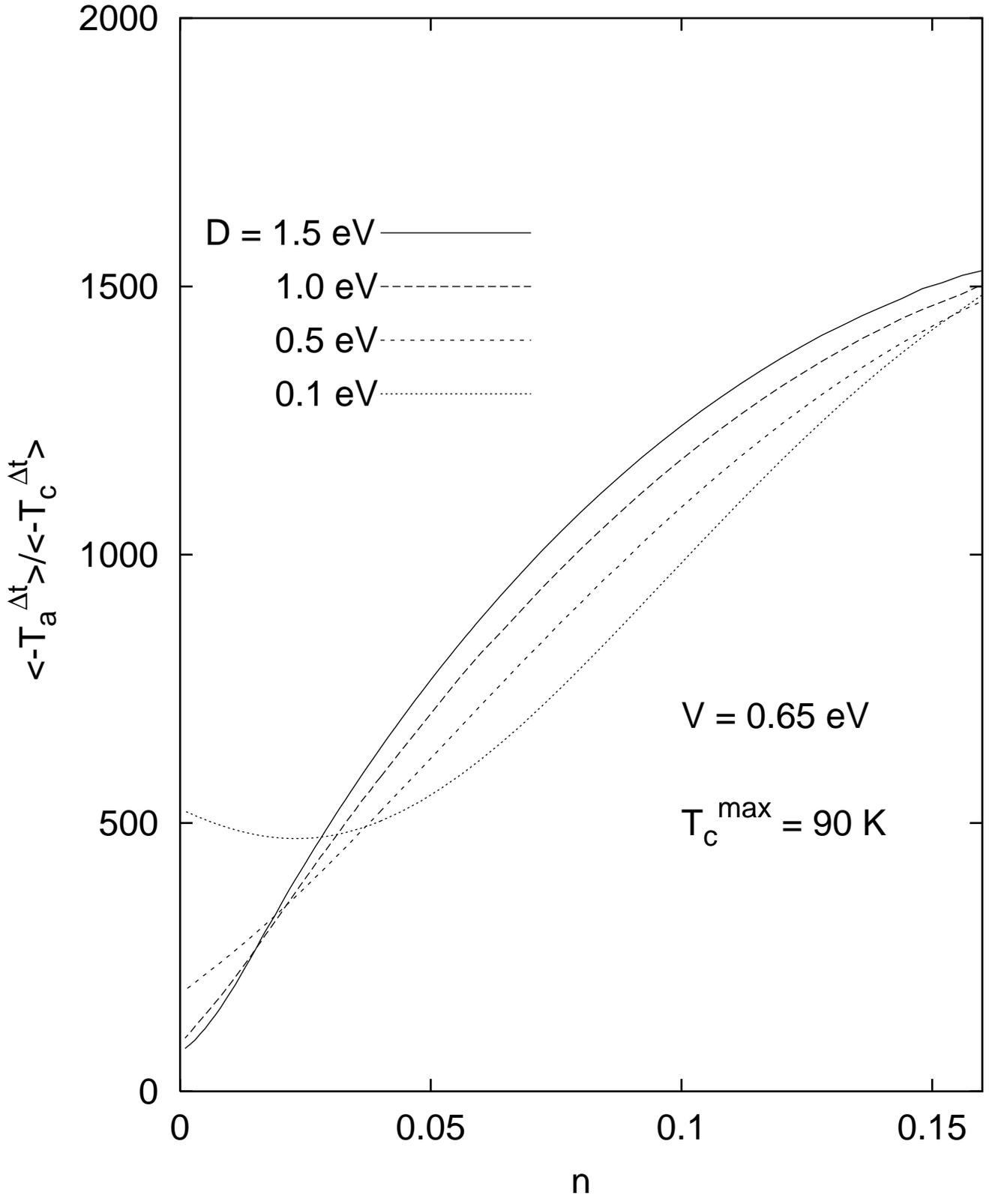
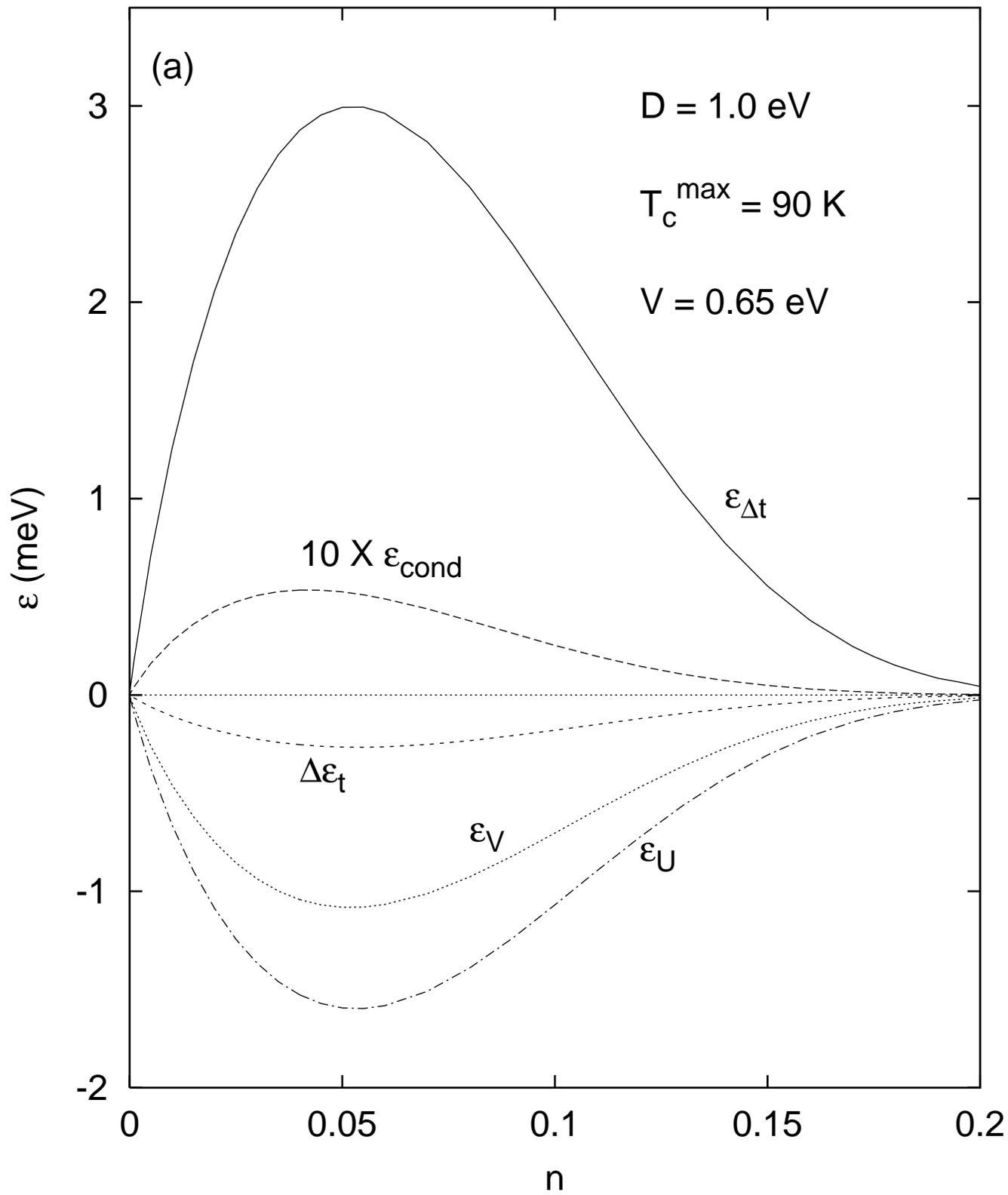
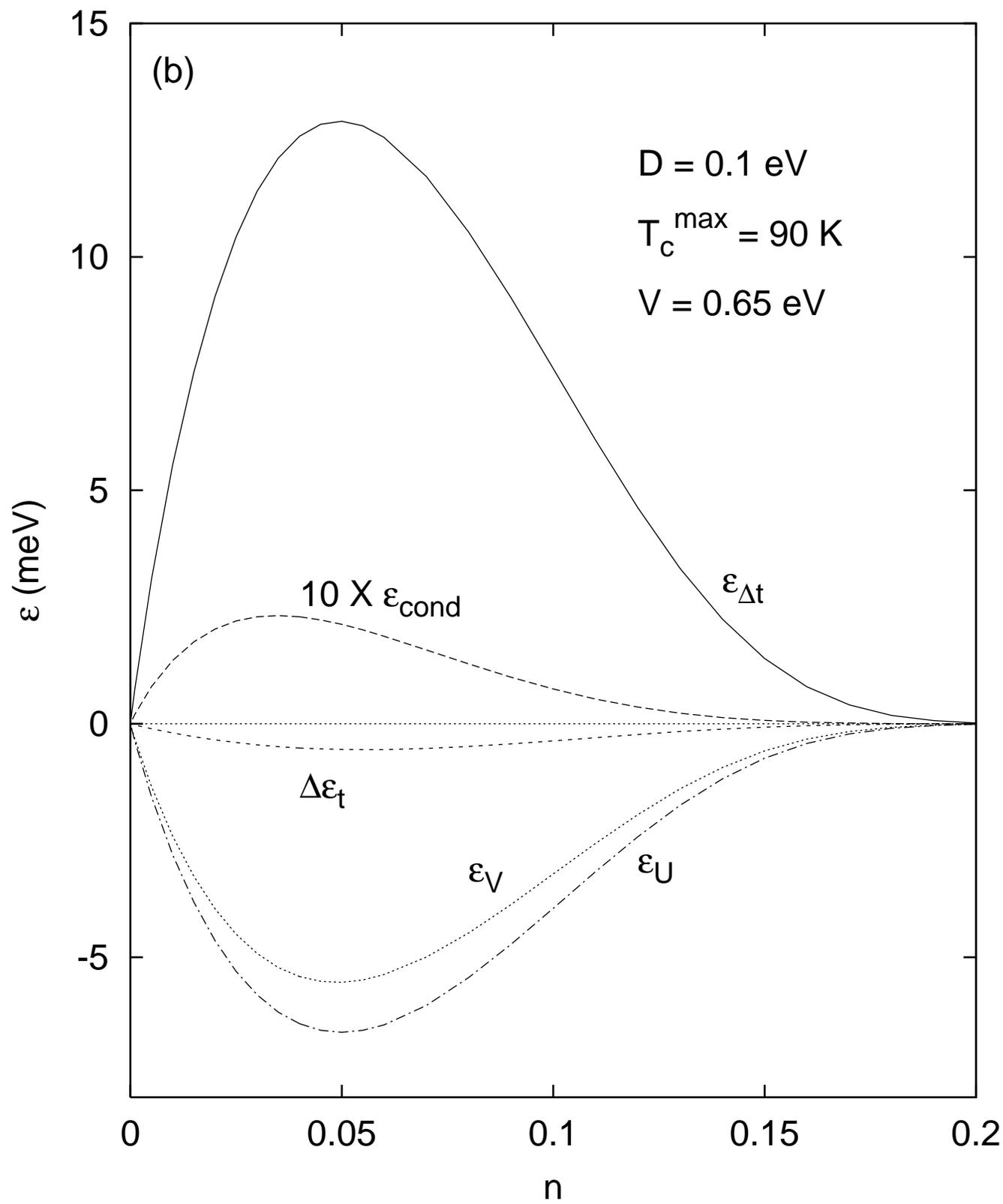
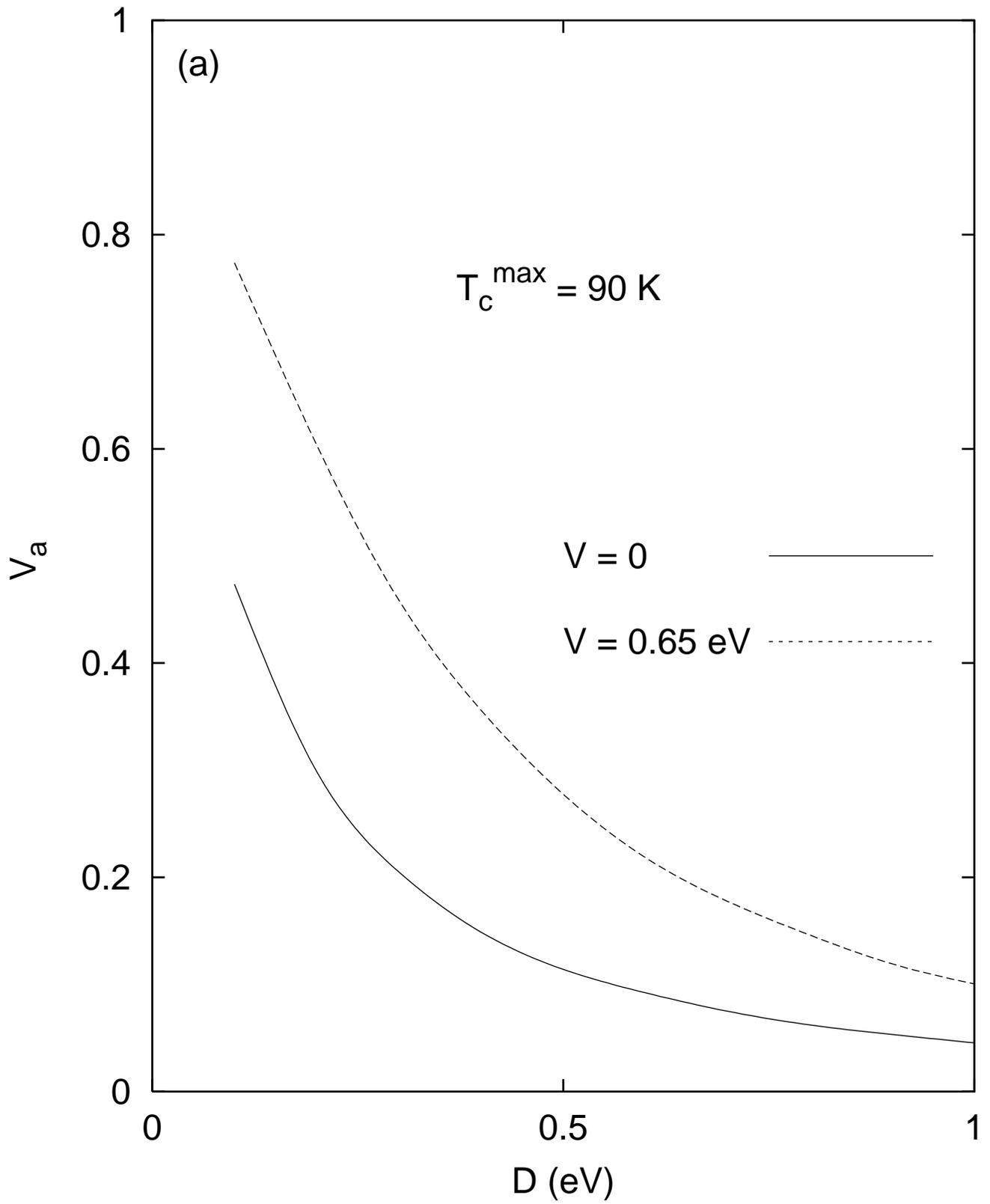
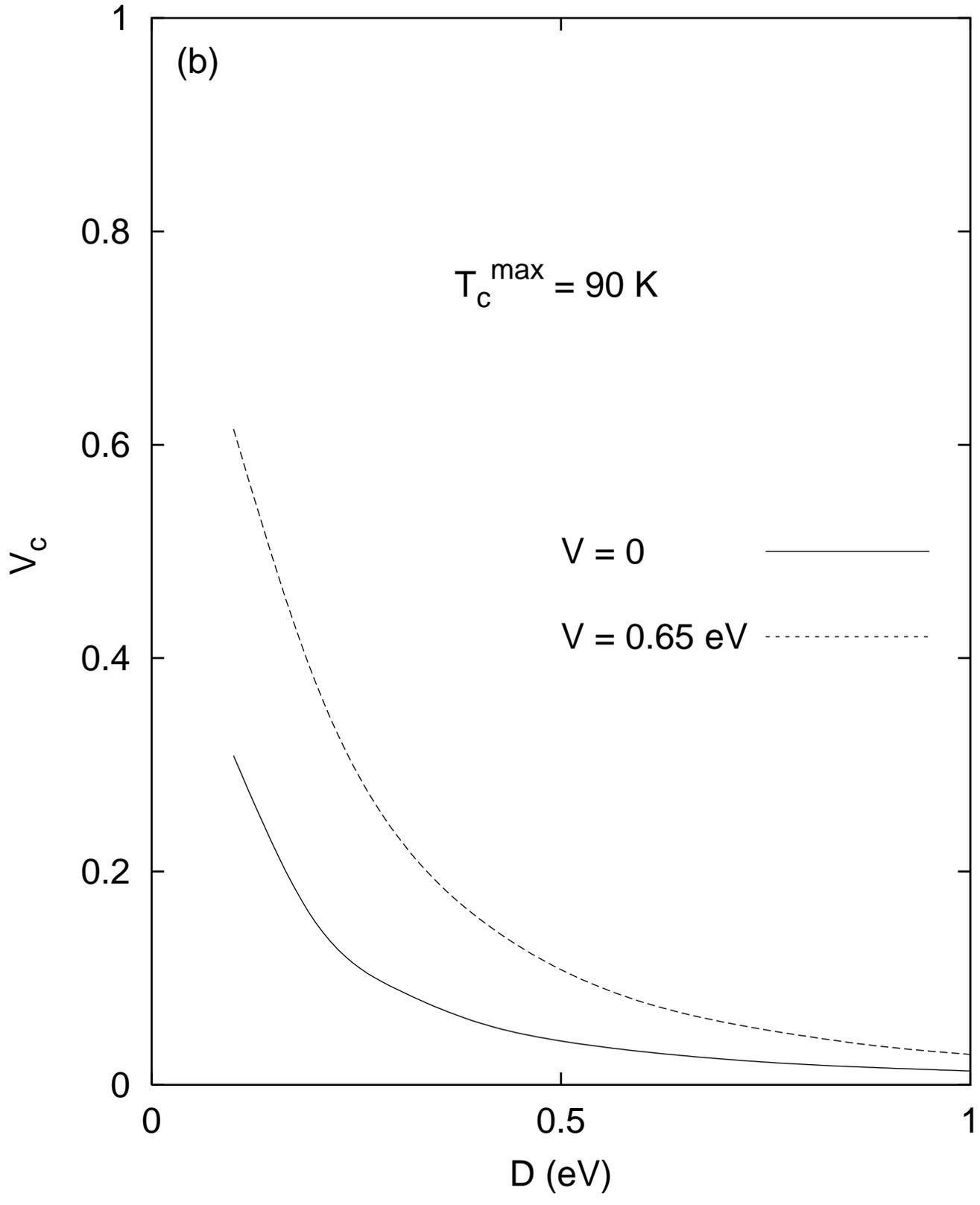
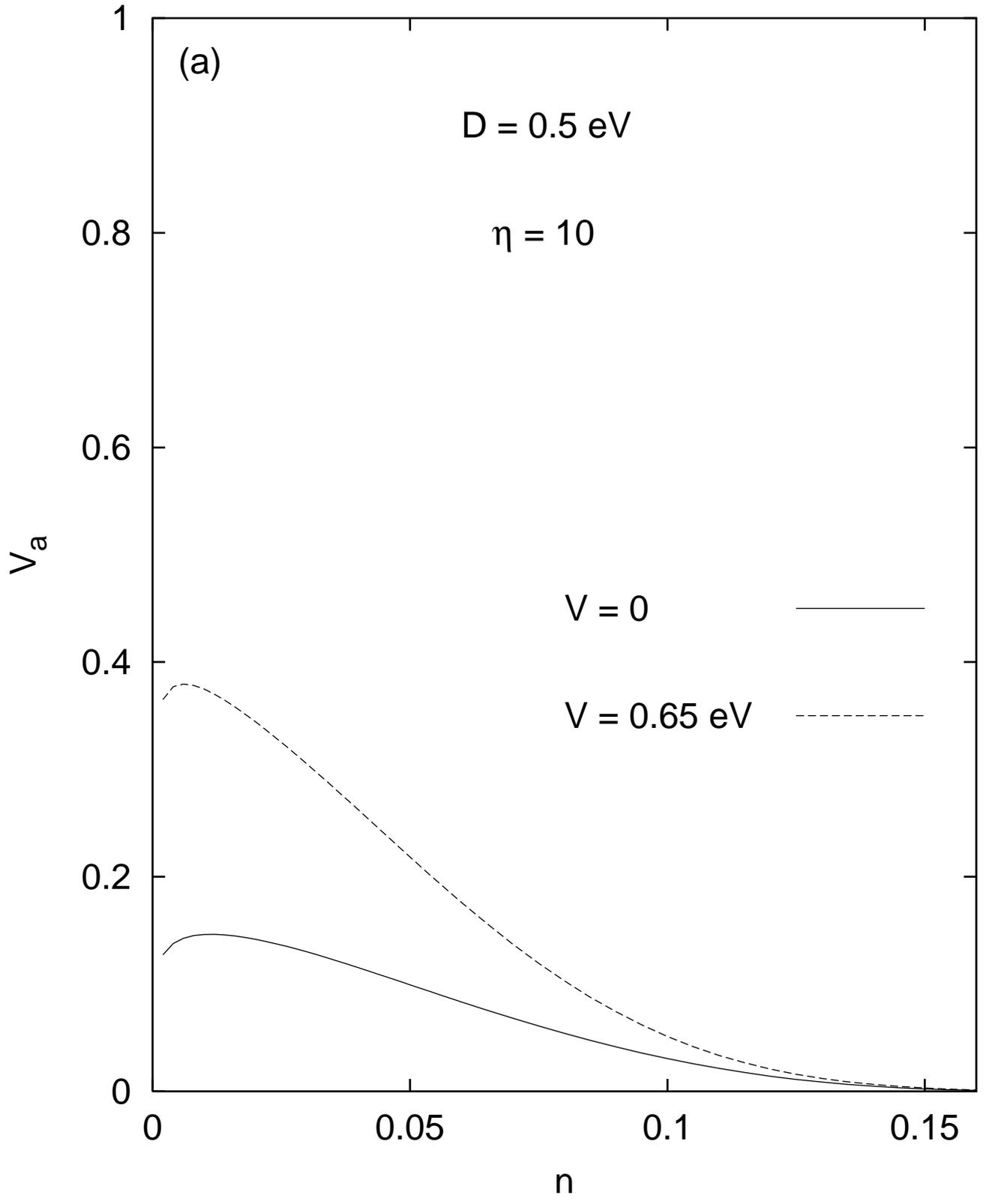
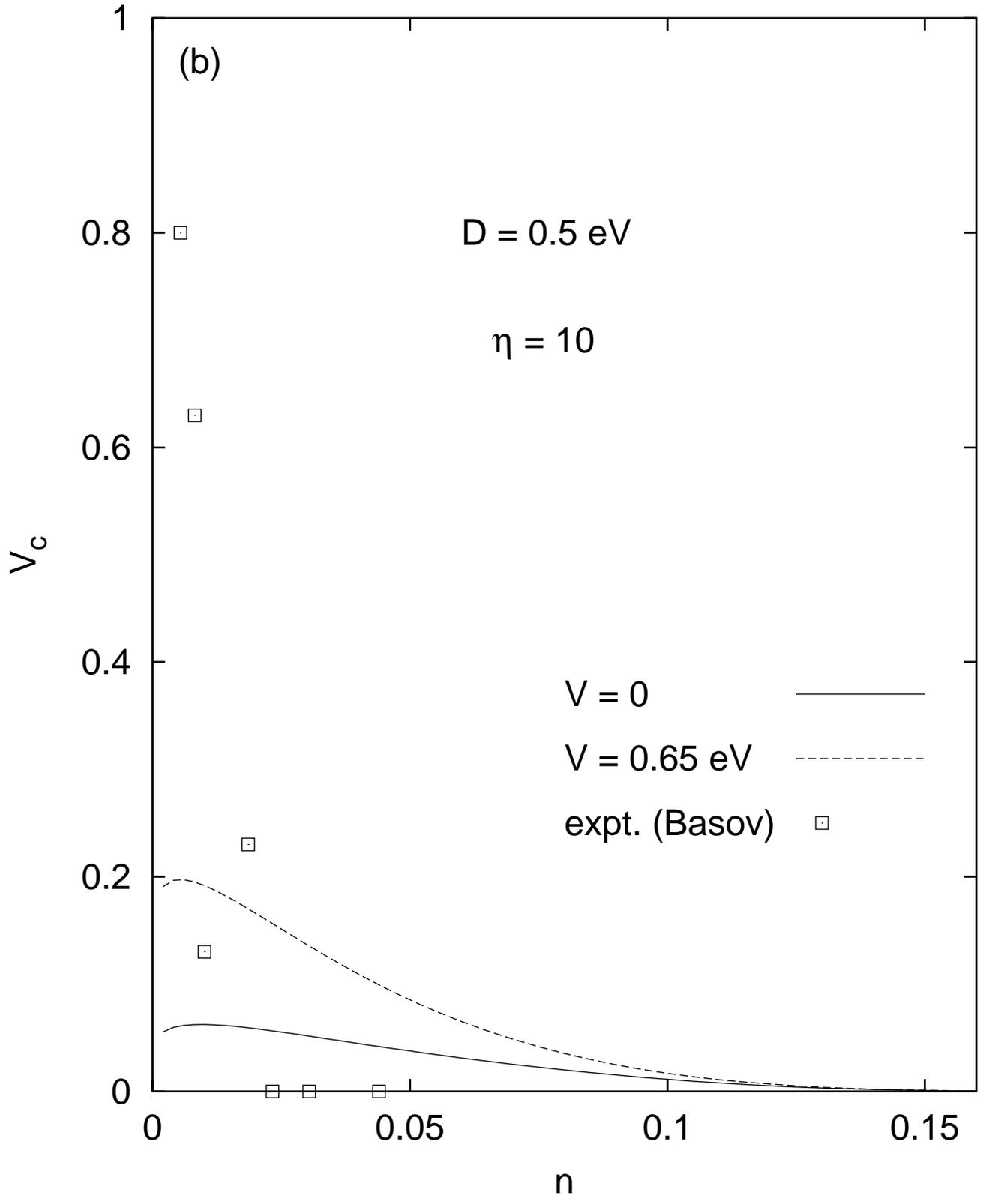
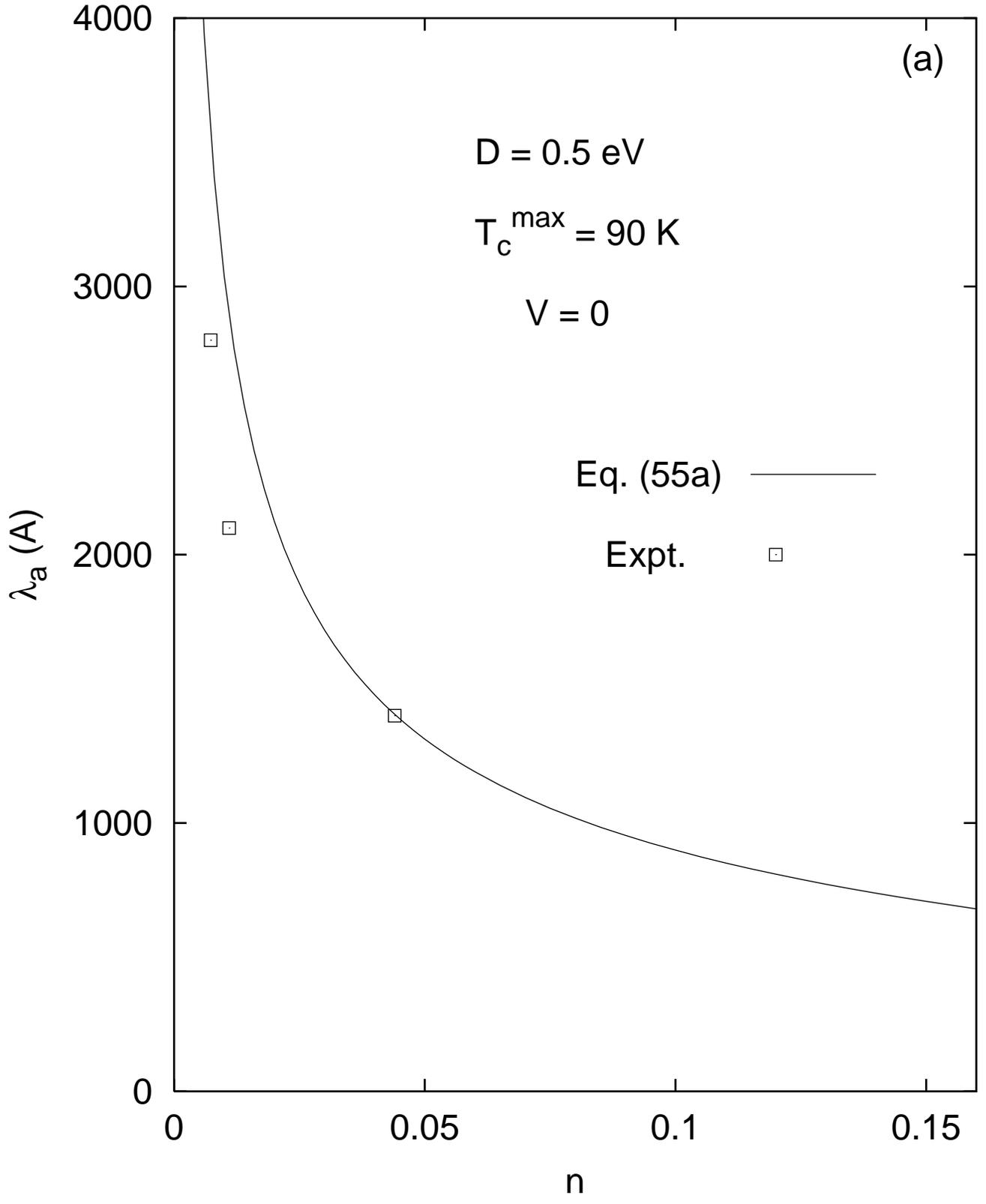
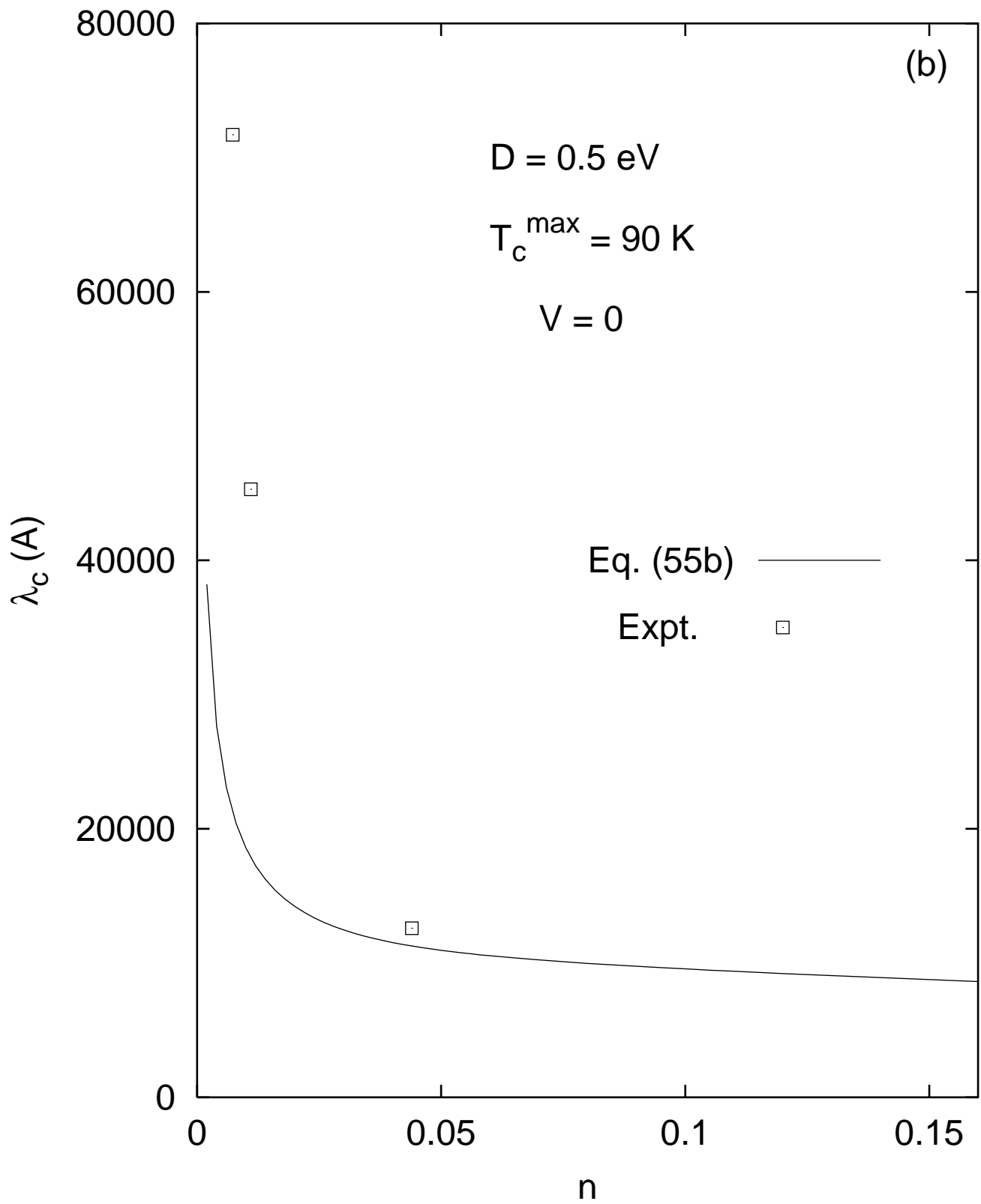
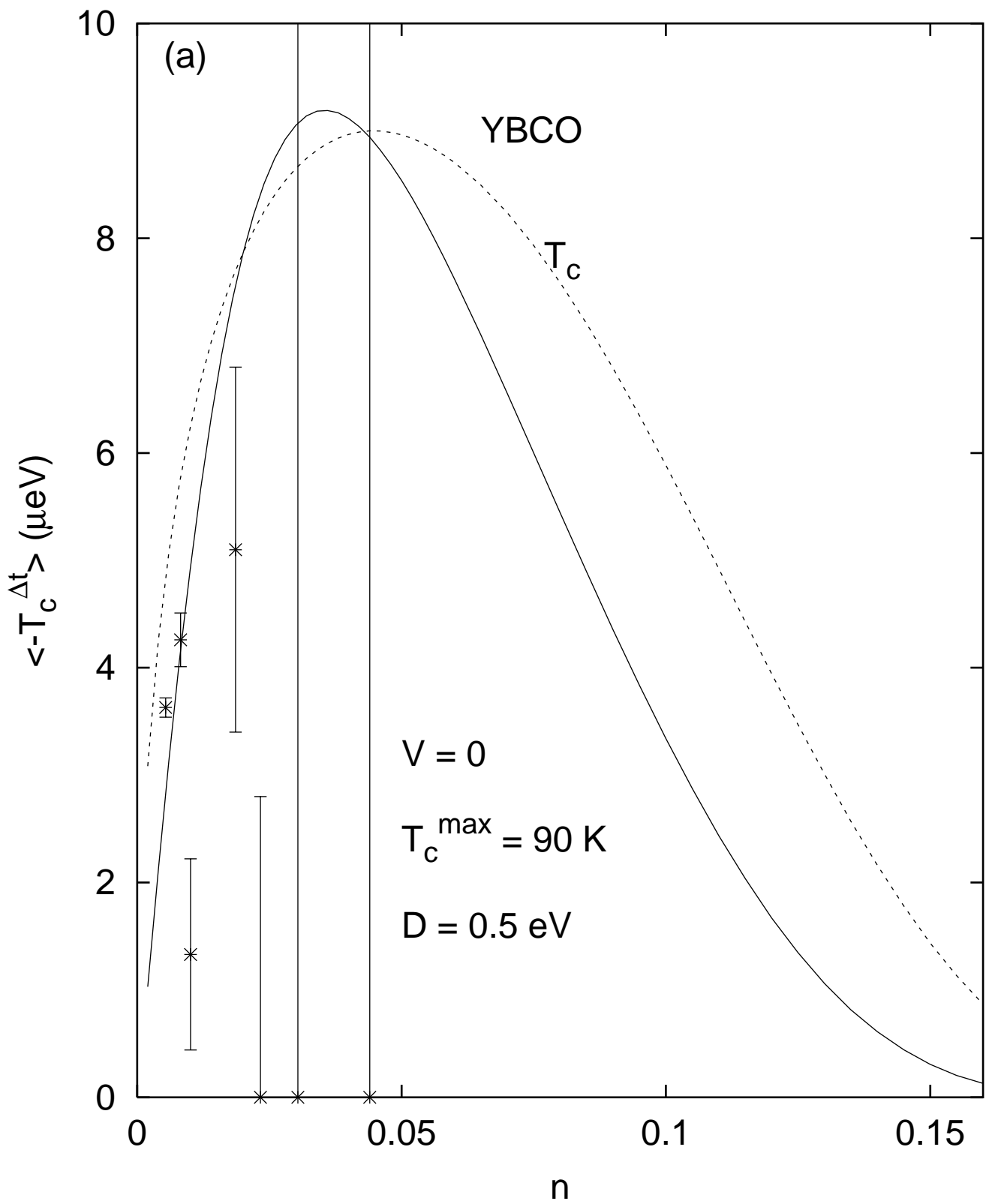
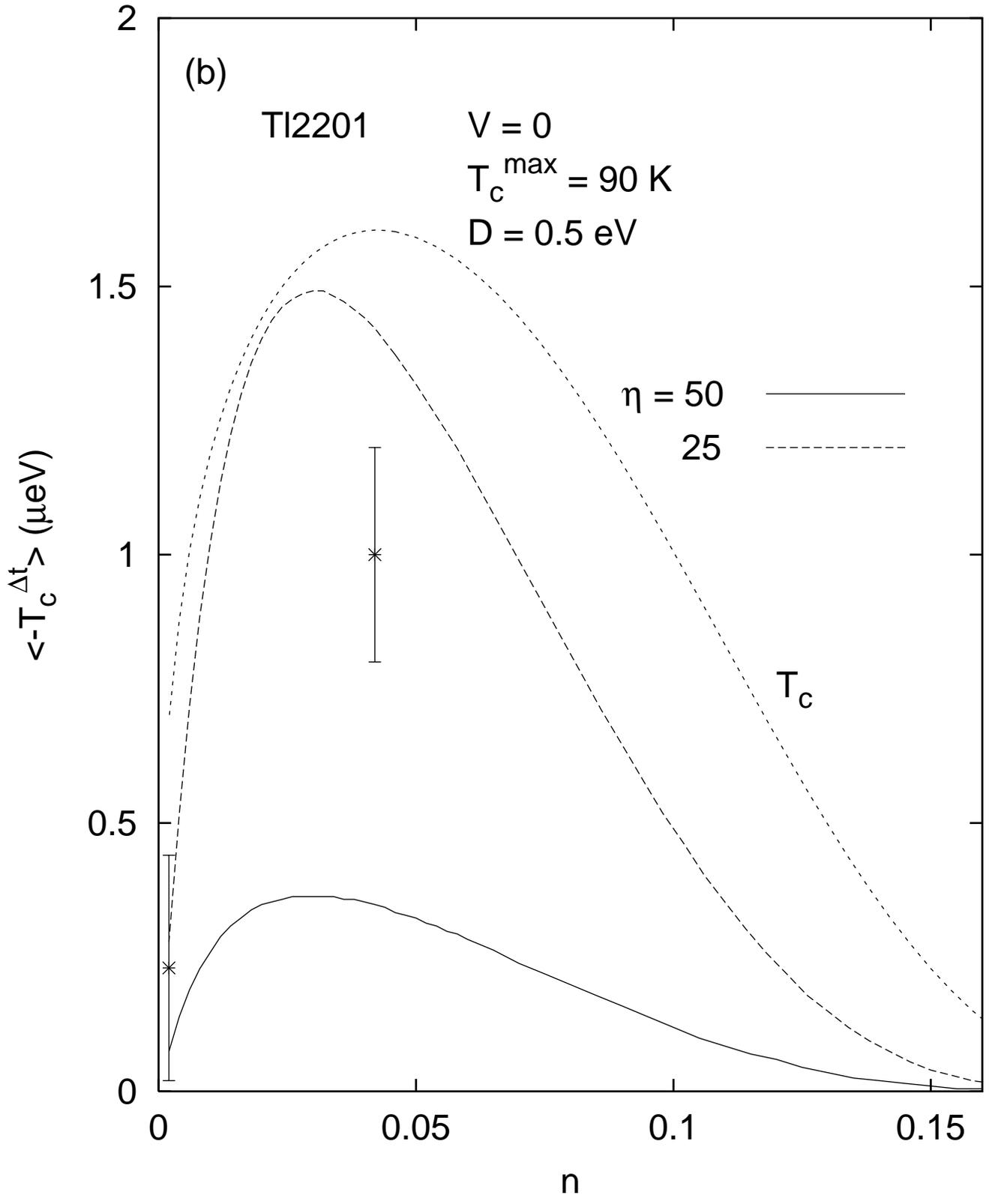
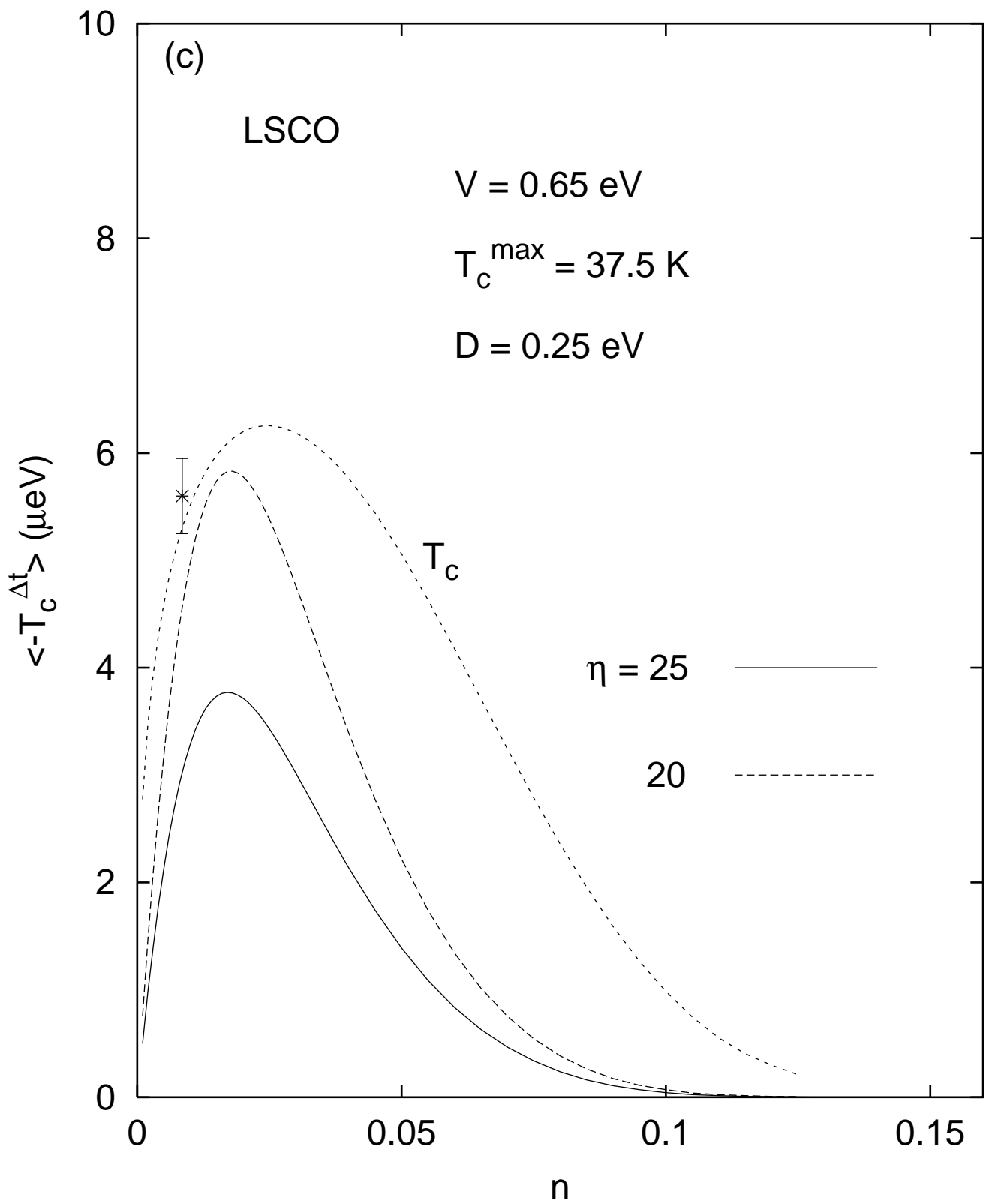
\end{document}